\documentclass[aps,prb,twocolumn]{revtex4}
\usepackage{bm}
\usepackage{tabularx}
\usepackage{ascmac}
\usepackage{amsmath}
\usepackage{wrapfig}
\usepackage{graphicx}
\usepackage{float}
\usepackage[FIGBOTCAP]{subfigure}
\usepackage{color}

\bmdefine{\bolds}{s}
\bmdefine{\boldi}{i}
\bmdefine{\boldj}{j}
\bmdefine{\boldtau}{\tau}
\bmdefine{\boldsigma}{\sigma}
\bmdefine{\boldl}{l}
\bmdefine{\boldlambda}{\lambda}
\bmdefine{\boldx}{x}
\bmdefine{\boldX}{X}
\bmdefine{\boldk}{k}
\bmdefine{\boldK}{K}
\bmdefine{\boldq}{q}
\bmdefine{\boldQ}{Q}
\bmdefine{\boldr}{r}
\bmdefine{\boldj}{j}
\bmdefine{\boldzero}{0}
\begin{document}

%\preprint{}

\title{
Fermi surface versus Fermi sea contributions 
to intrinsic anomalous and spin Hall effects 
of multiorbital metals in the presence of 
Coulomb interaction and spin-Coulomb drag
}


\author{Naoya Arakawa}
\email{arakawa@hosi.phys.s.u-tokyo.ac.jp} 
\affiliation{
Center for Emergent Matter Science (CEMS), 
RIKEN, Wako, Saitama 351-0198, Japan}
%\thanks{}
%\altaffiliation{a}


\begin{abstract}
Anomalous Hall effect (AHE) and spin Hall effect (SHE) 
are fundamental phenomena, 
and their potential for application is great. 
However, we understand the interaction effects unsatisfactorily, 
and should have clarified issues about 
the roles of the Fermi sea term and Fermi surface term 
of the conductivity of the intrinsic AHE or SHE 
of an interacting multiorbital metal 
and about 
the effects of spin-Coulomb drag on the intrinsic SHE. 
Here we resolve the first issue and 
provide the first step about the second issue 
by developing a general formalism in the linear response theory 
with appropriate approximations 
and using analytic arguments. 
The most striking result 
is that 
even without impurities 
the Fermi surface term, 
a non-Berry-curvature term, 
plays dominant roles at high or slightly low temperatures. 
In particular, 
this Fermi surface term causes 
the temperature dependence of 
the dc anomalous Hall or spin Hall conductivity 
due to the interaction-induced quasiparticle damping 
and the correction of the dc spin Hall conductivity 
due to the spin-Coulomb drag. 
Those results revise our understanding of the intrinsic AHE and SHE. 
We also find that 
the differences between the dc anomalous Hall and longitudinal conductivities 
arise from the difference in the dominant multiband excitations. 
This not only explains why the Fermi sea term such as the Berry-curvature term 
becomes important in clean and low-temperature case only for interband transports 
but also provides the useful principles on treating the electron-electron interaction 
in an interacting multiorbital metal for general formalism of transport coefficients. 
Several correspondences between our results and experiments 
are finally discussed.  
\end{abstract}

\date{\today}
\maketitle


\section{Introduction}
Anomalous Hall effect (AHE) and spin Hall effect (SHE) 
are fundamental phenomena and 
have great potential for application. 
The AHE~\cite{AHE-exp-review,Karplus-Luttinger,
Smit,AnomalousNernst,AHE-review-Nagaosa} 
causes a charge current perpendicular to an external electric field 
even without an external magnetic field, 
and its spin-current version is 
the SHE~\cite{Dyakonov-Perel,Hirsch-SHE,Murakami-SHE,Sinova-SHE,Kato-SHE,Saitoh-SHE,
Tinkham-SHE,SHE-review-Sinova}. 
Since the AHE and SHE are similar to usual Hall effect~\cite{Ashcroft-Mermin}, 
an understanding of their properties develops 
our fundamental understanding of transport phenomena. 
Then, since we can control the magnitude and direction of the charge current of the AHE and 
spin current of the SHE in principle,
the AHE and SHE may be utilized 
as useful devices~\cite{SHE-review-Science,SHE-review-Nature}. 

For the fundamental understanding and efficient utilization of 
the AHE or SHE, 
we need to understand how its response depends on 
the detail of the electronic structure. 
Since the response may be affected by the differences 
in the band structure, 
the structure of doped impurities, 
and the strength of the electron-electron interaction, 
an understanding of their dependence of the response is helpful 
to understand the fundamental properties and find a good material for application.  

The previous studies partially revealed the dependence of the response 
of the AHE or SHE on the detail of the electronic structure, 
and showed the potential of the intrinsic mechanism for a large response. 
First, 
the mechanisms of the AHE or SHE are categorized as 
either an intrinsic mechanism to the band structure~\cite{Karplus-Luttinger,
Murakami-SHE,Sinova-SHE,Kontani-AHE,SrRuO3-AHE,Kontani-SHE,SHE-Pt-Nagaosa,Mizoguchi-SHE} 
or an extrinsic mechanism due to the scattering of 
doped impurities~\cite{Smit,Dyakonov-Perel,Hirsch-SHE,Berger,Extrinsic}. 
Then, 
we can understand 
the intrinsic mechanisms for a lot of metals 
as acquiring the Aharanov-Bohm-type phase factor~\cite{JJSakurai} 
by using the onsite spin-orbit coupling (SOC) and 
several hopping integrals~\cite{Kontani-OrbitalAB} 
(for more details see Appendix A). 
On the other hand, 
we can understand several extrinsic mechanisms~\cite{Smit,Berger,Extrinsic} 
by considering a special scattering of doped nonmagnetic impurities. 
However, 
if their onsite scattering potential is small and 
the intrinsic term is non-negligible, 
the extrinsic term is less important than the intrinsic term. 
Actually, 
the extrinsic term completely vanishes 
in even-parity systems for the weak onsite scattering potential 
of dilute nonmagnetic impurities~\cite{Kontani-AHE,Kontani-SHE}. 
Furthermore, 
even in the absence of the inversion symmetry at an $ab$-plane, 
the extrinsic term remains very small 
if orbital degrees of freedom exist and 
the hopping induced by the inversion-symmetry breaking is not large~\cite{Mizoguchi-CVC}. 
Since a lot of multiorbital metals have 
finite intrinsic terms~\cite{Kontani-AHE,SrRuO3-AHE,Kontani-SHE,Kontani-OrbitalAB,Mizoguchi-SHE} 
and 
the typical value of the scattering potential estimated 
in a first-principle calculation~\cite{imp-1stprinciple} is 
of the order of magnitude $0.1$ eV, 
we may sufficiently analyze 
the AHE or SHE of a multiorbital metal 
by considering only the intrinsic mechanism. 
Actually, a systematic theoretical study~\cite{Kontani-SHE} about the intrinsic SHE 
can qualitatively reproduce 
a chemical trend of the experimental responses~\cite{SHE-sytematic-exp} 
in several 4$d$- or 5$d$-transition metals. 
Since 
a multiorbital metal is more suitable than a semiconductor 
to obtain a large response~\cite{Kimura-SHE}, 
a theoretical research on the intrinsic AHE or SHE of a multiorbital metal 
may develop our fundamental understanding and the possibilities of application. 

However, 
we have two issues about interaction effects, 
the effects of the electron-electron interaction, 
in the intrinsic AHE and SHE of a multiorbital metal. 

One is to clarify roles of the Fermi surface term 
and Fermi sea term of $\sigma_{xy}^{\textrm{C}}$ or $\sigma_{xy}^{\textrm{S}}$,  
the intrinsic anomalous Hall or spin Hall conductivity, 
in the presence of the electron-electron interaction. 
Let us begin with noninteracting case 
with the weak onsite scattering potential of dilute nonmagnetic impurities 
at zero temperature. 
In that case, 
$\sigma_{xy}^{\textrm{C}}$ or $\sigma_{xy}^{\textrm{S}}$ 
consists of the Fermi surface term and Fermi sea term 
in general~\cite{Kontani-AHE,Kontani-SHE,Streda}. 
The Fermi surface term describes the excitations near the Fermi level, 
and the Fermi sea term describes the excitations in the Fermi sea. 
Then, those terms are affected by the nonmagnetic impurity scattering 
through changing the quasiparticle (QP) damping 
in $\sigma_{xy}^{\textrm{C}}$ or $\sigma_{xy}^{\textrm{S}}$ 
even if the extrinsic term is negligible~\cite{Kontani-AHE,Kontani-SHE}. 
If that QP damping goes to zero, 
$\sigma_{xy}^{\textrm{C}}$ or $\sigma_{xy}^{\textrm{S}}$ is given by 
the Berry-curvature term~\cite{SrRuO3-AHE,SHE-Pt-Nagaosa,Onoda-Nagaosa}, 
part of the Fermi sea term~\cite{Kontani-AHE,Kontani-SHE}, 
because of the cancellation between the other part of the Fermi sea term 
and the Fermi surface term~\cite{Kontani-AHE,Kontani-SHE}. 
As the QP damping increases due to 
an increase of the impurity concentration $n_{\textrm{imp}}$, 
the dominant term of $\sigma_{xy}^{\textrm{C}}$ or $\sigma_{xy}^{\textrm{S}}$ 
becomes the Fermi surface term because of the cancellation between 
the two parts of the Fermi sea term~\cite{Kontani-AHE,Kontani-SHE}. 
This Fermi surface term qualitatively differs from 
the Berry-curvature term 
because only the former contains a retarded-advanced product of 
two single-particle Green's functions~\cite{Kontani-AHE,Kontani-SHE} 
[for the explicit comparison, for example, 
see Eqs. (\ref{eq:SigXYC0-I}) and (\ref{eq:SigXYC0-II})]. 
Thus, only the Berry-curvature term is insufficient, 
and the Fermi surface term and Fermi sea term play important roles 
in discussing the intrinsic AHE or SHE of a noninteracting multiorbital metal. 
However, for discussions at finite temperatures, 
we should consider the electron-electron interaction 
because that may affect 
$\sigma_{xy}^{\textrm{C}}$ or $\sigma_{xy}^{\textrm{S}}$ 
through the inelastic scattering. 
Thus, 
it remains a challenging issue to clarify 
the roles of the Fermi surface term and Fermi sea term 
in an interacting multiorbital metal. 
Although this issue was discussed by Haldane~\cite{Haldane-AHE}, 
his proposal~\cite{Haldane-AHE} did not resolve this 
because he assumed that 
only the Berry-curvature term is always dominant and 
did not analyze the roles of the non-Berry-curvature terms; 
his proposal is that part of the partial-integral term of the Berry-curvature term 
corresponds to the Fermi surface term which 
plays important roles in the Fermi liquid. 
Thus, we need to discuss this issue in a more elaborated method. 

\begin{figure}[tb]
\includegraphics[width=42mm]{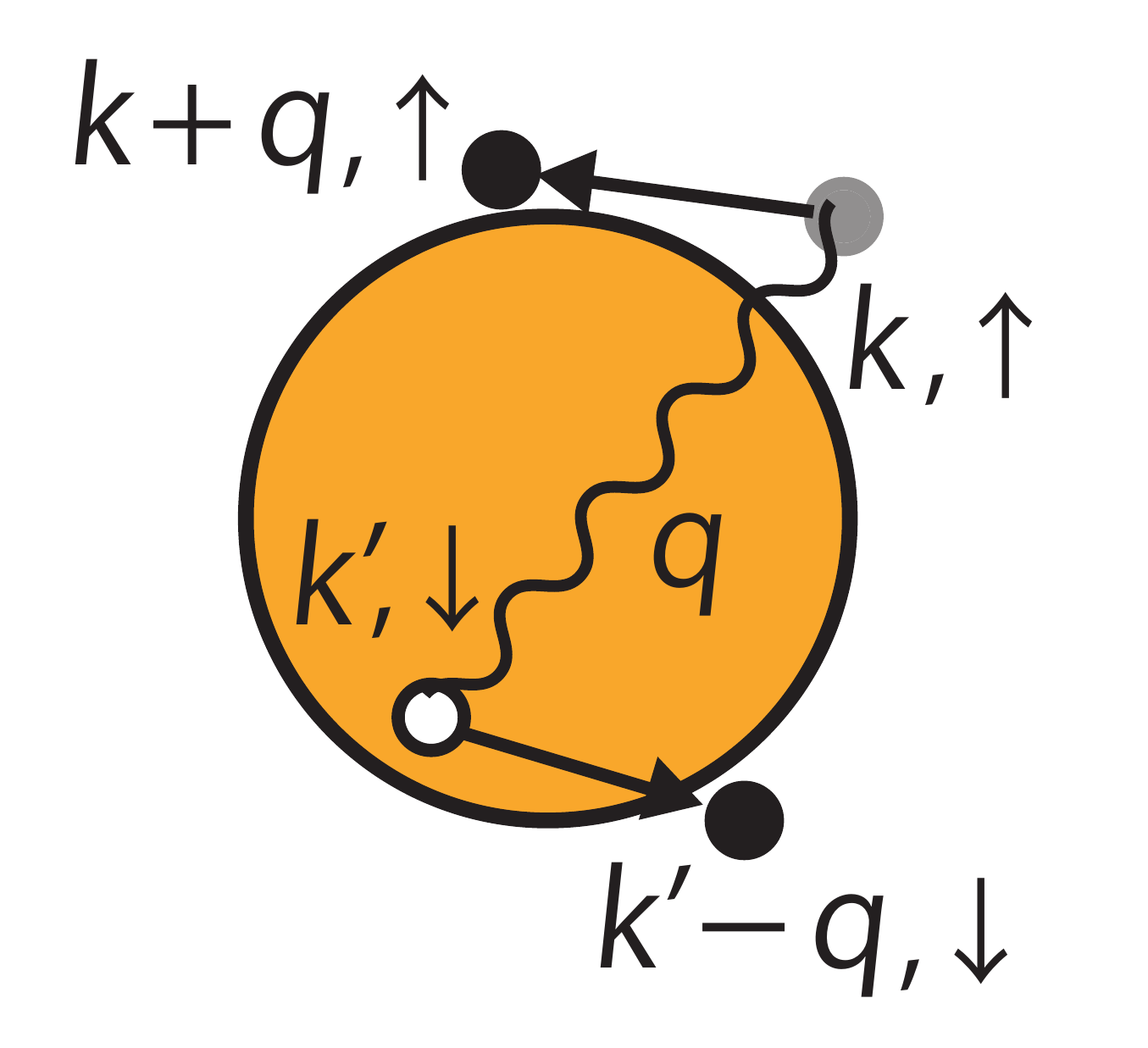}
\vspace{-6pt}
\caption{Schematic picture of the scattering between 
spin-up and spin-down electrons 
due to the electron-electron interaction with momentum transfer $\boldq$. 
The wavy line represents the electron-electron interaction, 
the black circles represent the electrons after the scattering, 
and the yellow circle represents the Fermi sphere. 
This scattering conserves the sum of the total momentums 
of the spin-up and the spin-down electrons 
(i.e., $\boldk+\boldk^{\prime}=\boldk+\boldq+\boldk^{\prime}-\boldq$), 
while the conservation of each total momentum is violated for $\boldq\neq \boldzero$ 
(i.e., $\boldk\neq \boldk+\boldq$ and $\boldk^{\prime}\neq \boldk^{\prime}-\boldq$).  
}
\label{fig:Fig1}
\end{figure}

The other issue is to clarify effects of spin-Coulomb drag (SCD) 
on the intrinsic SHE. 
If the electron-electron interaction causes the scattering between 
spin-up and spin-down electrons with finite momentum transfer, 
the spin-up and the spin-down component of the total momentum 
are not separately conserved~\cite{SCD-first,SCD-review} (see Fig. \ref{fig:Fig1}). 
This indicates the existence of the friction between 
spin-up and spin-down electrons, the SCD, 
even without the Umklapp scattering~\cite{SCD-first,SCD-review} 
because 
the momentum conservation results in 
the absence of the friction~\cite{Ziman,Ziman2}. 
This is in contrast to case of the charge current 
because in that case 
the Umklapp scattering is essential to obtain the friction, 
which results in the finite resistivity~\cite{Ziman,Ziman2,Yamada-Yosida}. 
Thus, 
the existence of the SCD is an important difference between 
spin transports and charge transports. 
Then, 
the SCD causes a correction~\cite{SCD-first,SCD-review}, 
which is different from the mass enhancement and Fermi-liquid correction, 
and that effect on the spin-diffusion constant is experimentally observed 
in a two-dimensional electron gas~\cite{SCD-exp}. 
In principle, 
the SCD may affect the intrinsic SHE~\cite{SCD-review}, 
and its effects may lead to some differences between the SHE and AHE. 
Furthermore, 
since in contrast to an electron gas 
a multiorbital metal has a multiband structure, 
an interacting multiorbital metal may be a good target 
to deduce multiband effects in the SCD. 
However, 
the effects of the SCD on the intrinsic SHE have not been studied 
and remain unclear~\cite{SCD-review}. 

To improve this situation, 
we develop a general formalism of $\sigma_{xy}^{\textrm{C}}$ or $\sigma_{xy}^{\textrm{S}}$ 
of an interacting multiorbital metal 
using the linear response theory~\cite{Kubo-formula} 
with approximations appropriate for such metal, 
clarify the roles of the Fermi surface term and Fermi sea term
and find a SCD-induced correction of $\sigma_{xy}^{\textrm{S}}$. 
The former result resolves the first issue, 
and the latter provides the first step towards 
the complete resolution of the second issue. 
In particular, 
we find an interaction-driven mechanism of 
the damping dependence of $\sigma_{xy}^{\textrm{C}}$ or $\sigma_{xy}^{\textrm{S}}$ 
and crossover from damping-dependent to damping-independent intrinsic AHE or SHE. 
This highlights the emergence of the temperature dependence 
in high-temperature region of the intrinsic AHE or SHE 
even for clean systems. 
We also propose several experiments related to those results. 
Then, 
we clarify the origin of the differences between $\sigma_{xy}^{\textrm{C}}$
and the longitudinal conductivity, $\sigma_{xx}^{\textrm{C}}$, 
and deduce the general principles 
in the formulations of transport coefficients 
including the interaction and the multiband effects. 
This origin is helpful to understand 
why the Fermi sea term such as the Berry-curvature term sometimes becomes 
important only for the interband transports such as the AHE, 
although only the Fermi surface term is always important 
for the intraband transports such as the resistivity. 
In addition, 
the obtained principles help guide further research of transports 
including the interaction effects and the multiband effects. 

\section{Method}
In this section, 
we explain the method to analyze the intrinsic AHE and SHE 
of an interacting multiorbital metal. 
First, we show the Hamiltonian of our model, 
and argue its validity for their realistic analysis. 
Second, we explain how to treat each term of the Hamiltonian, 
and deduce several consequences of this treatment about 
the self-energy, the QP damping, 
and the irreducible four-point vertex function. 
Third, 
we show the exact expressions of $\sigma_{xy}^{\textrm{C}}$ and $\sigma_{xy}^{\textrm{S}}$ 
within the linear response of an external electric field. 
In addition, we explain several advantages of the linear response theory 
and an important remark about taking the limits 
such as $\lim_{\omega\rightarrow 0}$ and $\lim_{\boldq\rightarrow \boldzero}$. 
In part of the derivations of those exact expressions, 
we use Appendix B. 

Hereafter we set $\hbar=c=k_{\textrm{B}}=1$. 

\subsection{Model}

We consider a $d$-orbital Hubbard model~\cite{MultiHubbard-Yanase} 
with the onsite SOC~\cite{Kontani-SHE} and 
the weak onsite scattering potential~\cite{Kontani-SHE} of dilute nonmagnetic impurities. 
Its Hamiltonian consists of four terms: 
\begin{align}
\hat{H}
=&\ \hat{H}_{0}+\hat{H}_{\textrm{LS}}+\hat{H}_{\textrm{int}}+\hat{H}_{\textrm{imp}}.\label{eq:H}
\end{align}
First, 
$\hat{H}_{0}$ represents the nonrelativistic noninteracting terms, 
\begin{align}
\hat{H}_{0}
=\sum\limits_{\boldk}
\sum\limits_{a,b}\sum\limits_{s=\uparrow,\downarrow} 
\epsilon_{ab}(\boldk)
\hat{c}^{\dagger}_{\boldk a s}\hat{c}_{\boldk b s},\label{eq:H0}
\end{align}
where $\hat{c}^{\dagger}_{\boldk a s}$ and $\hat{c}_{\boldk a s}$ 
are a creation and an annihilation operator, respectively, 
of an electron at momentum $\boldk$, orbital $a$, and spin $s$, 
and $\epsilon_{ab}(\boldk)$ is the noninteracting energy dispersion 
measuring from the chemical potential. 
Second, 
$\hat{H}_{\textrm{LS}}$ represents the onsite SOC~\cite{Kontani-SHE},
\begin{align}
\hat{H}_{\textrm{LS}}
=\xi_{\textrm{LS}}\sum\limits_{\boldj} 
\hat{\boldl}_{\boldj}\cdot \hat{\bolds}_{\boldj},\label{eq:HLS} 
\end{align}
where 
$\boldj$ is site index, 
$\hat{\boldl}_{\boldj}$ and $\hat{\bolds}_{\boldj}$ are 
an orbital and a spin angular momentum operator, respectively~\cite{Kontani-SHE}, 
and $\xi_{\textrm{LS}}$ is the coupling constant. 
Third, 
$\hat{H}_{\textrm{int}}$ represents 
the onsite multiorbital Hubbard interaction terms~\cite{MultiHubbard-Yanase}, 
\begin{align}
\hat{H}_{\textrm{int}}
=&\
 U 
\sum\limits_{\boldj}
\sum\limits_{a}
\hat{n}_{\boldj a \uparrow} \hat{n}_{\boldj a \downarrow}
+ U^{\prime}  
\sum\limits_{\boldj}
\sum\limits_{a}
\sum\limits_{b<a}
\hat{n}_{\boldj a} \hat{n}_{\boldj b}\notag\\
&- 
J_{\textrm{H}} 
\sum\limits_{\boldj}
\sum\limits_{a}
\sum\limits_{b<a}
( 
2 \hat{\bolds}_{\boldj a} \cdot 
\hat{\bolds}_{\boldj b} 
+ 
\frac{1}{2} \hat{n}_{\boldj a} \hat{n}_{\boldj b} 
)\notag\\
&+
J^{\prime} 
\sum\limits_{\boldj}
\sum\limits_{a}
\sum\limits_{b\neq a}
\hat{c}_{\boldj a \uparrow}^{\dagger} 
\hat{c}_{\boldj a \downarrow}^{\dagger} 
\hat{c}_{\boldj b \downarrow} 
\hat{c}_{\boldj b \uparrow},\label{eq:Hint}
\end{align}
where 
$\hat{n}_{\boldj a s}$ is $\hat{n}_{\boldj a s}=\hat{c}^{\dagger}_{\boldj a s}\hat{c}_{\boldj a s}$, 
$\hat{n}_{\boldj a}$ is $\hat{n}_{\boldj a}=\sum_{s} \hat{n}_{\boldj a s}$, 
$\hat{\bolds}_{\boldj a}$ is 
$\hat{\bolds}_{\boldj a}=\frac{1}{2}\sum_{s,s^{\prime}}\hat{c}^{\dagger}_{\boldj a s}
\boldsigma_{s,s^{\prime}}\hat{c}_{\boldj a s^{\prime}}$ 
with the Pauli matrices $\boldsigma_{s,s^{\prime}}$, 
$U$ is the intraorbital Coulomb interaction, 
$U^{\prime}$ is the interorbital Coulomb interaction, 
$J_{\textrm{H}}$ is the Hund's rule coupling, 
and $J^{\prime}$ is the pair hopping term.
Fourth, 
$\hat{H}_{\textrm{imp}}$ represents the onsite scattering potential 
of dilute nonmagnetic impurities~\cite{Kontani-SHE},
\begin{align}
\hat{H}_{\textrm{imp}}
=I_{\textrm{imp}}
\sum\limits_{\boldj}
\sum\limits_{a}
\sum\limits_{s}
\hat{c}^{\dagger}_{\boldj a s}\hat{c}_{\boldj a s},\label{eq:Himp} 
\end{align}
where $I_{\textrm{imp}}$ is the potential amplitude. 

This model is sufficient 
for a realistic analysis of the intrinsic AHE and SHE of an interacting metal 
because of the following four reasons.  
First, 
we can choose any form of $\epsilon_{ab}(\boldk)$ 
if $\epsilon_{ab}(\boldk)$ contains the interorbital hopping 
whose mirror symmetries for a $xz$ and a $yz$ plane are odd 
[e.g., the next-nearest-neighbor hopping between the $d_{yz}$ and $d_{xz}$ orbitals 
in Fig. \ref{fig:Supp-Fig1}(a) of Appendix A]; 
as we will see in Sec. III A 2, 
such interorbital hopping is necessary 
to obtain finite $\sigma_{xy}^{\textrm{C}}$ or 
$\sigma_{xy}^{\textrm{S}}$~\cite{Kontani-AHE,Kontani-OrbitalAB}. 
Second, 
among several possibilities of the SOCs, 
only the onsite SOC is sufficient 
because its effect is leading in a solid 
and because 
we can analyze the intrinsic AHE or SHE of a metal 
even without the inversion symmetry at an $ab$ plane 
by not using the nonlocal SOC~\cite{Mizoguchi-SHE,Mizoguchi-CVC}; 
the effect of that inversion symmetry breaking can be included 
in $\epsilon_{ab}(\boldk)$~\cite{Yanase-Rashba}.  
Third, 
we may sufficiently describe 
the screened short-ranged electron-electron interaction 
in an interacting multiorbital metal 
by the onsite multiorbital Hubbard interactions 
because those interactions have not only the intraorbital term 
but also the interorbital terms;  
our formalism can be easily extended if the interactions are short-ranged. 
Fourth, 
$\hat{H}_{\textrm{imp}}$ may be sufficient to include 
effects of dilute nonmagnetic impurities, 
which exist in a realistic situation, 
because 
the effects can be roughly described by 
their weak onsite scattering potential~\cite{AGD}. 

\subsection{Treatment of each Hamiltonian}

To analyze the intrinsic AHE or SHE of an interacting multiorbital metal, 
we use $\hat{H}_{0}+\hat{H}_{\textrm{LS}}$ as the nonperturbative Hamiltonian, 
\begin{align}
\hat{H}_{0}+\hat{H}_{\textrm{LS}}
=\sum\limits_{\boldk}
\sum\limits_{a,b}\sum\limits_{s,s^{\prime}} 
\bar{\epsilon}_{ab}^{ss^{\prime}}(\boldk)
\hat{c}^{\dagger}_{\boldk a s}\hat{c}_{\boldk b s^{\prime}},\label{eq:H0+HLS}
\end{align}
and $\hat{H}_{\textrm{int}}+\hat{H}_{\textrm{imp}}$ as the perturbative Hamiltonian. 
In particular, 
for a simple treatment of $\hat{H}_{\textrm{imp}}$, 
we assume both that nonmagnetic impurities are randomly distributed 
and that  
$I_{\textrm{imp}}$ is smaller than the bandwidth 
so as to satisfy $k_{\textrm{F}}l \gg 1$ (i.e., case away from the Mott-Ioffe-Regel limit), 
where $k_{\textrm{F}}$ is of the order of magnitude the Fermi momentum and 
$l$ is the mean free path. 
The first assumption is standard~\cite{AGD,Kontani-AHE,Kontani-SHE}, and 
the second is reasonable in several transition metals or transition-metal oxides. 
Then, 
because of the first assumption, 
we can use the averaging over each impurity position~\cite{AGD}; 
because of the second, 
we can neglect the combination terms of $\hat{H}_{\textrm{int}}$ and $\hat{H}_{\textrm{imp}}$ 
in the self-energy 
and sufficiently treat $\hat{H}_{\textrm{imp}}$ 
in the Born approximation~\cite{Kontani-review}. 

In this treatment, 
we can use simple treatments about several quantities. 
First, 
the retarded self-energy is given by 
\begin{align}
\Sigma_{ab}^{ss^{\prime}}(\tilde{k})
=\Sigma_{\textrm{el-el};ab}^{ss^{\prime}}(\tilde{k})
+\frac{n_{\textrm{imp}}I_{\textrm{imp}}^{2}}{N}
\sum\limits_{\boldk^{\prime}} G_{ab}^{ss^{\prime}}(\boldk^{\prime},i\epsilon_{m}),
\end{align}
where 
$\tilde{k}$ is $\tilde{k}\equiv (\boldk,i\epsilon_{m})$, 
$\Sigma_{\textrm{el-el};ab}^{ss^{\prime}}(\tilde{k})$ 
is the self-energy arising from $\hat{H}_{\textrm{int}}$ 
in the perturbation theory, 
and the second term is the self-energy arising from $\hat{H}_{\textrm{imp}}$ 
in the Born approximation~\cite{AGD,Kontani-SHE} 
with $N$, the number of lattice sites. 
Correspondingly, 
we obtain the QP damping arising from $\hat{H}_{\textrm{int}}$ and $\hat{H}_{\textrm{imp}}$ 
because the QP damping is defined as 
\begin{align}
\gamma_{\alpha}^{\ast}(\boldk)
=-z_{\alpha}(\boldk)\textrm{Im}\Sigma_{\alpha}^{(\textrm{R})}
(\boldk,\xi_{\alpha}^{\ast}(\boldk)),\label{eq:QPdamp-eq}
\end{align}
where $\alpha$ is the band index of a QP, 
$\xi_{\alpha}^{\ast}(\boldk)$ is the QP energy determined by the solution of 
det$|\epsilon \delta_{a,b}\delta_{s,s^{\prime}}
-\bar{\epsilon}_{ab}^{ss^{\prime}}(\boldk)
-\textrm{Re}\Sigma_{ab}^{(\textrm{R})ss^{\prime}}(k)|=0$, 
$\Sigma_{\alpha}^{(\textrm{R})}(k)$ with $k\equiv (\boldk,\epsilon)$ 
is the retarded self-energy of the QP band $\alpha$, 
and $z_{\alpha}(\boldk)$ is the QP weight,
\begin{align}
z_{\alpha}(\boldk)=
[1-\frac{\partial \textrm{Re}\Sigma_{\alpha}^{(\textrm{R})}(\boldk,\epsilon)}
{\partial \epsilon}|_{\epsilon \rightarrow \xi_{\alpha}^{\ast}(\boldk)}]^{-1}.
\end{align}
In general, 
$\gamma_{\alpha}^{\ast}(\boldk)$ depends on temperature 
because of the temperature dependence of 
$\Sigma_{\textrm{el-el};ab}^{(\textrm{R})ss^{\prime}}(\boldk,\epsilon)$, 
e.g., the $T^{2}$ dependence of 
$\gamma_{\alpha}^{\ast}(\boldk)$ near $\boldk=\boldk_{\textrm{F}}$ 
in the Fermi liquid~\cite{AGD}. 
Then, the irreducible four-point vertex function 
in Matsubara-frequency representation, 
$\Gamma_{\{a\}}^{(1)\{s_{1}\}}
(\tilde{k},\tilde{k^{\prime}};\tilde{q})
\equiv 
\Gamma_{abcd}^{(1)s_{1}s_{2}s_{3}s_{4}}
(\boldk+\boldq,i\epsilon_{m+n},\boldk,i\epsilon_{m},
\boldk^{\prime}+\boldq,i\epsilon_{m^{\prime}+n},\boldk^{\prime},i\epsilon_{m^{\prime}})$, 
is given by 
\begin{align}
&\Gamma_{\{a\}}^{(1)\{s_{1}\}}
(\tilde{k},\tilde{k^{\prime}};\tilde{q})\notag\\
=&\
\frac{\delta \Sigma_{\textrm{el-el};ab}^{ss^{\prime}}(\tilde{k})}
{\delta G_{cd}^{s^{\prime\prime}s^{\prime\prime\prime}}(\tilde{k^{\prime}})}
+n_{\textrm{imp}}I_{\textrm{imp}}^{2}
\delta_{a,c}\delta_{b,d}
\delta_{s,s^{\prime\prime}}\delta_{s^{\prime},s^{\prime\prime\prime}},\label{eq:Gam1-full}
\end{align}
where the first term is the irreducible four-point vertex function 
arising from $\hat{H}_{\textrm{int}}$, and 
the second term is the irreducible four-point vertex function arising from $\hat{H}_{\textrm{imp}}$ 
in the Born approximation~\cite{Kontani-review}.  
Because of this decomposition, 
$\hat{H}_{\textrm{imp}}$ causes no correction to the charge and the spin current 
for even-parity systems, 
resulting in the disappearance of the extrinsic terms of 
the dc anomalous Hall or spin Hall conductivity 
in the similar way 
in noninteracting case~\cite{Kontani-AHE,Kontani-SHE} 
[see the sentences below Eq. (\ref{eq:alpha-Gamma})]. 

\subsection{Linear response theory}

To formulate $\sigma_{xy}^{\textrm{C}}$ and $\sigma_{xy}^{\textrm{S}}$ 
as general as possible, 
we use the linear response theory~\cite{Kubo-formula}. 
This is 
because the linear response theory provides 
exact expressions of $\sigma_{xy}^{\textrm{C}}$ and $\sigma_{xy}^{\textrm{S}}$ 
within the linear response of an external electric field 
and because that theory with appropriate approximations 
has several advantages compared with the phenomenological theory. 

We can derive an exact expression of $\sigma_{xy}^{\textrm{C}}$ 
within the linear response of an external electric field from 
the Kubo formula~\cite{Kubo-formula} 
for the charge current perpendicular to it 
without an external magnetic field: 
\begin{align}
\sigma_{xy}^{\textrm{C}}
=&
\lim\limits_{\omega\rightarrow 0}
\lim\limits_{\boldq\rightarrow \boldzero}
\dfrac{\tilde{K}_{xy}^{\textrm{C}(\textrm{R})}(\boldq,\omega)
-\tilde{K}_{xy}^{\textrm{C}(\textrm{R})}(\boldq,0)}{i\omega},\label{eq:sigXY-C}
\end{align}
where $\tilde{K}_{xy}^{\textrm{C}(\textrm{R})}(\omega)\equiv 
\tilde{K}_{xy}^{\textrm{C}(\textrm{R})}(\boldzero,\omega)$ is
obtained by the analytic continuation of $\tilde{K}_{xy}^{\textrm{C}}(i\Omega_{n})$ 
with bosonic Matsubara frequency $\Omega_{n}=2\pi T n$: 
\begin{align}
\tilde{K}_{xy}^{\textrm{C}(\textrm{R})}(\omega)=
\tilde{K}_{xy}^{\textrm{C}}(i\Omega_{n})|_{i\Omega_{n} \rightarrow \omega+i0+},\label{eq:K-analytic}
\end{align} 
with 
\begin{align}
\tilde{K}_{xy}^{\textrm{C}}(i\Omega_{n})
&=
\dfrac{1}{N}
\lim\limits_{\boldq\rightarrow \boldzero}
\int^{T^{-1}}_{0}
\hspace{-14pt}d\tau e^{i\Omega_{n}\tau}
\langle \textrm{T}_{\tau}  
\hat{J}_{\boldq x}^{\textrm{C}}(\tau)
\hat{J}_{-\boldq y}^{\textrm{C}}(0)\rangle\notag\\
&= 
\frac{1}{N}
\sum\limits_{\boldk,\boldk^{\prime}}
\sum\limits_{\{a\}}
\sum\limits_{\{s\}}
\int^{T^{-1}}_{0}
\hspace{-10pt}d\tau e^{i\Omega_{n}\tau}
(-e)\delta_{s^{\prime},s}
(v_{\boldk x})_{ba}^{ss}\notag\\
&\times 
(-e)\delta_{s^{\prime\prime},s^{\prime\prime\prime}}
(v_{\boldk^{\prime} y})_{cd}^{s^{\prime\prime}s^{\prime\prime}}\notag\\
&\times 
\langle \textrm{T}_{\tau}  
\hat{c}_{\boldk b s^{\prime}}^{\dagger}(\tau)
\hat{c}_{\boldk a s}(\tau)
\hat{c}_{\boldk^{\prime} c s^{\prime\prime}}^{\dagger}
\hat{c}_{\boldk^{\prime} d s^{\prime\prime\prime}}\rangle\notag\\
&= 
\frac{(-e)^{2}}{N}
\sum\limits_{\boldk,\boldk^{\prime}}
\sum\limits_{\{a\}}
\sum\limits_{\{s\}}
\delta_{s^{\prime},s}
(v_{\boldk x})_{ba}^{ss}
\delta_{s^{\prime\prime},s^{\prime\prime\prime}}
(v_{\boldk^{\prime} y})_{cd}^{s^{\prime\prime}s^{\prime\prime}}\notag\\
&\times 
K_{abcd}^{ss^{\prime\prime}}(\boldk,\boldk^{\prime};i\Omega_{n}).\label{eq:Ktild}
\end{align}
(Note that 
we should carry out the integration about $\tau$ 
before carrying out $i\Omega_{n} \rightarrow \omega+i0+$~\cite{Takada-text-old}.) 
In Eq. (\ref{eq:Ktild}), 
$\textrm{T}_{\tau}$ is the time-ordering operator~\cite{AGD}, 
$\sum_{\{a\}}$ is $\sum_{\{a\}}\equiv \sum_{a,b,c,d}$, 
$\sum_{\{s\}}$ is $\sum_{\{s\}}\equiv\sum_{s,s^{\prime},s^{\prime\prime},s^{\prime\prime\prime}}$, 
the charge current operator is  
\begin{align}
\hat{J}_{\boldq \nu}^{\textrm{C}}
=(-e)
\sum\limits_{\boldk}
\sum\limits_{a,b}
\sum\limits_{s,s^{\prime}}
\delta_{s^{\prime},s}
(v_{\boldk \nu})_{ba}^{ss}
\hat{c}_{\boldk-\frac{\boldq}{2} b s^{\prime}}^{\dagger}
\hat{c}_{\boldk+\frac{\boldq}{2} a s},\label{eq:JC}
\end{align}
and the noninteracting group velocity is 
\begin{align}
(v_{\boldk \nu})_{ab}^{ss}
=\frac{\partial \epsilon_{ab}(\boldk)}{\partial k_{\nu}}.\label{eq:vk}
\end{align}
The noninteracting group velocity is not affected by the onsite SOC 
because that is independent of momentum~\cite{Kontani-SHE}. 

Also, we can exactly derive $\sigma_{xy}^{\textrm{S}}$ within the linear-response 
in the similar way for $\sigma_{xy}^{\textrm{C}}$ 
if we define the spin current operator. 
Let us use a standard definition~\cite{SHE-review-Sinova,JS-UsusalDef-Niu}: 
\begin{align}
\hat{J}_{\boldq \nu}^{\textrm{S}}
=
\frac{1}{2}
\sum\limits_{\boldk}
\sum\limits_{a,b}
\sum\limits_{s,s^{\prime}}
\textrm{sgn}(s)
\delta_{s^{\prime},s}
(v_{\boldk \nu})_{ba}^{ss}
\hat{c}_{\boldk-\frac{\boldq}{2} b s^{\prime}}^{\dagger}
\hat{c}_{\boldk+\frac{\boldq}{2} a s},\label{eq:JS}
\end{align}
with $\textrm{sgn}(\uparrow)=+1$ or $\textrm{sgn}(\downarrow)=-1$. 
In this definition, 
the spin current is the difference between 
the spin-up and the spin-down component 
of the charge current~\cite{Kontani-SHE}: 
\begin{align}
\hat{J}_{\boldq \nu}^{\textrm{S}}
=
\frac{1}{2(-e)}
[(\hat{J}_{\boldq \nu}^{\textrm{C}})_{\uparrow \uparrow}
-(\hat{J}_{\boldq \nu}^{\textrm{C}})_{\downarrow \downarrow}],
\end{align}
where $(\hat{J}_{\boldq \nu}^{\textrm{C}})_{ss}$ is defined by 
$\hat{J}_{\boldq \nu}^{\textrm{C}}=\sum_{s}(\hat{J}_{\boldq \nu}^{\textrm{C}})_{ss}$. 
Even if we use a different but single-body definition, 
we can carry out the general formulation in the similar way.
By adopting this definition Eq. (\ref{eq:JS}) to the Kubo formula for $\sigma_{xy}^{\textrm{S}}$, 
its exact expression is obtained: 
\begin{align}
\sigma_{xy}^{\textrm{S}}
=&
\lim\limits_{\omega\rightarrow 0}
\lim\limits_{\boldq\rightarrow \boldzero}
\dfrac{\tilde{K}_{xy}^{\textrm{S}(\textrm{R})}(\boldq,\omega)
-\tilde{K}_{xy}^{\textrm{S}(\textrm{R})}(\boldq,0)}{i\omega},\label{eq:sigXY-S}
\end{align}
with 
\begin{align}
\tilde{K}_{xy}^{\textrm{S}(\textrm{R})}(\omega)
\equiv 
\tilde{K}_{xy}^{\textrm{S}(\textrm{R})}(\boldzero,\omega)
=
\tilde{K}_{xy}^{\textrm{S}}(i\Omega_{n})|_{i\Omega_{n} \rightarrow \omega+i0+},\label{eq:KS-analytic}
\end{align}
and 
\begin{align}
\tilde{K}_{xy}^{\textrm{S}}(i\Omega_{n})
&=
\dfrac{1}{N}
\lim\limits_{\boldq\rightarrow \boldzero}
\int^{T^{-1}}_{0}
\hspace{-14pt}d\tau e^{i\Omega_{n}\tau}
\langle \textrm{T}_{\tau}  
\hat{J}_{\boldq x}^{\textrm{S}}(\tau)
\hat{J}_{-\boldq y}^{\textrm{C}}(0)\rangle\notag\\
&= 
\frac{1}{N}
\sum\limits_{\boldk,\boldk^{\prime}}
\sum\limits_{\{a\}}
\sum\limits_{\{s\}}
\int^{T^{-1}}_{0}
\hspace{-10pt}d\tau e^{i\Omega_{n}\tau}
\frac{1}{2}\textrm{sgn}(s)
\delta_{s^{\prime},s}
(v_{\boldk x})_{ba}^{ss}\notag\\
&\times 
(-e)\delta_{s^{\prime\prime},s^{\prime\prime\prime}}
(v_{\boldk^{\prime} y})_{cd}^{s^{\prime\prime}s^{\prime\prime}}\notag\\
&\times 
\langle \textrm{T}_{\tau}  
\hat{c}_{\boldk b s^{\prime}}^{\dagger}(\tau)
\hat{c}_{\boldk a s}(\tau)
\hat{c}_{\boldk^{\prime} c s^{\prime\prime}}^{\dagger}
\hat{c}_{\boldk^{\prime} d s^{\prime\prime\prime}}\rangle\notag\\
&= 
\frac{-e}{2N}
\sum\limits_{\boldk,\boldk^{\prime}}
\sum\limits_{\{a\}}
\sum\limits_{\{s\}}
\textrm{sgn}(s)
\delta_{s^{\prime},s}
(v_{\boldk x})_{ba}^{ss}\notag\\
&\times 
\delta_{s^{\prime\prime},s^{\prime\prime\prime}}
(v_{\boldk^{\prime} y})_{cd}^{s^{\prime\prime}s^{\prime\prime}}
K_{abcd}^{ss^{\prime\prime}}(\boldk,\boldk^{\prime};i\Omega_{n}).\label{eq:KStild}
\end{align}

Then, 
the linear response theory~\cite{Kubo-formula} has several advantages compared with 
the phenomenological theory such as 
the Boltzmann theory in the relaxation-time approximation~\cite{Ashcroft-Mermin}. 
The most important advantage is about the treatment of the dominant excitations. 
The linear response theory does not assume whether the dominant excitations 
are either Fermi surface or Fermi sea type; 
instead, 
the dominant excitations are naturally determined as a result 
of the treatment of the perturbation terms. 
On the other hand, 
the Boltzmann theory assumes the dominant excitations 
as a result of assuming the distribution function 
[e.g., $f_{\boldk}=f_{\boldk}^{0}-\Phi_{\boldk}\frac{\partial f_{\boldk}^{0}}{\partial \epsilon_{\boldk}}$ 
in Eq. (7.7.1) of Ref. \onlinecite{Ziman2}]. 
Thus, 
the linear response theory is suitable to analyze 
the roles of the Fermi surface and the Fermi sea term. 
Then, 
in the linear response theory we can analyze the interaction effects 
with keeping momentum conservation 
in combination with Baym-Kadanoff's conserving approximation~\cite{Kontani-review,NA-review,
Baym-Kadanoff}; 
in the relaxation-time approximation~\cite{Ashcroft-Mermin}, 
momentum conservation is violated because of the introduction of 
the momentum- and frequency-independent relaxation time~\cite{Kontani-review}. 
This is one of the advantages because 
the appropriate treatment of momentum conservation is vital 
to analyze transport phenomena. 
Actually, 
only if we use the appropriate treatment, 
we can obtain the disappearance of the resistivity without 
the lattice and impurities~\cite{Yamada-Yosida}; 
in the relaxation-time approximation, 
the resistivity remains finite. 
In addition, 
the linear response theory is useful 
to study a variety of material dependence 
because the material dependence arises from the differences 
in the electronic structure 
and because we can naturally include those differences 
in the linear response theory; 
in the relaxation-time approximation, 
it is difficult to include the differences 
in the interaction effects. 

In the remaining part of this section, 
we explain the derivation only for $\sigma_{xy}^{\textrm{C}}$. 
This is because the difference 
between $\sigma_{xy}^{\textrm{C}}$ and $\sigma_{xy}^{\textrm{S}}$ 
comes from the difference between 
$\hat{J}_{\boldq x}^{\textrm{C}}(\tau)$ and $\hat{J}_{\boldq x}^{\textrm{S}}(\tau)$ 
and because 
we obtain $\sigma_{xy}^{\textrm{S}}$ by replacing 
$(-e)\delta_{s^{\prime},s}(v_{\boldk x})_{ba}^{ss}$ 
in $\sigma_{xy}^{\textrm{C}}$ 
by $\frac{1}{2}$sgn$(s)\delta_{s^{\prime},s}(v_{\boldk x})_{ba}^{ss}$ 
[compare Eqs. (\ref{eq:Ktild}) and (\ref{eq:KStild})].  

Before the details of the derivation, 
we remark on the importance of the order of taking the limits. 
To obtain the finite observable current, 
we should take $\boldq\rightarrow \boldzero$ 
before taking $\omega\rightarrow 0$~\cite{NA-review}. 
Then, for dc conductivities, 
only after taking $\omega\rightarrow 0$, 
we can take $\gamma_{\alpha}^{\ast}(\boldk)\rightarrow 0$ 
because we should hold $\omega\tau_{\textrm{trans}}\ll 1$~\cite{Eliashberg,Fukuyama,NA-review}, 
where $\tau_{\textrm{trans}}$ is the transport relaxation time 
which is of the order of magnitude the inverse of the QP damping. 
Namely, 
to discuss the dc conductivities in the clean and zero-temperature limit, 
we should take $\lim_{\gamma_{\alpha}^{\ast}(\boldk)\rightarrow 0}
\lim_{\omega\rightarrow 0}\lim_{\boldq\rightarrow \boldzero}$ in this order. 
If we take 
$\gamma_{\alpha}^{\ast}(\boldk)\rightarrow 0$ 
before taking $\omega\rightarrow 0$, 
the results become unphysical. 
In particular, 
the order of those limits is crucial for interacting systems 
because the important difference between cases 
in $\omega\tau_{\textrm{trans}}\ll 1$ and $\omega\tau_{\textrm{trans}}\gg 1$ 
is known as the difference between 
the first and the zero sound in the Fermi liquid~\cite{Pines-Nozieres}. 
However, in noninteracting systems 
only in the clean and zero-temperature limit, 
the unphysical limit~\cite{Onoda-Nagaosa} 
leads to the same $\sigma_{xy}^{\textrm{C}}$ or $\sigma_{xy}^{\textrm{S}}$ 
as that in the physical limit~\cite{Kontani-AHE,Kontani-SHE}. 
Since we cannot expect such accidental agreement in interacting systems, 
we should care about the order of taking the limits. 

The derivation for $\sigma_{xy}^{\textrm{C}}$ consists of three steps. 

The first step is to express $K_{abcd}^{ss^{\prime\prime}}(\boldk,\boldk^{\prime};i\Omega_{n})$ 
in terms of the single-particle Green's functions and 
the reducible four-point vertex function~\cite{Nozieres}; 
the latter describes the multiple electron-hole scattering. 
We can carry out that procedure by the perturbative expansion 
using the Bloch-De Dominicis theorem~\cite{Eliashberg}: 
\begin{align}
&K_{abcd}^{ss^{\prime\prime}}(\boldk,\boldk^{\prime};i\Omega_{n})
=
-\delta_{\boldk,\boldk^{\prime}}
T\sum\limits_{m}
G_{ac}^{ss^{\prime\prime}}(\tilde{k_{+}})
G_{db}^{s^{\prime\prime}s}(\tilde{k})\notag\\
-&\frac{T^{2}}{N}
\sum\limits_{m,m^{\prime}}
\sum\limits_{\{A\}}
\sum\limits_{\{s_{1}\}}
G_{aA}^{ss_{1}}(\tilde{k_{+}})
G_{dD}^{s^{\prime\prime}s_{4}}(\tilde{k^{\prime}})
G_{Cc}^{s_{3}s^{\prime\prime}}(\tilde{k^{\prime}_{+}})
G_{Bb}^{s_{2}s}(\tilde{k})\notag\\
&\times
\Gamma_{\{A\}}^{\{s_{1}\}}(\tilde{k},\tilde{k^{\prime}};\boldzero,i\Omega_{n}),\label{eq:K}
\end{align}
with 
$\tilde{k_{+}}\equiv (\boldk,i\epsilon_{m}+i\Omega_{n})$, 
fermionic Matsubara frequency, $\epsilon_{m}=2\pi T(m+\frac{1}{2})$, 
and the reducible four-point vertex function in Matsubara-frequency representation, 
$\Gamma_{\{A\}}^{\{s_{1}\}}(\tilde{k},\tilde{k^{\prime}};\boldq,i\Omega_{n})
\equiv \Gamma_{ABCD}^{s_{1}s_{2}s_{3}s_{4}}(\boldk+\boldq,i\epsilon_{m+n},\boldk,i\epsilon_{m},
\boldk^{\prime}+\boldq,i\epsilon_{m^{\prime}+n},\boldk^{\prime},i\epsilon_{m^{\prime}})$.  
In principle, 
we can calculate $G_{ab}^{ss^{\prime}}(\tilde{k})$ 
from Dyson's equation using the self-energy, 
\begin{align}
&G_{ab}^{ss^{\prime}}(\tilde{k})
=\ 
G_{ab}^{(0)ss^{\prime}}(\tilde{k})\notag\\
&+
\sum\limits_{c,d}
\sum\limits_{s^{\prime\prime},s^{\prime\prime\prime}}
G_{ac}^{(0)ss^{\prime\prime}}(\tilde{k})
\Sigma_{cd}^{(0)s^{\prime\prime}s^{\prime\prime\prime}}(\tilde{k})
G_{db}^{s^{\prime\prime\prime}s^{\prime}}(\tilde{k}),\label{eq:G-Dyson}
\end{align}
with the noninteracting single-particle Green's function, 
$G_{ab}^{(0)ss^{\prime}}(\tilde{k})$, 
and $\Gamma_{\{a\}}^{\{s_{1}\}}(\tilde{k},\tilde{k^{\prime}};\boldq,i\Omega_{n})$ 
from the Bethe-Salpeter equation 
using the irreducible four-point vertex function~\cite{Nozieres,NA-full},
\begin{align}
&\Gamma_{\{a\}}^{\{s_{1}\}}(\tilde{k},\tilde{k^{\prime}};
\boldq,i\Omega_{n})
=\Gamma_{\{a\}}^{(1)\{s_{1}\}}(\tilde{k},\tilde{k^{\prime}};\boldq,i\Omega_{n})\notag\\
&+
\frac{T}{N}
\sum\limits_{\boldk^{\prime\prime}}
\sum\limits_{m^{\prime\prime}}
\sum\limits_{\{A\}}
\sum\limits_{\{s_{1}^{\prime}\}}
\Gamma_{abCD}^{(1)s_{1}s_{2}s_{3}^{\prime}s_{4}^{\prime}}
(\tilde{k},\tilde{k^{\prime\prime}};\boldq,i\Omega_{n})\notag\\
&\times 
G_{CA}^{s_{3}^{\prime}s_{1}^{\prime}}(\tilde{k^{\prime\prime}}+\tilde{q})
G_{BD}^{s_{2}^{\prime}s_{4}^{\prime}}(\tilde{k^{\prime\prime}})
\Gamma_{ABcd}^{s_{1}^{\prime}s_{2}^{\prime}s_{3}s_{4}}
(\tilde{k^{\prime\prime}},\tilde{k^{\prime}};\boldq,i\Omega_{n}).
\end{align} 

The second step is to carry out the analytic continuation 
of $\tilde{K}_{xy}^{\textrm{C}}(i\Omega_{n})$. 
This procedure is the same for $\sigma_{xx}^{\textrm{C}}$~\cite{NA-full} 
with $\hat{H}_{0}$ and $\hat{H}_{\textrm{int}}$ 
without $\hat{H}_{\textrm{LS}}$ and $\hat{H}_{\textrm{imp}}$ 
because the relevant parameters in this procedure 
are only frequencies~\cite{Eliashberg} and spin indices are irrelevant. 
In this procedure, we use 
the analytic properties~\cite{Eliashberg} 
of the single-particle Green's function 
and reducible four-point vertex function 
and rewrite the sum of the Matsubara frequency by the corresponding contour integral; 
$G_{ab}^{ss^{\prime}}(\boldk,\epsilon)$ is singular 
on the horizontal line $\textrm{Im}\epsilon=0$; 
$\Gamma_{\{a\}}^{\{s_{1}\}}(\boldk,\epsilon,\boldk^{\prime},\epsilon^{\prime};\boldzero,\omega)$ 
is singular 
on the horizontal lines 
$\textrm{Im}\epsilon=0$, $\textrm{Im}(\epsilon+\omega)=0$, 
$\textrm{Im}\epsilon^{\prime}=0$, $\textrm{Im}(\epsilon^{\prime}+\omega)=0$, 
$\textrm{Im}(\epsilon+\epsilon^{\prime}+\omega)=0$, 
and $\textrm{Im}(\epsilon-\epsilon^{\prime})=0$, 
where the horizontal line $\textrm{Im}\omega=0$ is excluded 
because we consider $\textrm{Im}\omega >0$ [see Eq. (\ref{eq:K-analytic})].  
As derived in Appendix B, 
we obtain 
\begin{align}
&\tilde{K}_{xy}^{\textrm{C}(\textrm{R})}(\omega)
=
-
\frac{(-e)^{2}}{2i}
\sum\limits_{k}
\sum\limits_{\boldk^{\prime}}
\sum\limits_{\{a\}}
\sum\limits_{\{s\}}
\delta_{s^{\prime},s}
(v_{\boldk x})_{ba}^{ss}\notag\\
&\times 
\delta_{s^{\prime\prime},s^{\prime\prime\prime}}
(v_{\boldk^{\prime} y})_{cd}^{s^{\prime\prime}s^{\prime\prime}}
\sum\limits_{l=1}^{3}
T_{l}(\epsilon,\omega)
K_{l;abcd}^{ss^{\prime\prime}}(\boldk,\boldk^{\prime};\epsilon;\omega),\label{eq:K-4VC}
\end{align}
where 
\begin{align}
\sum\limits_{k}\equiv & 
\frac{1}{N}\sum\limits_{\boldk}\int^{\infty}_{-\infty}\frac{d\epsilon}{2\pi},\\
T_{1}(\epsilon,\omega) =& \tanh \frac{\epsilon}{2T},\label{eq:T1}\\ 
T_{2}(\epsilon,\omega) =& \tanh \frac{\epsilon+\omega}{2T}
-\tanh \frac{\epsilon}{2T},\label{eq:T2}\\
T_{3}(\epsilon,\omega) =& -\tanh \frac{\epsilon+\omega}{2T},\label{eq:T3}
\end{align}
and 
\begin{align}
&K_{l;abcd}^{ss^{\prime\prime}}(\boldk,\boldk^{\prime};\epsilon;\omega)\notag\\
&=
\delta_{\boldk,\boldk^{\prime}}g_{l;acdb}^{ss^{\prime\prime}s^{\prime\prime}s}(k;\omega)
+
\frac{1}{N}
\int^{\infty}_{-\infty}\frac{d\epsilon^{\prime}}{4\pi i}
\sum\limits_{\{A\}}
\sum\limits_{\{s_{1}\}}
g_{l;aABb}^{ss_{1}s_{2}s}(k;\omega)\notag\\
&\ \ \ \ \times 
\sum\limits_{l^{\prime}=1}^{3}
\mathcal{J}_{ll^{\prime};\{A\}}^{\{s_{1}\}}(k,k^{\prime};\omega)
g_{l^{\prime};CcdD}^{s_{3}s^{\prime\prime}s^{\prime\prime}s_{4}}(k^{\prime};\omega),\label{eq:K-4VC-part}
\end{align}
with  
\begin{align}
g_{1;acdb}^{ss^{\prime\prime}s^{\prime\prime\prime}s^{\prime}}(k;\omega)
=&\ G_{ac}^{(\textrm{R})ss^{\prime\prime}}(\boldk,\epsilon+\omega)
G_{db}^{(\textrm{R})s^{\prime\prime\prime}s^{\prime}}(\boldk,\epsilon),\label{eq:g1}\\
g_{2;acdb}^{ss^{\prime\prime}s^{\prime\prime\prime}s^{\prime}}(k;\omega)
=&\ G_{ac}^{(\textrm{R})ss^{\prime\prime}}(\boldk,\epsilon+\omega)
G_{db}^{(\textrm{A})s^{\prime\prime\prime}s^{\prime}}(\boldk,\epsilon),\label{eq:g2}\\
g_{3;acdb}^{ss^{\prime\prime}s^{\prime\prime\prime}s^{\prime}}(k;\omega)
=&\ G_{ac}^{(\textrm{A})ss^{\prime\prime}}(\boldk,\epsilon+\omega)
G_{db}^{(\textrm{A})s^{\prime\prime\prime}s^{\prime}}(\boldk,\epsilon),\label{eq:g3}
\end{align}
and 
$\mathcal{J}_{ll^{\prime};\{a\}}^{\{s_{1}\}}(k,k^{\prime};\omega)$, 
connected with 
$\mathcal{J}_{ll^{\prime};\{a\}}^{(1)\{s_{1}\}}(k,k^{\prime};\omega)$ 
by the Bethe-Salpeter equation, 
\begin{align}
\mathcal{J}_{ll^{\prime};\{a\}}^{\{s_{1}\}}(k,k^{\prime};\omega)
&=\ 
\mathcal{J}_{ll^{\prime};\{a\}}^{(1)\{s_{1}\}}(k,k^{\prime};\omega)+
\frac{1}{2i}
\sum\limits_{k^{\prime\prime}}
\sum\limits_{\{A\}}
\sum\limits_{\{s_{1}^{\prime}\}}
\sum\limits_{l^{\prime\prime}=1}^{3}\notag\\
&\times 
\mathcal{J}_{ll^{\prime\prime};abCD}^{(1)s_{1}s_{2}s_{3}^{\prime}s_{4}^{\prime}}
(k,k^{\prime\prime};\omega)
g_{l^{\prime\prime};CABD}^{s_{3}^{\prime}s_{1}^{\prime}s_{2}^{\prime}s_{4}^{\prime}}(k^{\prime\prime};\omega)\notag\\
&\times 
\mathcal{J}_{l^{\prime\prime}l^{\prime};ABcd}^{s_{1}^{\prime}s_{2}^{\prime}s_{3}^{\prime}s_{4}^{\prime}}
(k^{\prime\prime},k^{\prime};\omega).\label{eq:GamRed-GamIrred}
\end{align}
Here $\mathcal{J}_{ll^{\prime};\{a\}}^{\{s_{1}\}}(k,k^{\prime};\omega)$ 
is connected with 
$\Gamma_{ll^{\prime};\{a\}}^{\{s_{1}\}}(k,k^{\prime};\omega)$, 
the reducible four-point vertex function in real-frequency representation, 
as shown in Eqs. (\ref{eq:4VC-11}){--}(\ref{eq:4VC-33}); 
for the connections between 
$\mathcal{J}_{ll^{\prime};\{a\}}^{(1)\{s_{1}\}}(k,k^{\prime};\omega)$ 
and $\Gamma_{ll^{\prime};\{a\}}^{(1)\{s_{1}\}}(k,k^{\prime};\omega)$, 
we should add the superscript $(1)$ to 
$\mathcal{J}_{ll^{\prime};\{a\}}^{\{s_{1}\}}(k,k^{\prime};\omega)$ 
and 
$\Gamma_{ll^{\prime};\{a\}}^{\{s_{1}\}}(k,k^{\prime};\omega)$ 
in those equations. 

The third step is to rewrite $\tilde{K}_{xy}^{\textrm{C}(\textrm{R})}(\omega)$ 
in a more compact form by using the vertex function of the charge current. 
The vertex function of the charge current 
in Matsubara-frequency representation, 
$\Lambda_{\nu;AB}^{\textrm{C};s^{\prime}s^{\prime\prime}}(\tilde{k}; \tilde{q})
\equiv \Lambda_{\nu;AB}^{\textrm{C};s^{\prime}s^{\prime\prime}}(\boldk+\boldq,i\omega_{m+n}, \boldk,i\omega_{m})$ 
($\nu=x,y$), is defined as follows~\cite{Takada-new}: 
\begin{align}
&\sum\limits_{A,B,s^{\prime},s^{\prime\prime}}
G_{aA}^{ss^{\prime}}(\tilde{k}+\tilde{q})
\Lambda_{\nu;AB}^{\textrm{C};s^{\prime}s^{\prime\prime}}(\tilde{k};\tilde{q})
G_{Bb}^{s^{\prime\prime}s}(\tilde{k})\notag\\
=&
\int^{T^{-1}}_{0}\hspace{-16pt}d\tau e^{i\omega_{m+n}\tau}
\int^{T^{-1}}_{0}\hspace{-16pt}d\tau^{\prime} e^{-i\Omega_{n}\tau^{\prime}}
\langle \textrm{T}_{\tau}  
\hat{c}_{\boldk+\boldq a s}(\tau)
\hat{J}_{-\boldq \nu}^{\textrm{C}}(\tau^{\prime})
\hat{c}_{\boldk b s}^{\dagger}
\rangle.\label{eq:def-3VC-Matsu}
\end{align}
Thus, 
$\Lambda_{\nu;AB}^{\textrm{C};s^{\prime}s^{\prime\prime}}(\tilde{k};\tilde{q})$ 
is connected with $\Gamma_{ABC^{\prime}D^{\prime}}^{s^{\prime}s^{\prime\prime}s_{3}s_{4}}
(\tilde{k},\tilde{k^{\prime}};\boldq,i\Omega_{n})$ 
through the Bethe-Salpeter equation,
\begin{align}
&\Lambda_{\nu;AB}^{\textrm{C};s^{\prime}s^{\prime\prime}}
(\tilde{k};\tilde{q})
=\ \delta_{s^{\prime},s^{\prime\prime}}(v_{\boldk \nu})_{AB}^{s^{\prime}s^{\prime}}\notag\\
&+\frac{T}{N}
\sum\limits_{\boldk^{\prime}}
\sum\limits_{m^{\prime}}
\sum\limits_{\{A^{\prime}\}}
\sum\limits_{\{s_{1}\}}
\Gamma_{ABC^{\prime}D^{\prime}}^{s^{\prime}s^{\prime\prime}s_{3}s_{4}}
(\tilde{k},\tilde{k^{\prime}};\boldq,i\Omega_{n})\notag\\
&\times 
G_{C^{\prime}A^{\prime}}^{s_{3}s_{1}}(\tilde{k^{\prime}}+\tilde{q})
G_{B^{\prime}D^{\prime}}^{s_{2}s_{4}}(\tilde{k^{\prime}})
\delta_{s_{1},s_{2}}(v_{\boldk^{\prime} \nu})_{A^{\prime}B^{\prime}}^{s_{1}s_{2}}.\label{eq:3VC-Matsu}
\end{align}
Then, to convert this relation into the relation 
in real-frequency representation, 
we should carry out the analytic continuation of 
$\Lambda_{\nu;AB}^{\textrm{C};s^{\prime}s^{\prime\prime}}(\tilde{k};\tilde{q})$. 
Since this procedure is similar for the second term of Eq. (\ref{eq:K}), 
we can carry out this procedure in the similar way. 
As a result, we obtain that connection, 
\begin{align}
&\Lambda_{\nu;l;cd}^{\textrm{C}; s^{\prime\prime}s^{\prime\prime\prime}}(k;\omega)
=\ \delta_{s^{\prime\prime},s^{\prime\prime\prime}}
(v_{\boldk y})_{cd}^{s^{\prime\prime}s^{\prime\prime\prime}}\notag\\
&+\sum\limits_{k^{\prime}}
\sum\limits_{A,B,s_{1},s_{2}} 
\hspace{-6pt}
\alpha_{l;cdAB}^{s^{\prime\prime}s^{\prime\prime\prime}s_{1}s_{2}}(k,k^{\prime};\omega)
\delta_{s_{1},s_{2}}
(v_{\boldk^{\prime} y})_{AB}^{s_{1}s_{1}},\label{eq:3VC-real}
\end{align}
with $\Lambda_{y;l;cd}^{\textrm{C}; s^{\prime\prime}s^{\prime\prime\prime}}(k;\omega)
\equiv \Lambda_{y;l;cd}^{\textrm{C}; s^{\prime\prime}s^{\prime\prime\prime}}
(\boldk,\epsilon+\omega,\boldk,\epsilon)$ and  
\begin{align}
\alpha_{l;cdAB}^{s^{\prime\prime}s^{\prime\prime\prime}s_{1}s_{2}}(k,k^{\prime};\omega)
=&\ 
\frac{1}{2i}
\sum\limits_{C,D,s_{3},s_{4}}
\sum\limits_{l^{\prime}=1}^{3}
\mathcal{J}_{ll^{\prime};cdCD}^{s^{\prime\prime}s^{\prime\prime\prime}s_{3}s_{4}}(k,k^{\prime};\omega)\notag\\
&\times 
g_{l^{\prime};CABD}^{s_{3}s_{1}s_{2}s_{4}}(k^{\prime};\omega).\label{eq:alpha-Gamma}
\end{align}
These equations with Eq. (\ref{eq:GamRed-GamIrred}) show that 
the correction terms to the noninteracting charge current 
come from the multiple electron-hole scattering, 
described by the reducible four-point vertex function~\cite{Nozieres}. 
Furthermore, 
we can show that 
the correction term arising from $\hat{H}_{\textrm{imp}}$ 
disappears 
for even-parity systems 
because we can rewrite Eq. (\ref{eq:3VC-real}) as
\begin{align}
&\Lambda_{\nu;l;cd}^{\textrm{C}; s^{\prime\prime}s^{\prime\prime\prime}}(k;\omega)
=\ \delta_{s^{\prime\prime},s^{\prime\prime\prime}}
(v_{\boldk y})_{cd}^{s^{\prime\prime}s^{\prime\prime\prime}}\notag\\
&+\frac{1}{2i}
\sum\limits_{k^{\prime}}
\sum\limits_{\{A\}} 
\sum\limits_{\{s_{1}\}} 
\sum\limits_{l^{\prime}=1}^{3}
\mathcal{J}_{ll^{\prime};cdCD}^{(1)s^{\prime\prime}s^{\prime\prime\prime}s_{3}s_{4}}(k,k^{\prime};\omega)\notag\\
&\times 
g_{l^{\prime};CABD}^{s_{3}s_{1}s_{2}s_{4}}(k^{\prime};\omega)
\Lambda_{\nu;l^{\prime};AB}^{\textrm{C}; s_{1}s_{2}}(k^{\prime};\omega),\label{eq:3VC-real-G0}
\end{align}
and because part of the above second term arising from $\hat{H}_{\textrm{imp}}$ 
exactly vanishes in even-parity systems 
due to the combination of the momentum-independent irreducible four-point vertex function 
in the Born approximation~\cite{Kontani-AHE,Kontani-SHE} [see Eq. (\ref{eq:Gam1-full})], 
the even-parity symmetry of the single-particle Green's functions, 
and the odd-parity symmetry of the noninteracting group velocity, 
which results in the odd-parity symmetry of the vertex function of 
the charge current. 
Namely, for even-parity systems with the weak onsite scattering potential of the impurities, 
the correction terms in $\Lambda_{\nu;l;cd}^{\textrm{C}; s^{\prime\prime}s^{\prime\prime\prime}}(k;\omega)$ 
arise from only $\hat{H}_{\textrm{int}}$. 
Rewriting part of Eq. (\ref{eq:K-4VC}) by using the relation, 
\begin{align}
&\sum\limits_{\boldk^{\prime}}
\sum\limits_{c,d,s^{\prime\prime},s^{\prime\prime\prime}}
K_{l;abcd}^{ss^{\prime\prime}}(\boldk,\boldk^{\prime};\epsilon;\omega)
\delta_{s^{\prime\prime},s^{\prime\prime\prime}}
(v_{\boldk^{\prime} y})_{cd}^{s^{\prime\prime}s^{\prime\prime}}\notag\\
=&
\sum\limits_{c,d,s^{\prime\prime},s^{\prime\prime\prime}}
g_{l;acdb}^{ss^{\prime\prime}s^{\prime\prime}s}(k;\omega)
\delta_{s^{\prime\prime},s^{\prime\prime\prime}}
(v_{\boldk y})_{cd}^{s^{\prime\prime}s^{\prime\prime}}\notag\\
&+
\sum\limits_{c,d,s^{\prime\prime},s^{\prime\prime\prime}}
\sum\limits_{A,B,s_{1},s_{2}}
g_{l;aABb}^{ss_{1}s_{2}s}(k;\omega)
\frac{1}{2i}\sum\limits_{k^{\prime}}
\sum\limits_{C,D,s_{3},s_{4}}
\sum\limits_{l^{\prime}=1}^{3}
\notag\\
&\ \times 
\mathcal{J}_{ll^{\prime};\{A\}}^{\{s_{1}\}}(k,k^{\prime};\omega)
g_{l^{\prime};CcdD}^{s_{3}s^{\prime\prime}s^{\prime\prime\prime}s_{4}}(k^{\prime};\omega)
\delta_{s^{\prime\prime},s^{\prime\prime\prime}}
(v_{\boldk^{\prime} y})_{cd}^{s^{\prime\prime}s^{\prime\prime}}\notag\\
=&\ 
\sum\limits_{c,d,s^{\prime\prime},s^{\prime\prime\prime}}
g_{l;acdb}^{ss^{\prime\prime}s^{\prime\prime\prime}s}(k;\omega)
\Lambda_{y;l;cd}^{\textrm{C};s^{\prime\prime}s^{\prime\prime\prime}}(k;\omega),
\end{align}
we can rewrite $\tilde{K}_{xy}^{\textrm{C}(\textrm{R})}(\omega)$ 
as follows: 
\begin{align}
\tilde{K}_{xy}^{\textrm{C}(\textrm{R})}(\omega)&
=
-
\frac{(-e)^{2}}{2i}
\sum\limits_{k}
\sum\limits_{\{a\}}
\sum\limits_{\{s\}}
\delta_{s^{\prime},s}
(v_{\boldk x})_{ba}^{ss}
\sum\limits_{l=1}^{3}
T_{l}(\epsilon,\omega)\notag\\
&
\times 
g_{l;acdb}^{ss^{\prime\prime}s^{\prime\prime\prime}s^{\prime}}(k;\omega)
\Lambda_{y;l;cd}^{\textrm{C};s^{\prime\prime}s^{\prime\prime\prime}}(k;\omega).\label{eq:KC-real}
\end{align}

A set of Eqs. (\ref{eq:sigXY-C}), (\ref{eq:T1}){--}(\ref{eq:g3}), (\ref{eq:3VC-real}), 
and (\ref{eq:alpha-Gamma}) provides 
a framework to obtain an exact expression of $\sigma_{xy}^{\textrm{C}}$ 
within the linear response of an external electric field. 

We also obtain an exact framework for $\sigma_{xy}^{\textrm{S}}$ 
by replacing 
$(-e)\delta_{s^{\prime},s}(v_{\boldk x})_{ba}^{ss}$ in $\sigma_{xy}^{\textrm{C}}$ 
by $\frac{1}{2}$sgn$(s)\delta_{s^{\prime},s}(v_{\boldk x})_{ba}^{ss}$. 
Namely, 
instead of Eqs. (\ref{eq:sigXY-C}) and (\ref{eq:KC-real}), 
we use Eq. (\ref{eq:sigXY-S}) and
\begin{align}
\tilde{K}_{xy}^{\textrm{S}(\textrm{R})}(\omega)
&=
-
\frac{(-e)}{2^{2}i}
\sum\limits_{k}
\sum\limits_{\{a\}}
\sum\limits_{\{s\}}
\delta_{s^{\prime},s}
\textrm{sgn}(s)
(v_{\boldk x})_{ba}^{ss}
\sum\limits_{l=1}^{3}\notag\\
&\times 
T_{l}(\epsilon,\omega)
g_{l;acdb}^{ss^{\prime\prime}s^{\prime\prime\prime}s^{\prime}}(k;\omega)
\Lambda_{y;l;cd}^{\textrm{C};s^{\prime\prime}s^{\prime\prime\prime}}(k;\omega).\label{eq:KS-real}
\end{align}

\section{Results}
Since it is difficult to solve 
the exact expressions of $\sigma_{xy}^{\textrm{C}}$ and $\sigma_{xy}^{\textrm{S}}$ 
in the linear response theory, 
we adopt the approximation appropriate for an interacting metal 
to these expressions, 
and analyze the interaction effects on $\sigma_{xy}^{\textrm{C}}$ and $\sigma_{xy}^{\textrm{S}}$. 
First, 
we derive the approximate expressions of 
$\sigma_{xy}^{\textrm{C}}$ and $\sigma_{xy}^{\textrm{S}}$
in \'{E}liashberg's approximation~\cite{Eliashberg}; 
in part of this derivation, we use Appendix C. 
This approximation is usually used to derive 
transport coefficients of an interacting metal 
microscopically~\cite{Eliashberg,Fukuyama,Kohno-Yamada,Kontani-review,NA-full}, 
and its result~\cite{Eliashberg} can reproduce 
the phenomenological transport equation in the Fermi liquid~\cite{Landau,Nozieres,AGD}; 
thus, this approximation may be appropriate 
if the terms included remain considerable. 
Comparing the derived $\sigma_{xy}^{\textrm{C}}$ or $\sigma_{xy}^{\textrm{S}}$ 
with the corresponding noninteracting result, 
we analyze the interaction effects on 
the derived $\sigma_{xy}^{\textrm{C}}$ or $\sigma_{xy}^{\textrm{S}}$. 
Second, 
we address the applicability of this approximation, 
and show its limit in clean and low-$T$ case. 
The correct understanding of this applicability 
is important to understand the difference between 
$\sigma_{xy}^{\textrm{C}}$ and $\sigma_{xx}^{\textrm{C}}$. 
Third, 
we introduce an approximation beyond \'{E}liashberg's approximation 
in order to describe the outside of the applicable region of \'{E}liashberg's approximation, 
and derive the approximate expressions of 
$\sigma_{xy}^{\textrm{C}}$ and $\sigma_{xy}^{\textrm{S}}$. 
We also analyze how the additional terms 
are affected by the electron-electron interaction. 

\subsection{\'{E}liashberg's approximation}
After reviewing the singular property of a retarded-advanced product 
of two single-particle Green's functions 
in the presence of the Fermi surface with several long-lived QPs, 
we derive the approximate expressions of $\sigma_{xy}^{\textrm{C}}$ and $\sigma_{xy}^{\textrm{S}}$ 
by utilizing this property. 
Then, let us argue the interaction effects due to the modifications 
from the noninteracting result. 
 
\subsubsection{Formulation}

We begin with the singular properties~\cite{AGD,Nozieres} of a retarded-advanced product of 
two single-particle Green's functions such as 
$G_{ac}^{(\textrm{R})s_{1}s_{3}}(k+\frac{q}{2})G_{db}^{(\textrm{A})s_{4}s_{2}}(k-\frac{q}{2})$ 
in the limits $q\rightarrow 0$ 
and $\gamma_{\alpha}^{\ast}(\boldk_{\textrm{F}})/T\rightarrow 0$ 
in the presence of the Fermi surface. 
In the presence of the Fermi surface, 
we can well define QPs with the long-lived lifetime 
for at least several Fermi momenta~\cite{Sr2RuO4-APRES,
FMQCP-exp-QPspectrum,Mott-exp-QPspectrum}. 
These QPs are well described by 
the coherent part of the single-particle Green's function~\cite{AGD,Nozieres,NA-review}, 
given by  
\begin{align}
G_{\textrm{coh};ab}^{(\textrm{R})ss^{\prime}}(k)
=
\sum\limits_{\alpha}
(U_{\boldk})_{a\alpha}^{s}
\frac{z_{\alpha}(\boldk)}{\epsilon-\xi^{\ast}_{\alpha}(\boldk)+i\gamma_{\alpha}^{\ast}(\boldk)}
(U_{\boldk}^{\dagger})_{\alpha b}^{s^{\prime}},\label{eq:coherent-G}
\end{align}
where $(U_{\boldk})_{a\alpha}^{s}$ is the unitary matrix used to obtain $\xi_{\alpha}^{\ast}(\boldk)$. 
Then, 
for analyses of the limiting properties of the products of 
two single-particle Green's functions, 
it is sufficient to consider only the coherent parts~\cite{Nozieres}. 
This is because in the limits under consideration 
the incoherent part [i.e., 
$G_{ab}^{(\textrm{R})ss^{\prime}}(k)-G_{\textrm{coh};ab}^{(\textrm{R})ss^{\prime}}(k)$]
is well defined 
and only the product of the coherent parts can be singular 
due to the merging of their poles~\cite{Nozieres}. 
Such singular behavior is obtained 
only for a retarded-advanced product 
because the poles of the coherent parts merge 
only if one of the poles crosses over the Fermi surface 
and because such crossing occurs 
only for a retarded-advanced product~\cite{Nozieres}. 
As a result, 
a retarded-advanced product gives the leading dependence 
on external momentum and frequency and the QP damping, 
and the dependence of a retarded-retarded or an advanced-advanced product 
is approximately negligible~\cite{Eliashberg}. 
This treatment remains reasonable 
even for finite $\gamma_{\alpha}^{\ast}(\boldk_{\textrm{F}})/T$ 
if $\gamma_{\alpha}^{\ast}(\boldk_{\textrm{F}})/T$ satisfies 
$\gamma_{\alpha}^{\ast}(\boldk_{\textrm{F}})/T< 1$ 
because this treatment is regarded 
as a lowest-order expansion in terms of $\gamma_{\alpha}^{\ast}(\boldk_{\textrm{F}})/T$. 

Utilizing the singular property of 
a retarded-advanced product of two single-particle Green's functions, 
we derive approximate expressions of $\sigma_{xy}^{\textrm{C}}$ and $\sigma_{xy}^{\textrm{S}}$ 
in \'{E}liashberg's approximation. 
(Because of the same reason for the exact formulation in the linear-response theory, 
we explain the derivation for $\sigma_{xy}^{\textrm{C}}$ in detail.) 
To utilize the singular property, 
we introduce two quantities, $\mathcal{J}_{ll^{\prime};\{a\}}^{(0)\{s_{1}\}}(k,k^{\prime};\omega)$ 
and $\Lambda_{\nu;l;ab}^{\textrm{C}(0)ss^{\prime}}(k;\omega)$: 
\begin{align}
&\mathcal{J}_{ll^{\prime};\{a\}}^{(0)\{s_{1}\}}(k,k^{\prime};\omega)
=\ 
\mathcal{J}_{ll^{\prime};\{a\}}^{(1)\{s_{1}\}}(k,k^{\prime};\omega)\notag\\
&+
\frac{1}{2i}\sum\limits_{k^{\prime\prime}}
\sum\limits_{\{A\}}
\sum\limits_{\{s_{1}^{\prime}\}}
\sum\limits_{l^{\prime}=1,3}
\mathcal{J}_{ll^{\prime\prime};abCD}^{(1)s_{1}s_{2}s_{3}^{\prime}s_{4}^{\prime}}(k,k^{\prime\prime};\omega)\notag\\
&\times 
g_{l^{\prime\prime};CABD}^{s_{3}^{\prime}s_{1}^{\prime}s_{2}^{\prime}s_{4}^{\prime}}(k^{\prime\prime};\omega)
\mathcal{J}_{l^{\prime\prime}l^{\prime};ABcd}^{(0)s_{1}^{\prime}s_{2}^{\prime}s_{3}s_{4}}
(k^{\prime\prime},k^{\prime};\omega),\label{eq:BSeq-0-1}
\end{align}
and 
\begin{align}
&\Lambda_{\nu;l;ab}^{\textrm{C}(0)ss^{\prime}}(k;\omega)
=\ \delta_{s,s^{\prime}}(v_{\boldk \nu})_{ab}^{ss}\notag\\
&+\sum\limits_{k^{\prime}}
\sum\limits_{A,B,s_{1},s_{2}} 
\alpha_{l;abAB}^{(0)ss^{\prime}s_{1}s_{2}}(k,k^{\prime};\omega)
\delta_{s_{1},s_{2}}
(v_{\boldk^{\prime} \nu})_{AB}^{s_{1}s_{1}},\label{eq:JC0all-symbolic}
\end{align}
where $\alpha_{l;abAB}^{(0)ss^{\prime}s_{1}s_{2}}(k,k^{\prime};\omega)$ is 
\begin{align}
\alpha_{l;abAB}^{(0)ss^{\prime}s_{1}s_{2}}(k,k^{\prime};\omega)
=&\ 
\frac{1}{2i}
\sum\limits_{C,D,s_{3},s_{4}}
\sum\limits_{l^{\prime}=1,3}
\mathcal{J}_{ll^{\prime};abCD}^{(0)ss^{\prime}s_{3}s_{4}}(k,k^{\prime};\omega)\notag\\
&\times 
g_{l^{\prime}; CABD}^{s_{3}s_{1}s_{2}s_{4}}(k^{\prime};\omega).\label{eq:alpha0-def}
\end{align}
Equations (\ref{eq:BSeq-0-1}) and (\ref{eq:JC0all-symbolic}) 
show that 
$\mathcal{J}_{ll^{\prime};\{a\}}^{(0)\{s_{1}\}}(k,k^{\prime};\omega)$ 
and $\Lambda_{\nu;l;ab}^{\textrm{C}(0)ss^{\prime}}(k;\omega)$ 
do not include a retarded-advanced product of two single-particle Green's functions. 
Thus, 
those quantities can be used to exclude 
the terms including at least a retarded-advanced product 
from the terms of $\tilde{K}_{xy}^{\textrm{C}(\textrm{R})}(\omega)$ in Eq. (\ref{eq:KC-real}). 
Among those terms, 
we need to decompose the terms for $l=1$ and $3$ in $\tilde{K}_{xy}^{\textrm{C}(\textrm{R})}(\omega)$ 
into the terms without and with the retarded-advanced product. 
This is because $\Lambda_{\nu;l;ab}^{\textrm{C}(0)ss^{\prime}}(k;\omega)$ is connected 
with $\Lambda_{\nu;l;ab}^{\textrm{C};ss^{\prime}}(k;\omega)$ through the Bethe-Salpeter equation, 
\begin{align}
&\Lambda_{\nu;l;ab}^{\textrm{C}; ss^{\prime}}(k;\omega)
=\ \Lambda_{\nu;l;ab}^{\textrm{C}(0) ss^{\prime}}(k;\omega)\notag\\
&+\frac{1}{2i}
\sum\limits_{k^{\prime}}
\sum\limits_{\{A\}}
\sum\limits_{\{s_{1}\}}
\mathcal{J}_{l2;abCD}^{(0)ss^{\prime}s_{3}s_{4}}(k,k^{\prime};\omega)\notag\\
&\times 
g_{2;CABD}^{s_{3}s_{1}s_{2}s_{4}}(k^{\prime};\omega)
\Lambda_{\nu;2;AB}^{\textrm{C}; s_{1}s_{2}}(k^{\prime};\omega).\label{eq:relation_L0-L}
\end{align}
After the decomposition of the terms for $l=1$ and $3$ 
in $\tilde{K}_{xy}^{\textrm{C}(\textrm{R})}(\omega)$ in Eq. (\ref{eq:KC-real}), 
explained in Appendix C, 
we obtain  
\begin{align}
\tilde{K}_{xy}^{\textrm{C}(\textrm{R})}(\omega)
&=
-\frac{(-e)^{2}}{2i}
\sum\limits_{k}
\sum\limits_{\{a\}}
\sum\limits_{\{s\}}
\delta_{s^{\prime},s}
(v_{\boldk x})_{ba}^{ss}
\sum\limits_{l=1,3}
\notag\\
&\times
\textrm{sgn}(2-l)
T_{l}(\epsilon,\omega)g_{l;acdb}^{ss^{\prime\prime}s^{\prime\prime\prime}s^{\prime}}(k;\omega)
\Lambda_{y;l;cd}^{\textrm{C}(0)s^{\prime\prime}s^{\prime\prime\prime}}(k;\omega)\notag\\
& 
-
\frac{(-e)^{2}}{2i}
\sum\limits_{k}
\sum\limits_{\{a\}}
\sum\limits_{\{s\}}
\Lambda_{x;2;ba}^{\textrm{C}(0)s^{\prime}s}(\boldk,\epsilon,\boldk,\epsilon+\omega)\notag\\
&\times 
T_{2}(\epsilon,\omega)g_{2;acdb}^{ss^{\prime\prime}s^{\prime\prime\prime}s^{\prime}}(k;\omega)
\Lambda_{y;2;cd}^{\textrm{C};s^{\prime\prime}s^{\prime\prime\prime}}(k;\omega).\label{eq:KC-real2}
\end{align}
This equation shows that 
only the second term includes 
a retarded-advanced product of two single-particle Green's functions. 
Since we assume in \'{E}liashberg's approximation~\cite{Eliashberg} 
that the leading terms of $\tilde{K}_{xy}^{\textrm{C}(\textrm{R})}(\omega)$ 
come from the most divergent terms in $q\rightarrow 0$ 
and $\gamma_{\alpha}^{\ast}(\boldk_{\textrm{F}})/T\rightarrow 0$, 
we obtain $\sigma_{xy}^{\textrm{C}}$ in this approximation, 
$\sigma_{xy}^{\textrm{C}}=\sigma_{xy}^{\textrm{C}(\textrm{I})}$, 
\begin{align}
\sigma_{xy}^{\textrm{C}(\textrm{I})}=&
(-e)^{2}
\sum\limits_{k}
\sum\limits_{\{a\}}
\sum\limits_{\{s\}}
\Bigl(-\dfrac{\partial f(\epsilon)}{\partial \epsilon}\Bigr)
\Lambda_{x;2;ba}^{\textrm{C}(0)s^{\prime}s}(k;0)\notag\\
&\times 
G_{ac}^{(\textrm{R})ss^{\prime\prime}}(k)
G_{db}^{(\textrm{A})s^{\prime\prime\prime}s^{\prime}}(k)
\Lambda_{y;2;cd}^{\textrm{C}; s^{\prime\prime}s^{\prime\prime\prime}}(k;0),\label{eq:SigXYC-I} 
\end{align}
where $f(\epsilon)$ is the Fermi distribution function. 
We also obtain $\sigma_{xy}^{\textrm{S}}$ in this approximation, 
$\sigma_{xy}^{\textrm{S}}=\sigma_{xy}^{\textrm{S}(\textrm{I})}$, 
\begin{align}
\sigma_{xy}^{\textrm{S}(\textrm{I})}=&
\frac{(-e)}{2}
\sum\limits_{k}
\sum\limits_{\{a\}}
\sum\limits_{\{s\}}
\Bigl(-\dfrac{\partial f(\epsilon)}{\partial \epsilon}\Bigr)
\Lambda_{x;2;ba}^{\textrm{S}(0)s^{\prime}s}(k;0)\notag\\
&\times 
G_{ac}^{(\textrm{R})ss^{\prime\prime}}(k)
G_{db}^{(\textrm{A})s^{\prime\prime\prime}s^{\prime}}(k)
\Lambda_{y;2;cd}^{\textrm{C}; s^{\prime\prime}s^{\prime\prime\prime}}(k;0),\label{eq:SigXYS-I} 
\end{align}
by adopting the same argument to Eqs. (\ref{eq:sigXY-S}) and (\ref{eq:KS-real}) 
and introducing the vertex function of the spin current, 
\begin{align}
&\Lambda_{\nu;l;ab}^{\textrm{S}(0)ss^{\prime}}(k;\omega)
=\ \delta_{s,s^{\prime}}\textrm{sgn}(s)(v_{\boldk \nu})_{ab}^{ss}\notag\\
&+\sum\limits_{k^{\prime}}
\sum\limits_{A,B,s_{1},s_{2}} 
\alpha_{l;abAB}^{(0)ss^{\prime}s_{1}s_{2}}(k,k^{\prime};\omega)
\delta_{s_{1},s_{2}}
\textrm{sgn}(s_{1})
(v_{\boldk^{\prime} \nu})_{AB}^{s_{1}s_{1}}.\label{eq:JS0all-symbolic}
\end{align} 
By using that vertex function, 
$\tilde{K}_{xy}^{\textrm{S}(\textrm{R})}(\omega)$ 
can be exactly rewritten as follows: 
\begin{align}
\tilde{K}_{xy}^{\textrm{S}(\textrm{R})}(\omega)
&=
-\frac{(-e)}{2^{2}i}
\sum\limits_{k}
\sum\limits_{\{a\}}
\sum\limits_{\{s\}}
\delta_{s^{\prime},s}
\textrm{sgn}(s)
(v_{\boldk x})_{ba}^{ss}\notag\\
&\times 
\sum\limits_{l=1,3}
\textrm{sgn}(2-l)
T_{l}(\epsilon,\omega)g_{l;acdb}^{ss^{\prime\prime}s^{\prime\prime\prime}s^{\prime}}(k;\omega)\notag\\
&\times 
\Lambda_{y;l;cd}^{\textrm{C}(0)s^{\prime\prime}s^{\prime\prime\prime}}(k;\omega)\notag\\
& 
-
\frac{(-e)}{2^{2}i}
\sum\limits_{k}
\sum\limits_{\{a\}}
\sum\limits_{\{s\}}
\Lambda_{x;2;ba}^{\textrm{S}(0)s^{\prime}s}(\boldk,\epsilon,\boldk,\epsilon+\omega)\notag\\
&\times 
T_{2}(\epsilon,\omega)g_{2;acdb}^{ss^{\prime\prime}s^{\prime\prime\prime}s^{\prime}}(k;\omega)
\Lambda_{y;2;cd}^{\textrm{C};s^{\prime\prime}s^{\prime\prime\prime}}(k;\omega).\label{eq:KS-real2}
\end{align}
Because of the same reason for $\Lambda_{y;2;cd}^{\textrm{C};s^{\prime\prime}s^{\prime\prime\prime}}(k;\omega)$, 
$\Lambda_{x;2;ba}^{\textrm{C}(0)s^{\prime}s}(k;\omega)$ 
and 
$\Lambda_{x;2;ba}^{\textrm{S}(0)s^{\prime}s}(k;\omega)$ 
include the corrections to the noninteracting charge and spin currents, respectively, 
due to the multiple electron-hole scattering arising from $\hat{H}_{\textrm{int}}$ 
and such corrections arising from $\hat{H}_{\textrm{imp}}$ completely vanish 
in even-parity systems for the weak onsite scattering potential of dilute nonmagnetic impurities. 
Note that 
in the similar way for $\Lambda_{y;2;cd}^{\textrm{C};s^{\prime\prime}s^{\prime\prime\prime}}(k;\omega)$, 
we can rewrite Eqs. (\ref{eq:JC0all-symbolic}) and (\ref{eq:JS0all-symbolic}) 
using Eq. (\ref{eq:BSeq-0-1}) as follows:  
\begin{align}
&\Lambda_{\nu;l;ab}^{\textrm{C}(0)ss^{\prime}}(k;\omega)
=\ \delta_{s,s^{\prime}}(v_{\boldk \nu})_{ab}^{ss}\notag\\
&+
\frac{1}{2i}\sum\limits_{k^{\prime}}
\sum\limits_{\{A\}} 
\sum\limits_{\{s_{1}\}} 
\sum\limits_{l^{\prime}=1,3}
\mathcal{J}_{ll^{\prime};abCD}^{(1)ss^{\prime}s_{3}s_{4}}(k,k^{\prime};\omega)\notag\\
&\times 
g_{l^{\prime}; CABD}^{s_{3}s_{1}s_{2}s_{4}}(k^{\prime};\omega)
\Lambda_{\nu;l^{\prime};AB}^{\textrm{C}(0)s_{1}s_{2}}(k^{\prime};\omega),\label{eq:JC0all-symbolic-G1}
\end{align} 
and 
\begin{align}
&\Lambda_{\nu;l;ab}^{\textrm{S}(0)ss^{\prime}}(k;\omega)
=\ \delta_{s,s^{\prime}}\textrm{sgn}(s)(v_{\boldk \nu})_{ab}^{ss}\notag\\
&+
\frac{1}{2i}\sum\limits_{k^{\prime}}
\sum\limits_{\{A\}} 
\sum\limits_{\{s_{1}\}} 
\sum\limits_{l^{\prime}=1,3}
\mathcal{J}_{ll^{\prime};abCD}^{(1)ss^{\prime}s_{3}s_{4}}(k,k^{\prime};\omega)\notag\\
&\times
g_{l^{\prime}; CABD}^{s_{3}s_{1}s_{2}s_{4}}(k^{\prime};\omega)
\Lambda_{\nu;l^{\prime};AB}^{\textrm{S}(0)s_{1}s_{2}}(k^{\prime};\omega).\label{eq:JS0all-symbolic-G1}
\end{align} 

\subsubsection{Interaction effects}
\begin{figure}[tb]
\includegraphics[width=70mm]{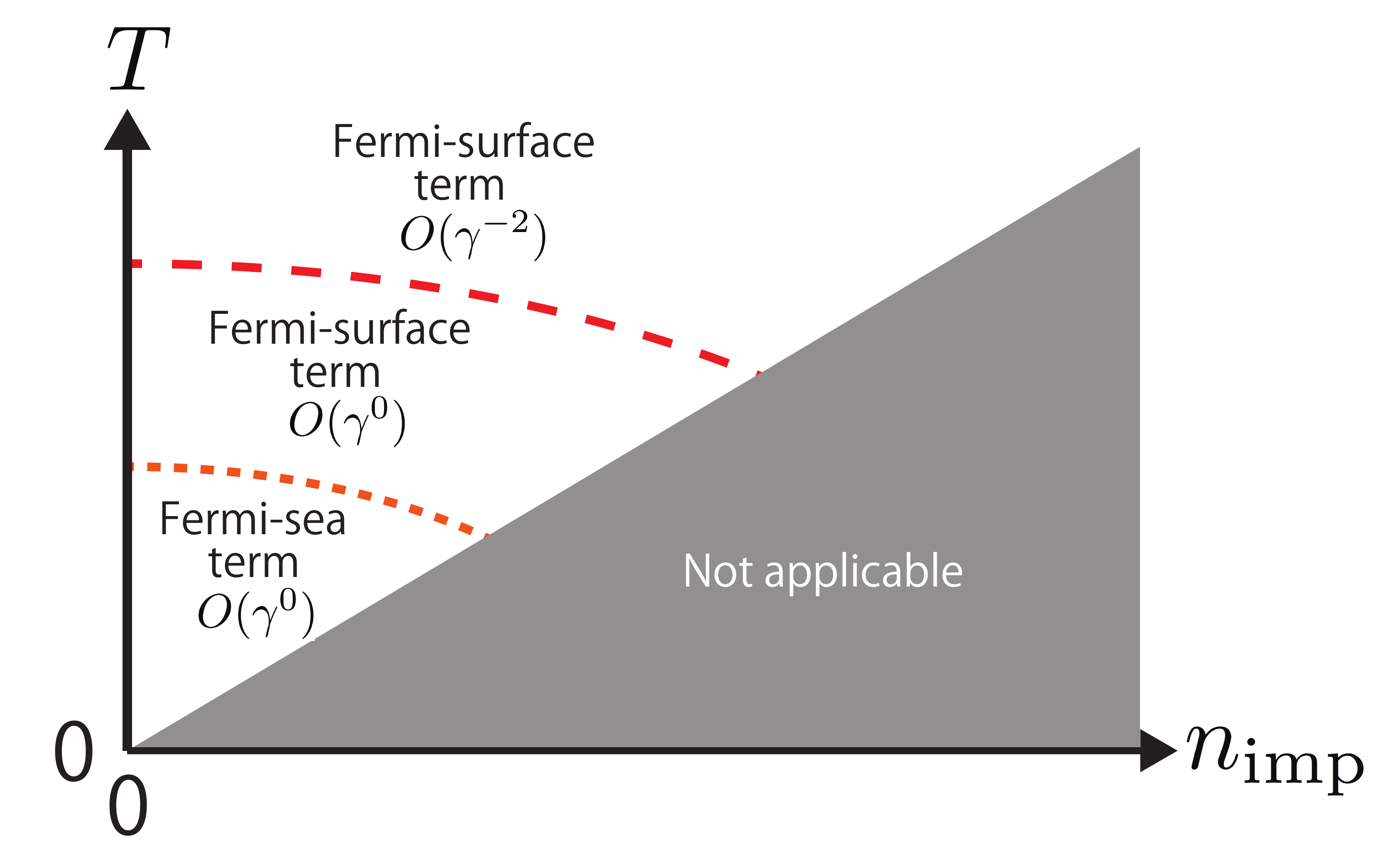}
\vspace{-6pt}
\caption{ 
Schematic diagram about the dominant term 
and damping dependence of $\sigma_{xy}^{\textrm{C}}$ or $\sigma_{xy}^{\textrm{S}}$. 
Our formalism is applicable outside the gray triangle region 
because that region satisfies $\gamma_{\alpha}^{\ast}(\boldk_{\textrm{F}})/T \geq 1$ 
due to the impurity-induced QP damping. 
The crossovers occur at the red and the orange dotted line. 
The form of the red dotted line depends strongly on 
the temperature dependence of the QP damping. 
}
\label{fig:Fig2}
\end{figure}

Since the comparison between the derived Fermi surface term 
and the noninteracting Fermi surface term 
is useful to deduce the interaction effects on the Fermi surface term, 
we show the noninteracting Fermi surface terms~\cite{Kontani-AHE,Kontani-SHE} 
of the intrinsic AHE and SHE, $\sigma_{xy}^{\textrm{C}(0;\textrm{I})}$ 
and $\sigma_{xy}^{\textrm{S}(0;\textrm{I})}$, 
\begin{align}
\sigma_{xy}^{\textrm{C}(0;\textrm{I})}&=
(-e)^{2}
\sum\limits_{k}
\sum\limits_{\{a\}}
\sum\limits_{\{s\}}
\Bigl(-\dfrac{\partial f(\epsilon)}{\partial \epsilon}\Bigr)
\delta_{s^{\prime},s}
(v_{\boldk x})_{ba}^{ss}\notag\\
&\times 
G_{ac}^{(0;\textrm{R})ss^{\prime\prime}}(k)
G_{db}^{(0;\textrm{A})s^{\prime\prime\prime}s^{\prime}}(k)
\delta_{s^{\prime\prime},s^{\prime\prime\prime}}
(v_{\boldk^{\prime} y})_{cd}^{s^{\prime\prime}s^{\prime\prime}},\label{eq:SigXYC0-I} 
\end{align}
and 
\begin{align}
\sigma_{xy}^{\textrm{S}(0;\textrm{I})}&=
\frac{(-e)}{2}
\sum\limits_{k}
\sum\limits_{\{a\}}
\sum\limits_{\{s\}}
\Bigl(-\dfrac{\partial f(\epsilon)}{\partial \epsilon}\Bigr)
\delta_{s^{\prime},s}
(v_{\boldk x})_{ba}^{ss}\notag\\
&\times 
G_{ac}^{(0;\textrm{R})ss^{\prime\prime}}(k)
G_{db}^{(0;\textrm{A})s^{\prime\prime\prime}s^{\prime}}(k)
\delta_{s^{\prime\prime},s^{\prime\prime\prime}}
(v_{\boldk^{\prime} y})_{cd}^{s^{\prime\prime}s^{\prime\prime}}.\label{eq:SigXYS0-I} 
\end{align}

Comparing Eq. (\ref{eq:SigXYC-I}) or (\ref{eq:SigXYS-I}) 
with Eq. (\ref{eq:SigXYC0-I}) or (\ref{eq:SigXYS0-I}), respectively, 
we see the electron-electron interaction causes three modifications. 
First, 
the $x$ component of the noninteracting charge or spin current 
becomes, respectively, 
$(-e)\Lambda_{x;2;ba}^{\textrm{C}(0)s^{\prime}s}(k;0)$ in $\sigma_{xy}^{\textrm{C}(\textrm{I})}$ 
or 
$\frac{1}{2}\Lambda_{x;2;ba}^{\textrm{S}(0)s^{\prime}s}(k;0)$ in $\sigma_{xy}^{\textrm{S}(\textrm{I})}$. 
Second,
the two single-particle Green's functions change from noninteracting 
to interacting. 
Third, 
the $y$ component of the noninteracting charge current becomes 
$(-e)\Lambda_{y;2;cd}^{\textrm{C}; s^{\prime\prime}s^{\prime\prime\prime}}(k;0)$. 

Let us begin with the interaction effect due to 
the replacement of the single-particle Green's functions.  
Since the interaction effects on the single-particle Green's function 
arise from the self-energy [Eq. (\ref{eq:G-Dyson})]
and the self-energy causes the QP damping [Eq. (\ref{eq:QPdamp-eq})],  
we analyze the damping dependence of 
$\sigma_{xy}^{\textrm{C}(\textrm{I})}$ or $\sigma_{xy}^{\textrm{S}(\textrm{I})}$. 
For that purpose, 
we need to analyze the damping dependence of 
$g_{2;acdb}^{ss^{\prime\prime}s^{\prime\prime\prime}s^{\prime}}(k;0)$ 
in $\sigma_{xy}^{\textrm{C}(\textrm{I})}$ or $\sigma_{xy}^{\textrm{S}(\textrm{I})}$ 
because the others are $O(\gamma^{0})$, 
where $\gamma$ is of the order of magnitude the QP damping~\cite{Eliashberg}.  
As explained, 
$g_{2;acdb}^{ss^{\prime\prime}s^{\prime\prime\prime}s^{\prime}}(k;0)$ has 
the leading damping dependence among several products of 
two single-particle Green's functions 
because of the limiting property of the product of the coherent parts 
of the retarded and the advanced single-particle Green's function. 
That leading damping dependence is given by~\cite{NA-review} 
\begin{align}
g_{2;acdb}^{ss^{\prime\prime}s^{\prime\prime\prime}s^{\prime}}(k;0)&\approx \
i2\pi \sum\limits_{\alpha,\beta}
u_{a\alpha c;d\beta b}^{ss^{\prime\prime}s^{\prime\prime\prime}s^{\prime}}(\boldk)
z_{\alpha}(\boldk)z_{\beta}(\boldk)\notag\\
&\times 
\dfrac{\delta(\epsilon-\xi_{\alpha}^{\ast}(\boldk))}
{\Delta \xi_{\beta \alpha}^{\ast}(\boldk)
+i[\gamma_{\alpha}^{\ast}(\boldk)+\gamma_{\beta}^{\ast}(\boldk)]},\label{eq:g2-approx-pre}
\end{align}
with $\Delta \xi_{\beta\alpha}^{\ast}(\boldk)\equiv \xi^{\ast}_{\beta}(\boldk)-\xi_{\alpha}^{\ast}(\boldk)$ 
and $u_{a\alpha c; d\beta b}^{ss^{\prime\prime}s^{\prime\prime\prime}s^{\prime}}(\boldk)\equiv 
(U_{\boldk})_{a\alpha}^{s}(U_{\boldk}^{\dagger})_{\alpha c}^{s^{\prime\prime\prime}}
(U_{\boldk})_{d\beta}^{s^{\prime\prime\prime}}(U_{\boldk}^{\dagger})_{\beta b}^{s^{\prime}}$.  
In deriving Eq. (\ref{eq:g2-approx-pre}), 
we have used Eq. (\ref{eq:coherent-G}) and the identity, 
\begin{align}
\lim\limits_{\delta\rightarrow 0+}[
\frac{1}{z-X+i\delta}-\frac{1}{z-X-i\delta}]=-2\pi i\delta(z-X).
\end{align}
Equation (\ref{eq:g2-approx-pre}) can be also rewritten as 
\begin{align}
g_{2;acdb}^{ss^{\prime\prime}s^{\prime\prime\prime}s^{\prime}}(k;0)
&\approx \
i\pi \sum\limits_{\alpha,\beta}
u_{a\alpha c;d\beta b}^{ss^{\prime\prime}s^{\prime\prime\prime}s^{\prime}}(\boldk)
z_{\alpha}(\boldk)z_{\beta}(\boldk)\notag\\
&\times 
[\delta(\epsilon-\xi_{\alpha}^{\ast}(\boldk))+\delta(\epsilon-\xi_{\beta}^{\ast}(\boldk))]\notag\\
&\times 
\dfrac{\Delta \xi_{\beta \alpha}^{\ast}(\boldk)
-i[\gamma_{\alpha}^{\ast}(\boldk)+\gamma_{\beta}^{\ast}(\boldk)]}
{\Delta \xi_{\beta \alpha}^{\ast}(\boldk)^{2}
+[\gamma_{\alpha}^{\ast}(\boldk)+\gamma_{\beta}^{\ast}(\boldk)]^{2}},
\label{eq:g2-approx} 
\end{align}
by using two equalities, 
\begin{align}
G_{ab}^{(\textrm{R})ss^{\prime}}(k)=G_{ba}^{(\textrm{A})s^{\prime}s}(k)^{\ast},
\end{align}
and 
\begin{align}
u_{a\alpha c; d\beta b}^{ss^{\prime\prime}s^{\prime\prime\prime}s^{\prime}}(\boldk)
=u_{b\beta d;c\alpha a}^{s^{\prime}s^{\prime\prime\prime}s^{\prime\prime}s}(\boldk)^{\ast}.
\end{align}
To see the finite components of $g_{2;acdb}^{ss^{\prime\prime}s^{\prime\prime\prime}s^{\prime}}(k;0)$ 
in $\sigma_{xy}^{\textrm{C}(\textrm{I})}$ or $\sigma_{xy}^{\textrm{S}(\textrm{I})}$, 
we should detect 
the terms odd with respect to $k_{x}$ and $k_{y}$.  
This is because $\Lambda_{x;2;ba}^{\textrm{C}(0)s^{\prime}s}(k;0)$ 
and $\Lambda_{x;2;ba}^{\textrm{S}(0)s^{\prime}s}(k;0)$ 
are odd with respect to $k_{x}$ due to the $k_{x}$ derivative 
in $(v_{\boldk x})_{ba}^{ss}$ [see Eqs. (\ref{eq:JC0all-symbolic}) 
and (\ref{eq:JS0all-symbolic})]
and $\Lambda_{y;2;cd}^{\textrm{C}; s^{\prime\prime}s^{\prime\prime\prime}}(k;0)$ 
is odd with respect to $k_{y}$ due to the $k_{y}$ derivative in $(v_{\boldk y})_{ba}^{ss}$
[see Eq. (\ref{eq:3VC-real})], i.e.
the terms other than $\Lambda_{x;2;ba}^{\textrm{C}(0)s^{\prime}s}(k;0)$ 
[$\Lambda_{x;2;ba}^{\textrm{S}(0)s^{\prime}s}(k;0)$] and 
$\Lambda_{y;2;cd}^{\textrm{C}; s^{\prime\prime}s^{\prime\prime\prime}}(k;0)$ 
in $\sigma_{xy}^{\textrm{C}(\textrm{I})}$ [$\sigma_{xy}^{\textrm{S}(\textrm{I})}$] 
should be odd with respect to $k_{x}$ and $k_{y}$ 
to obtain finite terms 
after taking the $\boldk$ summation. 
Note that 
an integrand of the $\boldk$ summation should be even about each $k_{\eta}$ 
to obtain the finite value. 
Since such odd terms arise from the terms proportional to 
$u_{a\alpha c;d\beta b}^{ss^{\prime\prime}s^{\prime\prime\prime}s^{\prime}}(\boldk)
\Delta \xi_{\beta \alpha}^{\ast}(\boldk)$ $(\alpha\neq \beta)$ 
in Eq. (\ref{eq:g2-approx-pre}), 
the dominant multiband excitations for $\sigma_{xy}^{\textrm{C}(\textrm{I})}$ 
or $\sigma_{xy}^{\textrm{S}(\textrm{I})}$ are interband; 
to obtain finite odd terms arising from those terms, 
the hopping integral with the odd mirror symmetry is necessary. 
For further argument, 
let us consider a simple but sufficient situation: 
the finite terms of $\sigma_{xy}^{\textrm{C}}$ or $\sigma_{xy}^{\textrm{S}}$ 
come from 
the interband excitations only at $\boldk=\boldk_{0}$. 
In this situation, 
the leading terms of $g_{2;acdb}^{ss^{\prime\prime}s^{\prime\prime\prime}s^{\prime}}(k;0)$ 
in $\sigma_{xy}^{\textrm{C}(\textrm{I})}$ or $\sigma_{xy}^{\textrm{S}(\textrm{I})}$ become 
$O(\gamma^{-2})$ 
in $|\Delta \xi_{\beta\alpha}^{\ast}(\boldk_{0})|\ll 
[\gamma_{\alpha}^{\ast}(\boldk_{0})+\gamma_{\beta}^{\ast}(\boldk_{0})]$, 
and $O(\gamma^{0})$ in $|\Delta \xi_{\beta\alpha}^{\ast}(\boldk_{0})|\gg 
[\gamma_{\alpha}^{\ast}(\boldk_{0})+\gamma_{\beta}^{\ast}(\boldk_{0})]$. 
As a result, 
$\sigma_{xy}^{\textrm{C}}$ or $\sigma_{xy}^{\textrm{S}}$ becomes 
$O(\gamma^{-2})$ in the former limit 
and $O(\gamma^{0})$ in the later limit. 
More precisely, the leading terms of 
$\sigma_{xy}^{\textrm{C}(\textrm{I})}$ 
and $\sigma_{xy}^{\textrm{S}(\textrm{I})}$ 
in $|\Delta \xi_{\beta\alpha}^{\ast}(\boldk_{0})|\ll 
[\gamma_{\alpha}^{\ast}(\boldk_{0})+\gamma_{\beta}^{\ast}(\boldk_{0})]$ 
are given by
\begin{align}
\sigma_{xy}^{\textrm{C}(\textrm{I})}&\approx 
\frac{-(-e)^{2}}{2N}
\sum\limits_{\alpha}
\sum\limits_{\beta\neq \alpha}
\frac{\Delta \xi_{\beta\alpha}^{\ast}(\boldk_{0})}
{[\gamma_{\alpha}^{\ast}(\boldk_{0})+\gamma_{\beta}^{\ast}(\boldk_{0})]^{2}}\notag\\
&\times 
\{
\textrm{Im}[\tilde{\Lambda}_{x;2;\beta\alpha}^{\textrm{C}(0)}(\boldk_{0},\xi_{\alpha}^{\ast}(\boldk_{0}))
\tilde{\Lambda}_{y;2;\alpha\beta}^{\textrm{C}}(\boldk_{0},\xi_{\alpha}^{\ast}(\boldk_{0}))]\notag\\
+
&\textrm{Im}[\tilde{\Lambda}_{x;2;\beta\alpha}^{\textrm{C}(0)}(\boldk_{0},\xi_{\beta}^{\ast}(\boldk_{0}))
\tilde{\Lambda}_{y;2;\alpha\beta}^{\textrm{C}}(\boldk_{0},\xi_{\beta}^{\ast}(\boldk_{0}))]
\},\label{eq:sigCXY1-limit1}
\end{align}
and 
\begin{align}
\sigma_{xy}^{\textrm{S}(\textrm{I})}&\approx 
\frac{-(-e)}{2^{2}N}
\sum\limits_{\alpha}
\sum\limits_{\beta\neq \alpha}
\frac{\Delta \xi_{\beta\alpha}^{\ast}(\boldk_{0})}
{[\gamma_{\alpha}^{\ast}(\boldk_{0})+\gamma_{\beta}^{\ast}(\boldk_{0})]^{2}}\notag\\
&\times 
\{
\textrm{Im}[\tilde{\Lambda}_{x;2;\beta\alpha}^{\textrm{S}(0)}(\boldk_{0},\xi_{\alpha}^{\ast}(\boldk_{0}))
\tilde{\Lambda}_{y;2;\alpha\beta}^{\textrm{C}}(\boldk_{0},\xi_{\alpha}^{\ast}(\boldk_{0}))]\notag\\
+
&\textrm{Im}[\tilde{\Lambda}_{x;2;\beta\alpha}^{\textrm{S}(0)}(\boldk_{0},\xi_{\beta}^{\ast}(\boldk_{0}))
\tilde{\Lambda}_{y;2;\alpha\beta}^{\textrm{C}}(\boldk_{0},\xi_{\beta}^{\ast}(\boldk_{0}))]
\},\label{eq:sigSXY1-limit1}
\end{align}
respectively; in $|\Delta \xi_{\beta\alpha}^{\ast}(\boldk_{0})|\gg 
[\gamma_{\alpha}^{\ast}(\boldk_{0})+\gamma_{\beta}^{\ast}(\boldk_{0})]$, 
$\sigma_{xy}^{\textrm{C}(\textrm{I})}$ and $\sigma_{xy}^{\textrm{S}(\textrm{I})}$ are given by
\begin{align}
\sigma_{xy}^{\textrm{C}(\textrm{I})}&\approx 
\frac{-(-e)^{2}}{2N}
\sum\limits_{\alpha}
\sum\limits_{\beta\neq \alpha}
\frac{1}{\Delta \xi_{\beta\alpha}^{\ast}(\boldk_{0})}\notag\\
&\times 
\{
\textrm{Im}[\tilde{\Lambda}_{x;2;\beta\alpha}^{\textrm{C}(0)}(\boldk_{0},\xi_{\alpha}^{\ast}(\boldk_{0}))
\tilde{\Lambda}_{y;2;\alpha\beta}^{\textrm{C}}(\boldk_{0},\xi_{\alpha}^{\ast}(\boldk_{0}))]\notag\\
+
&\textrm{Im}[\tilde{\Lambda}_{x;2;\beta\alpha}^{\textrm{C}(0)}(\boldk_{0},\xi_{\beta}^{\ast}(\boldk_{0}))
\tilde{\Lambda}_{y;2;\alpha\beta}^{\textrm{C}}(\boldk_{0},\xi_{\beta}^{\ast}(\boldk_{0}))]
\},\label{eq:sigCXY1-limit2}
\end{align}
and 
\begin{align}
\sigma_{xy}^{\textrm{S}(\textrm{I})}&\approx 
\frac{-(-e)}{2^{2}N}
\sum\limits_{\alpha}
\sum\limits_{\beta\neq \alpha}
\frac{1}{\Delta \xi_{\beta\alpha}^{\ast}(\boldk_{0})}\notag\\
&\times 
\{
\textrm{Im}[\tilde{\Lambda}_{x;2;\beta\alpha}^{\textrm{S}(0)}(\boldk_{0},\xi_{\alpha}^{\ast}(\boldk_{0}))
\tilde{\Lambda}_{y;2;\alpha\beta}^{\textrm{C}}(\boldk_{0},\xi_{\alpha}^{\ast}(\boldk_{0}))]\notag\\
+
&\textrm{Im}[\tilde{\Lambda}_{x;2;\beta\alpha}^{\textrm{S}(0)}(\boldk_{0},\xi_{\beta}^{\ast}(\boldk_{0}))
\tilde{\Lambda}_{y;2;\alpha\beta}^{\textrm{C}}(\boldk_{0},\xi_{\beta}^{\ast}(\boldk_{0}))]
\},\label{eq:sigSXY1-limit2}
\end{align}
respectively. 
In those equations, 
we have introduced three quantities, 
\begin{align}
&\tilde{\Lambda}_{\nu;2;\beta\alpha}^{\textrm{C}(0)}(\boldk,\epsilon)
=
\sqrt{z_{\beta}(\boldk)z_{\alpha}(\boldk)}\notag\\
&\times 
\sum\limits_{a,b,s,s^{\prime}}
(U_{\boldk}^{\dagger})_{\beta b}^{s^{\prime}}
\Lambda_{\nu;2;ba}^{\textrm{C}(0)s^{\prime}s}(k;0)
(U_{\boldk})_{a\alpha}^{s},\label{eq:LambC0-tild}\\
&\tilde{\Lambda}_{\nu;2;\alpha\beta}^{\textrm{C}}(\boldk,\epsilon)
=
\sqrt{z_{\alpha}(\boldk)z_{\beta}(\boldk)}\notag\\
&\times 
\sum\limits_{c,d,s^{\prime\prime},s^{\prime\prime\prime}}
(U_{\boldk}^{\dagger})_{\alpha c}^{s^{\prime\prime}}
\Lambda_{\nu;2;cd}^{\textrm{C};s^{\prime\prime}s^{\prime\prime\prime}}(k;0)
(U_{\boldk})_{d\beta}^{s^{\prime\prime\prime}},\label{eq:LambC-tild}
\end{align} 
and 
\begin{align}
&\tilde{\Lambda}_{\nu;2;\beta\alpha}^{\textrm{S}(0)}(\boldk,\epsilon)
=
\sqrt{z_{\beta}(\boldk)z_{\alpha}(\boldk)}\notag\\
&\times 
\sum\limits_{a,b,s,s^{\prime}}
(U_{\boldk}^{\dagger})_{\beta b}^{s^{\prime}}
\Lambda_{\nu;2;ba}^{\textrm{S}(0)s^{\prime}s}(k;0)
(U_{\boldk})_{a\alpha}^{s}.\label{eq:LambS0-tild}
\end{align}
For more complex situations 
with the interband excitations at $\boldk=\boldk_{0}$, $\boldk_{1}$,$\cdots$, $\boldk_{K-1}$, 
we need to apply the above argument for the simple situation 
to each term of the interband excitations 
at $\boldk_{j}$ and combine each other's damping dependences: 
if at least one of the interband excitations 
satisfies $|\Delta \xi_{\beta\alpha}^{\ast}(\boldk_{j})|\ll 
[\gamma_{\alpha}^{\ast}(\boldk_{j})+\gamma_{\beta}^{\ast}(\boldk_{j})]$, 
$\sigma_{xy}^{\textrm{C}(\textrm{I})}$ or $\sigma_{xy}^{\textrm{S}(\textrm{I})}$ 
becomes damping-dependent; 
on the other hand, 
if all the interband excitations satisfy 
$|\Delta \xi_{\beta\alpha}^{\ast}(\boldk_{j})|\gg 
[\gamma_{\alpha}^{\ast}(\boldk_{j})+\gamma_{\beta}^{\ast}(\boldk_{j})]$, 
$\sigma_{xy}^{\textrm{C}(\textrm{I})}$ or $\sigma_{xy}^{\textrm{S}(\textrm{I})}$ 
is damping-independent. 
Thus, 
the electron-electron interaction 
causes the finite damping dependences of $\sigma_{xy}^{\textrm{C}}$ and $\sigma_{xy}^{\textrm{S}}$ 
at high temperatures even without impurities. 
Furthermore, 
since the interaction-induced QP damping 
decreases with decreasing temperature~\cite{AGD}, 
the electron-electron interaction causes 
the emergence of the temperature dependences of 
$\sigma_{xy}^{\textrm{C}}$ and $\sigma_{xy}^{\textrm{S}}$ and 
a crossover from damping-dependent to damping-independent 
$\sigma_{xy}^{\textrm{C}}$ or $\sigma_{xy}^{\textrm{S}}$ 
even without impurities (see Fig. \ref{fig:Fig2}). 

Then, we see the interaction effect 
due to the replacement of the spin current for $\sigma_{xy}^{\textrm{S}(\textrm{I})}$. 
This is related to the effects of the SCD 
because the difference between 
$\sigma_{xy}^{\textrm{C}(\textrm{I})}$ and $\sigma_{xy}^{\textrm{S}(\textrm{I})}$ 
comes from 
the difference between $(-e)\Lambda_{x;2;ba}^{\textrm{C}(0)s^{\prime}s}(k;0)$ 
and $\frac{1}{2}\Lambda_{x;2;ba}^{\textrm{S}(0)s^{\prime}s}(k;0)$ 
[see Eqs. (\ref{eq:SigXYC-I}) and (\ref{eq:SigXYS-I})]. 
Actually, 
rewriting $\Lambda_{2;x;ba}^{\textrm{S}(0)s^{\prime}s}(k;0)$ 
by using $\Lambda_{2;x;ba}^{\textrm{C}(0)s^{\prime}s}(k;0)$ as
\begin{align}
&\Lambda_{2;x;ba}^{\textrm{S}(0)s^{\prime}s}(k;0)
=\
\textrm{sgn}(s)
\Lambda_{2;x;ba}^{\textrm{C}(0)s^{\prime}s}(k;0)\notag\\
&-
2
\textrm{sgn}(s)
\sum\limits_{k^{\prime}}
\sum\limits_{A,B}
\alpha_{2;baAB}^{(0)s^{\prime}s-s-s}(k,k^{\prime};0)(v_{\boldk^{\prime}x})_{AB}^{-s-s},\label{eq:JS-SCD}
\end{align}
and substituting Eq. (\ref{eq:JS-SCD}) into Eq. (\ref{eq:SigXYS-I}), 
we can show that 
the second term of Eq. (\ref{eq:JS-SCD}) 
leads to a SCD-induced correction of $\sigma_{xy}^{\textrm{S}(\textrm{I})}$, 
$\Delta \sigma_{xy}^{(\textrm{SCD})}$: 
\begin{align}
\sigma_{xy}^{\textrm{S}(\textrm{I})}
=&\
\frac{1}{2(-e)}
\sum\limits_{\{s \}}
\textrm{sgn}(s)
(-e)^{2}
\sum\limits_{k}
\sum\limits_{\{a \}}
\Bigl(-\dfrac{\partial f(\epsilon)}{\partial \epsilon}\Bigr)\notag\\
&\times 
\Lambda_{x;2;ba}^{\textrm{C}(0)s^{\prime}s}(k;0)
g_{2;acdb}^{ss^{\prime\prime}s^{\prime\prime\prime}s^{\prime}}(k;0)
\Lambda_{y;2;cd}^{\textrm{C}; s^{\prime\prime}s^{\prime\prime\prime}}(k;0)\notag\\
&-
\frac{(-e)}{N}
\sum\limits_{\{s \}}
\textrm{sgn}(s)
\sum\limits_{k}
\sum\limits_{\{a \}}
\Bigl(-\dfrac{\partial f(\epsilon)}{\partial \epsilon}\Bigr)\notag\\
&\times 
\sum\limits_{k^{\prime}}
\sum\limits_{A,B}
\alpha_{2;baAB}^{(0)s^{\prime}s-s-s}(k,k^{\prime};0)
(v_{\boldk^{\prime} x})_{AB}^{-s-s}\notag\\
&\times 
g_{2;acdb}^{ss^{\prime\prime}s^{\prime\prime\prime}s^{\prime}}(k;0)
\Lambda_{y;2;cd}^{\textrm{C}; s^{\prime\prime}s^{\prime\prime\prime}}(k;0)\notag\\
=&\ 
\frac{1}{2(-e)}
\sum\limits_{\{s\}}
\textrm{sgn}(s)(\sigma_{xy}^{\textrm{C}(\textrm{I})})^{s^{\prime}ss^{\prime\prime}s^{\prime\prime\prime}}
+\Delta \sigma_{xy}^{(\textrm{SCD})},\label{eq:SigXYS-SCD}
\end{align} 
with spin-decomposed component of $\sigma_{xy}^{\textrm{C}}$, 
$(\sigma_{xy}^{\textrm{C}})^{s^{\prime}ss^{\prime\prime}s^{\prime\prime\prime}}$, 
defined as $\sigma_{xy}^{\textrm{C}(\textrm{I})}=
\sum_{\{s\}}(\sigma_{xy}^{\textrm{C}(\textrm{I})})^{s^{\prime}ss^{\prime\prime}s^{\prime\prime\prime}}$. 

I believe this interpretation is appropriate because of the following arguments. 
Since the SCD~\cite{SCD-first,SCD-review} affects only spin transports, 
it is reasonable to suppose that 
the difference between $\sigma_{xy}^{\textrm{S}(\textrm{I})}$ and 
$\sigma_{xy}^{\textrm{C}(\textrm{I})}$ is related to the effects of the SCD 
on the Fermi surface term. 
In addition, 
it is consistent with the general property of the SCD in metals 
to suppose that 
the second term of Eq. (\ref{eq:JS-SCD}) causes the correction due to the SCD 
because the second term represents the correction of the spin current 
due to the multiple scattering of the electron-electron interaction between different spins 
(see the second term for $s^{\prime}=s$). 
Here the general property is that 
only such multiple scattering causes the SCD in metals 
because for the onsite bare electron-electron interactions such as the Hubbard interactions  
the multiple scattering is necessary to obtain 
the finite momentum transfer. 
Note that 
this general property of metals 
indicates the importance of the momentum dependence of the self-energy 
due to the electron-electron interaction in discussing the SCD in metals 
because that momentum dependence is necessary to obtain 
finite second term of Eq. (\ref{eq:JS-SCD}). 

Finally, let us see the interaction effects due to the other modifications, 
i.e. the replacement of the $x$ component of the charge current 
in $\sigma_{xy}^{\textrm{C}(\textrm{I})}$ 
and the replacement of the $y$ component of the charge current 
in $\sigma_{xy}^{\textrm{C}(\textrm{I})}$ or $\sigma_{xy}^{\textrm{S}(\textrm{I})}$. 
First, 
the former replacement causes 
a magnitude decrease of $\sigma_{xy}^{\textrm{C}(\textrm{I})}$ 
from a noninteracting value. 
This is 
because the correction term in $\Lambda_{x;2;ba}^{\textrm{C}(0)s^{\prime}s}(k;0)$, 
the second term of Eq. (\ref{eq:JC0all-symbolic-G1}) for $\omega=0$, 
is related to the $k_{x}$ derivative of the real part of the self-energy 
due to a Ward identity~\cite{Nozieres,Kuroda-Nagi} 
and because 
its effect on the charge current, the renormalization of the group velocity, 
reduces a magnitude of the charge current~\cite{Eliashberg}. 
Then, 
the latter replacement 
maybe changes not only the magnitude of 
$\sigma_{xy}^{\textrm{C}(\textrm{I})}$ or $\sigma_{xy}^{\textrm{S}(\textrm{I})}$ 
but also its sign in some cases near an antiferromagnetic quantum-critical point 
due to the similar mechanism for the weak-field usual Hall effect~\cite{Kontani-review}. 
For the weak-field usual Hall effect without the onsite SOC, 
the angle change of the charge current can be induced 
near the antiferromagnetic quantum-critical point 
due to the momentum dependence of the irreducible four-point vertex function, 
and that angle change causes the sign change of the usual-Hall conductivity~\cite{Kontani-review}. 
To check this possibility for the intrinsic AHE or SHE, 
we need a numerical calculation 
for $\sigma_{xy}^{\textrm{C}(\textrm{I})}$ or $\sigma_{xy}^{\textrm{S}(\textrm{I})}$ 
by applying an approximation appropriate near the antiferromagnetic quantum-critical point 
to the particular band structure. 
Since that is a next step, 
it remains an important issue to clarify the interaction effects of that replacement 
on $\sigma_{xy}^{\textrm{C}(\textrm{I})}$ or $\sigma_{xy}^{\textrm{S}(\textrm{I})}$. 

\begin{figure*}[tb]
\includegraphics[width=150mm]{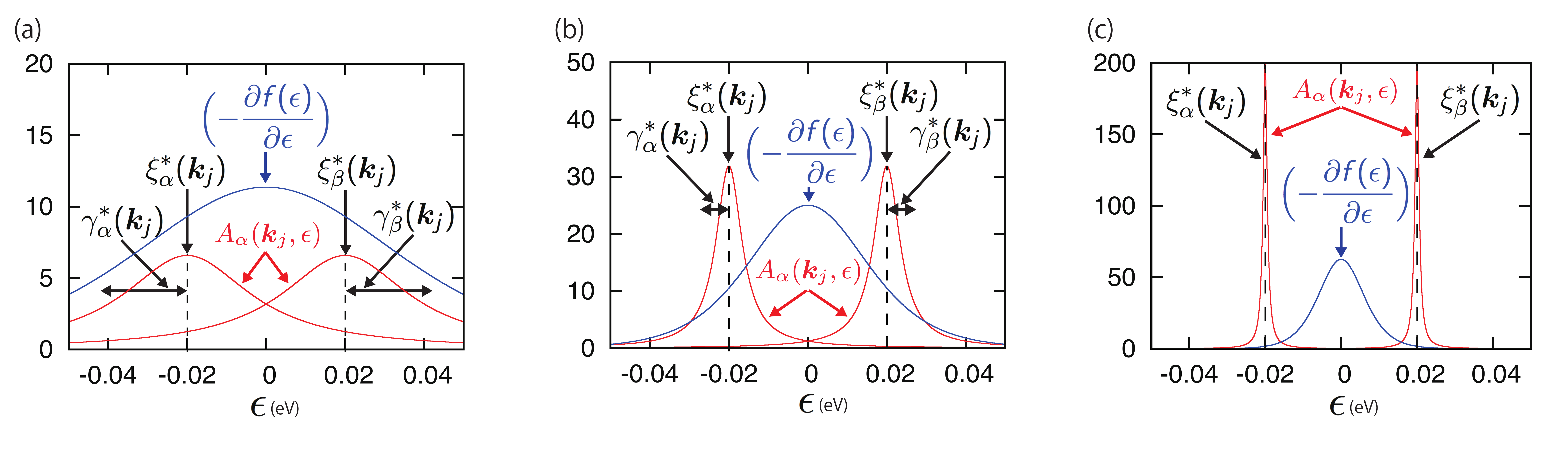}
\vspace{-6pt}
\caption{ 
QP spectral function, 
$A_{\alpha}(\boldk,\epsilon)$, and $(-\frac{\partial f(\epsilon)}{\partial \epsilon})$ 
for the interband excitation at $\boldk=\boldk_{j}$ 
in (a) the high-$T$ case, (b) the intermediate case, 
and (c) the low-$T$ case. 
In those panels, 
the QP spectral function is given by 
$A_{\alpha}(\boldk_{j},\epsilon)=\frac{z_{\alpha}(\boldk)}{\pi}
\frac{\gamma_{\alpha}^{\ast}(\boldk)}{[\epsilon-\xi_{\alpha}^{\ast}(\boldk)]^{2}
+\gamma_{\alpha}^{\ast}(\boldk)^{2}}$;  
the parameters are chosen as $z_{\alpha}(\boldk_{j})=z_{\beta}(\boldk_{j})=0.4$, 
$\xi_{\alpha}^{\ast}(\boldk_{j})=-\xi_{\beta}^{\ast}(\boldk_{j})=0.02$ eV, 
and $\gamma_{\alpha}^{\ast}(\boldk_{j})=\gamma_{\beta}^{\ast}(\boldk_{j})=40T^{2}$; 
$T$ in panels (a), (b), and (c) are $0.004$, $0.01$, and $0.022$ (eV), respectively. 
}
\label{fig:Fig3}
\end{figure*}
\subsection{Applicability of \'{E}liashberg's approximation}

We turn to applicability of \'{E}liashberg's approximation~\cite{Eliashberg,NA-review,NA-full} 
for $\sigma_{xy}^{\textrm{C}}$ or $\sigma_{xy}^{\textrm{S}}$. 
First, 
we should restrict arguments to cases for $\gamma_{\alpha}^{\ast}(\boldk_{\textrm{F}})/T < 1$ 
because 
\'{E}liashberg's approximation is reasonable 
only for $\gamma_{\alpha}^{\ast}(\boldk_{\textrm{F}})/T < 1$ (see Section III A).
Thus, 
the gray triangle region in Fig. \ref{fig:Fig2} 
is the outside of the applicable region.  
Then, 
there are two key factors 
to argue 
whether $\sigma_{xy}^{\textrm{C}(\textrm{I})}$ or $\sigma_{xy}^{\textrm{S}(\textrm{I})}$ 
become finite or not, 
i.e.  
the broadening of the QP spectra due to the QP damping 
and the broadening of $(-\frac{\partial f(\epsilon)}{\partial \epsilon})$ due to temperature. 
This is because of the following three facts: 
$\sigma_{xy}^{\textrm{C}(\textrm{I})}$ or $\sigma_{xy}^{\textrm{S}(\textrm{I})}$ 
consists of the integral of $(-\frac{\partial f(\epsilon)}{\partial \epsilon})$ 
for interband excitations; 
such integral becomes finite only for the finite overlap 
between the QP spectra of the two bands; 
that overlap arises from the above two key factors. 
Thus, 
in the range of $\gamma_{\alpha}^{\ast}(\boldk_{\textrm{F}})/T < 1$, 
we have three distinct cases, 
i.e. high-$T$ case, intermediate-$T$ case, and low-$T$ case: 
in the high-$T$ case, 
both of the two factors 
lead to a finite overlap between 
the QP spectra of the two bands 
for at least an interband excitation; 
in the intermediate-$T$ case, 
the finite overlap arises only from 
the broadening of $(-\frac{\partial f(\epsilon)}{\partial \epsilon})$; 
in the low-$T$ case, 
the overlap becomes negligible. 
For example, 
those three cases for the interband excitation at $\boldk=\boldk_{j}$ 
are shown schematically in Figs. \ref{fig:Fig3}(a), \ref{fig:Fig3}(b), 
and \ref{fig:Fig3}(c). 
As we see from Figs. \ref{fig:Fig3}(a) and \ref{fig:Fig3}(b), 
\'{E}liashberg's approximation gives 
finite $\sigma_{xy}^{\textrm{C}(\textrm{I})}$ or $\sigma_{xy}^{\textrm{S}(\textrm{I})}$ 
in the high-$T$ case and the intermediate-$T$ case. 
However, in the low-$T$ case, 
corresponding to Fig. \ref{fig:Fig3}(c), 
$\sigma_{xy}^{\textrm{C}(\textrm{I})}$ or $\sigma_{xy}^{\textrm{S}(\textrm{I})}$ 
become very small; 
thus, \'{E}liashberg's approximation becomes insufficient. 
Since the high-$T$ case, the intermediate-$T$ case, 
and the low-$T$ case correspond to, respectively, 
the upper, the middle, and the lower region 
of the left triangle of Fig. \ref{fig:Fig2}, 
\'{E}liashberg's approximation 
is sufficient to analyze the intrinsic AHE and SHE 
in the upper and the middle region; 
for the analysis in the lower region, 
we need to take \'{E}liashberg's approximation 
a step further. 

\subsection{Approximation beyond \'{E}liashberg's approximation}

Starting to explain the points missing in \'{E}liashberg's approximation 
and being important in the low-$T$ case, 
we construct an approximation beyond \'{E}liashberg's approximation 
and derive the approximate expressions of $\sigma_{xy}^{\textrm{C}}$ and $\sigma_{xy}^{\textrm{S}}$ 
in this approximation. 
Then, 
we see the damping dependence and the effects of the SCD in this approximation. 
Furthermore, 
by comparison with the noninteracting Fermi sea term, 
we deduce how the electron-electron interaction affects the Fermi sea term 
of $\sigma_{xy}^{\textrm{C}}$ or $\sigma_{xy}^{\textrm{S}}$. 

\subsubsection{Formulaiton}
As we see in Section III B, 
in the low-$T$ case, where 
temperature is low and the QP damping is small, 
the term of $\sigma_{xy}^{\textrm{C}}$ or $\sigma_{xy}^{\textrm{S}}$ 
considered in \'{E}liashberg's approximation 
becomes very small. 
For analyses in such case, 
we need to use an appropriate approximation 
beyond \'{E}liashberg's approximation. 
In particular, 
we should take account of the terms 
of the interband excitations including $f(\epsilon)$ 
because those terms remain 
finite even in clearn and low-$T$ case. 
Since \'{E}liashberg's approximation 
has succeeded in getting reasonable descriptions 
of several transports of interacting metals 
(e.g., the resistivity~\cite{Eliashberg,Yamada-Yosida} 
and the weak-field usual Hall effect~\cite{Fukuyama,Kohno-Yamada}), 
I suppose that 
\'{E}liashberg's approximation 
is not so bad even for the description of the intrinsic AHE or SHE, 
and that 
an approximation appropriate 
for analyses in the low-$T$ case can be obtained 
by extending \'{E}liashberg's approximation. 

On the basis of those suppositions, 
we construct an approximation beyond \'{E}liashberg's approximation 
by going back to the exact expression of 
$\tilde{K}_{xy}^{\textrm{C}(\textrm{R})}(\omega)$ 
or
$\tilde{K}_{xy}^{\textrm{S}(\textrm{R})}(\omega)$ 
[Eq. (\ref{eq:KC-real2}) or (\ref{eq:KS-real2})] 
and taking account of not only the terms considered in \'{E}liashberg's approximation 
but also the terms leading among 
the terms of the Fermi sea integral. 
Such leading terms come from the terms 
proportional to the $\omega$-linear term 
of $g_{l;acdb}^{ss^{\prime\prime}s^{\prime\prime\prime}s^{\prime}}(k;\omega)$
or $\Lambda_{y;l;cd}^{\textrm{C}(0)s^{\prime\prime}s^{\prime\prime\prime}}(k;\omega)$ ($l=1,3$) 
in the first term of Eq. (\ref{eq:KC-real2}) or (\ref{eq:KS-real2}) 
because we need to take the leading $\omega$-linear terms 
to obtain $\sigma_{xy}^{\textrm{C}}$ or $\sigma_{xy}^{\textrm{S}}$ 
[see Eq. (\ref{eq:sigXY-C}) or (\ref{eq:sigXY-S})]. 
Thus, the terms leading among the terms of the Fermi sea integral 
leads to additional terms of $\sigma_{xy}^{\textrm{C}}$ and $\sigma_{xy}^{\textrm{S}}$. 

As a result, 
$\sigma_{xy}^{\textrm{C}}$ and $\sigma_{xy}^{\textrm{S}}$ 
in this approximation become 
$\sigma_{xy}^{\textrm{C}}=\sigma_{xy}^{\textrm{C}(\textrm{I})}+\sigma_{xy}^{\textrm{C}(\textrm{II})}$ 
and $\sigma_{xy}^{\textrm{S}}=\sigma_{xy}^{\textrm{S}(\textrm{I})}+\sigma_{xy}^{\textrm{S}(\textrm{II})}$, 
respectively, 
where $\sigma_{xy}^{\textrm{C}(\textrm{II})}$ is  
\begin{align}
&\sigma_{xy}^{\textrm{C}(\textrm{II})}
=
(-e)^{2}
\sum\limits_{k}
\sum\limits_{\{a \}}
\sum\limits_{\{s\}}
f(\epsilon)
\delta_{s^{\prime},s}(v_{\boldk x})_{ba}^{ss}
\sum\limits_{l=1,3}
\notag\\
\times &
\textrm{sgn}(l-2)\lim\limits_{\omega\rightarrow 0}
\frac{\partial}{\partial \omega}
[g_{l;acdb}^{ss^{\prime\prime}s^{\prime\prime\prime}s^{\prime}}(k;\omega)
\Lambda_{y;l;cd}^{\textrm{C}(0) s^{\prime\prime}s^{\prime\prime\prime}}(k;\omega)],\label{eq:SigXYC-II}
\end{align}
and $\sigma_{xy}^{\textrm{S}(\textrm{II})}$ is 
\begin{align}
&\sigma_{xy}^{\textrm{S}(\textrm{II})}
=
\frac{(-e)}{2}
\sum\limits_{k}
\sum\limits_{\{a \}}
\sum\limits_{\{s\}}
f(\epsilon)
\textrm{sgn}(s)
\delta_{s^{\prime},s}(v_{\boldk x})_{ba}^{ss}
\sum\limits_{l=1,3}
\notag\\
\times &
\textrm{sgn}(l-2)\lim\limits_{\omega\rightarrow 0}
\frac{\partial}{\partial \omega}
[g_{l;acdb}^{ss^{\prime\prime}s^{\prime\prime\prime}s^{\prime}}(k;\omega)
\Lambda_{y;l;cd}^{\textrm{C}(0) s^{\prime\prime}s^{\prime\prime\prime}}(k;\omega)].\label{eq:SigXYS-II}
\end{align}
For the direct comparison with the noninteracting Fermi-sea term, 
we rewrite part of the terms proportional to 
the $\omega$ derivative of $g_{l;acdb}^{ss^{\prime\prime}s^{\prime\prime\prime}s^{\prime}}(k;\omega)$ 
$(l=1,3)$ in Eqs. (\ref{eq:SigXYC-II}) and (\ref{eq:SigXYS-II}) as follows: 
\begin{align}
&\sum\limits_{l=1,3}\textrm{sgn}(l-2)
\lim\limits_{\omega\rightarrow 0}
\frac{\partial g_{l;acdb}^{ss^{\prime\prime}s^{\prime\prime\prime}s^{\prime}}(k;\omega)}{\partial \omega}
\Lambda_{y;l;cd}^{\textrm{C}(0) s^{\prime\prime}s^{\prime\prime\prime}}(k;0)\notag\\
=
&-\frac{\partial G_{ac}^{(\textrm{R})ss^{\prime\prime}}(k)}{\partial \epsilon}
G_{db}^{(\textrm{R})s^{\prime\prime\prime}s^{\prime}}(k)
\Lambda_{y;1;cd}^{\textrm{C}(0)s^{\prime\prime}s^{\prime\prime\prime}}(k;0)\notag\\
&+\frac{\partial G_{ac}^{(\textrm{A})ss^{\prime\prime}}(k)}{\partial \epsilon}
G_{db}^{(\textrm{A})s^{\prime\prime\prime}s^{\prime}}(k)
\Lambda_{y;3;cd}^{\textrm{C}(0)s^{\prime\prime}s^{\prime\prime\prime}}(k;0),
\end{align}
where we use the identity, 
\begin{align}
\lim\limits_{\omega\rightarrow 0}
\frac{\partial F(\epsilon+\omega)}{\partial \omega}
=\frac{\partial F(\epsilon)}{\partial \epsilon}.
\end{align}
Namely, 
Eqs. (\ref{eq:SigXYC-II}) and (\ref{eq:SigXYS-II}) become 
\begin{align}
&\sigma_{xy}^{\textrm{C}(\textrm{II})}
=
-(-e)^{2}
\sum\limits_{k}
\sum\limits_{\{a \}}
\sum\limits_{\{s\}}
f(\epsilon)
\delta_{s^{\prime},s}(v_{\boldk x})_{ba}^{ss}
\notag\\
\times &
\Bigl[
\frac{\partial G_{ac}^{(\textrm{R})ss^{\prime\prime}}(k)}{\partial \epsilon}
G_{db}^{(\textrm{R})s^{\prime\prime\prime}s^{\prime}}(k)
\Lambda_{y;1;cd}^{\textrm{C}(0) s^{\prime\prime}s^{\prime\prime\prime}}
(k;0)\notag\\
&-\frac{\partial G_{ac}^{(\textrm{A})ss^{\prime\prime}}(k)}{\partial \epsilon}
G_{db}^{(\textrm{A})s^{\prime\prime\prime}s^{\prime}}(k)
\Lambda_{y;3;cd}^{\textrm{C}(0) s^{\prime\prime}s^{\prime\prime\prime}}
(k;0)\notag\\
&+G_{ac}^{(\textrm{R})ss^{\prime\prime}}(k)
G_{db}^{(\textrm{R})s^{\prime\prime\prime}s^{\prime}}(k)
\lim\limits_{\omega\rightarrow 0}
\frac{\partial \Lambda_{y;1;cd}^{\textrm{C}(0) s^{\prime\prime}s^{\prime\prime\prime}}
(k;\omega)}{\partial \omega}\notag\\
&-G_{ac}^{(\textrm{A})ss^{\prime\prime}}(k)
G_{db}^{(\textrm{A})s^{\prime\prime\prime}s^{\prime}}(k)
\lim\limits_{\omega\rightarrow 0}
\frac{\partial \Lambda_{y;3;cd}^{\textrm{C}(0) s^{\prime\prime}s^{\prime\prime\prime}}
(k;\omega)}{\partial \omega}
\Bigr],\label{eq:SigXYC-II-direct}
\end{align}
and 
\begin{align}
&\sigma_{xy}^{\textrm{S}(\textrm{II})}
=
-\frac{(-e)}{2}
\sum\limits_{k}
\sum\limits_{\{a \}}
\sum\limits_{\{s\}}
f(\epsilon)
\textrm{sgn}(s)
\delta_{s^{\prime},s}(v_{\boldk x})_{ba}^{ss}
\notag\\
\times &
\Bigl[
\frac{\partial G_{ac}^{(\textrm{R})ss^{\prime\prime}}(k)}{\partial \epsilon}
G_{db}^{(\textrm{R})s^{\prime\prime\prime}s^{\prime}}(k)
\Lambda_{y;1;cd}^{\textrm{C}(0) s^{\prime\prime}s^{\prime\prime\prime}}
(k;0)\notag\\
&-\frac{\partial G_{ac}^{(\textrm{A})ss^{\prime\prime}}(k)}{\partial \epsilon}
G_{db}^{(\textrm{A})s^{\prime\prime\prime}s^{\prime}}(k)
\Lambda_{y;3;cd}^{\textrm{C}(0) s^{\prime\prime}s^{\prime\prime\prime}}
(k;0)\notag\\
&+G_{ac}^{(\textrm{R})ss^{\prime\prime}}(k)
G_{db}^{(\textrm{R})s^{\prime\prime\prime}s^{\prime}}(k)
\lim\limits_{\omega\rightarrow 0}
\frac{\partial \Lambda_{y;1;cd}^{\textrm{C}(0) s^{\prime\prime}s^{\prime\prime\prime}}
(k;\omega)}{\partial \omega}\notag\\
&-G_{ac}^{(\textrm{A})ss^{\prime\prime}}(k)
G_{db}^{(\textrm{A})s^{\prime\prime\prime}s^{\prime}}(k)
\lim\limits_{\omega\rightarrow 0}
\frac{\partial \Lambda_{y;3;cd}^{\textrm{C}(0) s^{\prime\prime}s^{\prime\prime\prime}}
(k;\omega)}{\partial \omega}
\Bigr],\label{eq:SigXYS-II-direct}
\end{align}
respectively. 

\subsubsection{Interaction effects}

Before comparing the derived Fermi sea term with the noninteracting Fermi sea term, 
we analyze the damping dependence of 
$\sigma_{xy}^{\textrm{C}(\textrm{II})}$ or $\sigma_{xy}^{\textrm{S}(\textrm{II})}$ 
and the effects of the SCD on $\sigma_{xy}^{\textrm{S}(\textrm{II})}$ 
in order to clarify how the two important properties obtained in 
\'{E}liashberg's approximation modify in the low-$T$ case. 
Those properties are 
the crossover from the damping-dependent to the damping-independent 
$\sigma_{xy}^{\textrm{C}}$ or $\sigma_{xy}^{\textrm{S}}$ 
with decreasing temperature 
and the correction term of $\sigma_{xy}^{\textrm{S}}$, as shown in Section III A 2. 

First,  
$\sigma_{xy}^{\textrm{C}(\textrm{II})}$ and $\sigma_{xy}^{\textrm{S}(\textrm{II})}$ 
become O$(\gamma^{0})$ 
in $T\rightarrow 0$ and $\gamma_{\alpha}^{\ast}(\boldk_{\textrm{F}})/T\rightarrow 0$ 
because we can neglect the damping dependence of 
$g_{1;acdb}^{ss^{\prime\prime}s^{\prime\prime\prime}s^{\prime}}(k;0)$ or 
$g_{3;acdb}^{ss^{\prime\prime}s^{\prime\prime\prime}s^{\prime}}(k;0)$~\cite{Eliashberg}. 
Since that result remains qualitatively the same in the low-$T$ case, 
$\sigma_{xy}^{\textrm{C}}$ and $\sigma_{xy}^{\textrm{S}}$ 
become 
$\sigma_{xy}^{\textrm{C}}\approx \sigma_{xy}^{\textrm{C}(\textrm{II})}=O(\gamma^{0})$ 
and $\sigma_{xy}^{\textrm{S}}\approx \sigma_{xy}^{\textrm{S}(\textrm{II})}=O(\gamma^{0})$, 
respectively. 
In addition, 
another crossover occurs at the orange line in Fig. \ref{fig:Fig2} 
because the dominant term changes from the Fermi-surface term to the Fermi-sea term 
with decreasing temperature. 
It should be noted that 
$\sigma_{xy}^{\textrm{C}(\textrm{II})}$ or $\sigma_{xy}^{\textrm{S}(\textrm{II})}$ 
becomes negligible compared with $\sigma_{xy}^{\textrm{C}(\textrm{I})}$ or 
$\sigma_{xy}^{\textrm{S}(\textrm{I})}$, respectively, 
if the QP damping is larger than the energy of the interband excitation 
which gives the finite contribution to  
$\sigma_{xy}^{\textrm{C}(\textrm{II})}$ or $\sigma_{xy}^{\textrm{S}(\textrm{II})}$. 
This is because in $\sigma_{xy}^{\textrm{C}(\textrm{II})}$ or $\sigma_{xy}^{\textrm{S}(\textrm{II})}$ 
we neglect the dependence on the QP damping 
as a result of the leading-term expansion of 
the products of the two single-particle Green's functions 
in terms of $\gamma_{\alpha}^{\ast}(\boldk_{\textrm{F}})/T\rightarrow 0$, 
while we consider the dependence on the energy of the interband excitation. 
Thus, 
when the QP damping is larger, 
$\sigma_{xy}^{\textrm{C}(\textrm{II})}$ or $\sigma_{xy}^{\textrm{S}(\textrm{II})}$ 
becomes less dominant than $\sigma_{xy}^{\textrm{C}(\textrm{I})}$ or $\sigma_{xy}^{\textrm{S}(\textrm{I})}$, 
respectively, 
because only $\sigma_{xy}^{\textrm{C}(\textrm{I})}$ or $\sigma_{xy}^{\textrm{S}(\textrm{I})}$ 
has the leading dependence on the QP damping. 

Next, 
since the spin current in $\sigma_{xy}^{\textrm{S}(\textrm{II})}$ 
is the same as the noninteracting one, 
$\sigma_{xy}^{\textrm{S}(\textrm{II})}$ is not affected by the SCD. 
Thus, 
the SCD affects $\sigma_{xy}^{\textrm{S}}$ except at low temperatures. 

Then, 
to understand how the electron-electron interaciton 
affects the Fermi sea term, 
we compare Eqs. (\ref{eq:SigXYC-II-direct}) and (\ref{eq:SigXYS-II-direct}) 
with the noninteracting Fermi sea terms~\cite{Kontani-AHE,Kontani-SHE} 
of $\sigma_{xy}^{\textrm{C}}$ and $\sigma_{xy}^{\textrm{S}}$, 
respectively, 
$\sigma_{xy}^{\textrm{C}(0;\textrm{II})}$ and $\sigma_{xy}^{\textrm{S}(0;\textrm{II})}$, 
and deduce the interaction effects on the Fermi sea terms. 
$\sigma_{xy}^{\textrm{C}(0;\textrm{II})}$ and $\sigma_{xy}^{\textrm{S}(0;\textrm{II})}$ 
are given~\cite{Kontani-AHE,Kontani-SHE} by 
\begin{align}
\sigma_{xy}^{\textrm{C}(0;\textrm{II})}
=&
-(-e)^{2}
\sum\limits_{k}
\sum\limits_{\{a \}}
\sum\limits_{\{s\}}
f(\epsilon)
\delta_{s^{\prime},s}(v_{\boldk x})_{ba}^{ss}
\notag\\
\times &
\Bigl[
\frac{\partial G_{ac}^{(0;\textrm{R})ss^{\prime\prime}}(k)}{\partial \epsilon}
G_{db}^{(0;\textrm{R})s^{\prime\prime\prime}s^{\prime}}(k)\notag\\
&-\frac{\partial G_{ac}^{(0;\textrm{A})ss^{\prime\prime}}(k)}{\partial \epsilon}
G_{db}^{(0;\textrm{A})s^{\prime\prime\prime}s^{\prime}}(k)
\Bigr]\delta_{s^{\prime\prime},s^{\prime\prime\prime}}
(v_{\boldk y})_{cd}^{s^{\prime\prime}s^{\prime\prime}},\label{eq:SigXYC0-II}
\end{align}
and 
\begin{align}
\sigma_{xy}^{\textrm{S}(0;\textrm{II})}
=&
-\frac{(-e)}{2}
\sum\limits_{k}
\sum\limits_{\{a \}}
\sum\limits_{\{s\}}
f(\epsilon)
\textrm{sgn}(s)
\delta_{s^{\prime},s}(v_{\boldk x})_{ba}^{ss}
\notag\\
\times &
\Bigl[
\frac{\partial G_{ac}^{(0;\textrm{R})ss^{\prime\prime}}(k)}{\partial \epsilon}
G_{db}^{(0;\textrm{R})s^{\prime\prime\prime}s^{\prime}}(k)\notag\\
&-\frac{\partial G_{ac}^{(0;\textrm{A})ss^{\prime\prime}}(k)}{\partial \epsilon}
G_{db}^{(0;\textrm{A})s^{\prime\prime\prime}s^{\prime}}(k)
\Bigr]\delta_{s^{\prime\prime},s^{\prime\prime\prime}}
(v_{\boldk y})_{cd}^{s^{\prime\prime}s^{\prime\prime}},\label{eq:SigXYS0-II}
\end{align}
respectively. 
After carrying out the $\epsilon$ integral 
in Eqs. (\ref{eq:SigXYC0-II}) and (\ref{eq:SigXYS0-II}), 
$\sigma_{xy}^{\textrm{C}(0;\textrm{II})}$ or $\sigma_{xy}^{\textrm{S}(0;\textrm{II})}$ 
is decomposed into the Berry-curvature term and the other part of 
the Fermi sea term~\cite{Kontani-AHE,Kontani-SHE}. 
Comparing Eqs. (\ref{eq:SigXYC-II-direct}) and (\ref{eq:SigXYS-II-direct}) 
with Eqs. (\ref{eq:SigXYC0-II}) and (\ref{eq:SigXYS0-II}), respectively, 
we find that 
each of $\sigma_{xy}^{\textrm{C}(\textrm{II})}$ and $\sigma_{xy}^{\textrm{S}(\textrm{II})}$ 
has three modifications due to the electron-electron interaction. 
Those modifications are 
the replacement of the single-particle Green's functions by the interacting ones, 
the replacement of the $y$ component of the charge current 
by its vertex function, 
and the appearance of the $\omega$ derivative term of 
the $y$ component of the vertex functions of the charge current. 

Each of those modifications affects $\sigma_{xy}^{\textrm{C}(\textrm{II})}$ 
and $\sigma_{xy}^{\textrm{S}(\textrm{II})}$ as follows. 
First, the replacement of the single-particle Green's functions 
will little affect $\sigma_{xy}^{\textrm{C}(\textrm{II})}$ and $\sigma_{xy}^{\textrm{S}(\textrm{II})}$ 
because 
the QP damping of the retarded-retarded or advanced-advanced product 
is negligible~\cite{Eliashberg} 
and because the effects of $z_{\alpha}(\boldk)$ in the numerator and the denominator 
of the coherent parts of that product for finite $\epsilon$ 
are nearly cancelled out each other 
when the band dependence of $z_{\alpha}(\boldk)$ is not strong. 
Second, the effects 
of the replacement of the $y$ component of the charge current 
on $\sigma_{xy}^{\textrm{C}(\textrm{II})}$ or $\sigma_{xy}^{\textrm{S}(\textrm{II})}$ 
may be also not large 
because, as described in Section III A 2, 
the difference between 
$\Lambda_{y;l;cd}^{\textrm{C}(0)s^{\prime\prime}s^{\prime\prime\prime}}(\boldk,\epsilon;0)$ 
and $\delta_{s^{\prime\prime},s^{\prime\prime\prime}}(v_{\boldk y})_{cd}^{s^{\prime\prime}s^{\prime\prime}}$ 
just causes the renormalization of the group velocity. 
Third, 
the modification about the appearance 
of the $\omega$ derivative term of the charge current 
may lead to the finite correction term  
if the dynamical effects of the electron-electron interaction 
are considerable. 
If the effects of the electron-electron interaction can be 
either neglected or treated in a mean-field approximation, 
the $\omega$ derivative is exactly zero. 
Actual estimations of those three interaction effects 
by numerical calculations are remaining issues for a future study.

\section{Discussion}
In this section, 
we discuss the origin of the differences 
between $\sigma_{xy}^{\textrm{C}}$ and $\sigma_{xx}^{\textrm{C}}$, 
the differences between the present formalism and Haldane's formalism, 
and the correspondences between our results and experiments. 

Before discussing the origin of the differences between 
$\sigma_{xy}^{\textrm{C}}$ and $\sigma_{xx}^{\textrm{C}}$, 
we show $\sigma_{xx}^{\textrm{C}}$ 
in \'{E}liashberg's approximation 
for $\hat{H}$, 
and see its properties 
about the dominant multiband excitations, 
the damping dependence, 
and applicability of \'{E}liashberg's approximation. 
Since we obtain the exact expression of $\sigma_{xx}^{\textrm{C}}$ 
in the linear-response theory 
by replacing $\hat{J}_{-\boldq y}^{\textrm{C}}(0)$ in $\tilde{K}_{xy}^{\textrm{C}}(i\Omega_{n})$ 
in Eq. (\ref{eq:sigXY-C}) by $\hat{J}_{-\boldq x}^{\textrm{C}}(0)$, 
we can derive $\sigma_{xx}^{\textrm{C}}$ in \'{E}liashberg's approximation 
in the similar way for $\sigma_{xy}^{\textrm{C}}$. 
Thus, 
$\sigma_{xx}^{\textrm{C}}$ in this approximation 
becomes $\sigma_{xx}^{\textrm{C}}=\sigma_{xx}^{\textrm{C}(\textrm{I})}$ with 
\begin{align}
\sigma_{xx}^{\textrm{C}(\textrm{I})}&=
(-e)^{2}
\sum\limits_{k}
\sum\limits_{\{a\}}
\sum\limits_{\{s\}}
\Bigl(-\dfrac{\partial f(\epsilon)}{\partial \epsilon}\Bigr)
\Lambda_{x;2;ba}^{\textrm{C}(0)s^{\prime}s}(k;0)\notag\\
&\times 
G_{ac}^{(0;\textrm{R})ss^{\prime\prime}}(k;0)
G_{db}^{(0;\textrm{A})s^{\prime\prime\prime}s^{\prime}}(k;0)
\Lambda_{x;2;cd}^{\textrm{C}; s^{\prime\prime}s^{\prime\prime\prime}}(k;0).\label{eq:SigXXC-I} 
\end{align}
The difference between $\sigma_{xx}^{\textrm{C}(\textrm{I})}$ and $\sigma_{xy}^{\textrm{C}(\textrm{I})}$ 
is the difference between 
$\Lambda_{x;2;cd}^{\textrm{C}; s^{\prime\prime}s^{\prime\prime\prime}}(k;0)$ and 
$\Lambda_{y;2;cd}^{\textrm{C}; s^{\prime\prime}s^{\prime\prime\prime}}(k;0)$. 
In addition, 
from the similar argument for $\sigma_{xy}^{\textrm{C}(\textrm{I})}$, 
we can deduce  
several properties of $\sigma_{xx}^{\textrm{C}}$. 
First, 
because of the same reason for $\sigma_{xy}^{\textrm{C}(\textrm{I})}$, 
we can determine 
the dominant multiband excitations and damping dependence of $\sigma_{xx}^{\textrm{C}(\textrm{I})}$ 
by analyzing the leading terms of 
$g_{2;acdb}^{ss^{\prime\prime}s^{\prime\prime\prime}s^{\prime}}(k;0)$ 
which give the finite terms of $\sigma_{xx}^{\textrm{C}(\textrm{I})}$. 
Since $\sigma_{xx}^{\textrm{C}(\textrm{I})}$ includes two $k_{x}$ derivatives 
arising from the $k_{x}$ derivatives of the noninteracting group velocity 
in $\Lambda_{x;2;ba}^{\textrm{C}(0)s^{\prime}s}(k;0)$ and 
$\Lambda_{x;2;cd}^{\textrm{C}; s^{\prime\prime}s^{\prime\prime\prime}}(k;0)$ 
[see Eqs. (\ref{eq:JC0all-symbolic}) and (\ref{eq:3VC-real})], 
the terms in $\sigma_{xx}^{\textrm{C}(\textrm{I})}$ other than those 
should be even with respect to $k_{x}$ and $k_{y}$. 
Due to this property, 
the terms proportional to $-i[\gamma_{\alpha}^{\ast}(\boldk)
+\gamma_{\beta}^{\ast}(\boldk)]$ 
in the leading terms of $g_{2;acdb}^{ss^{\prime\prime}s^{\prime\prime\prime}s^{\prime}}(k;0)$ 
give the finite terms of $\sigma_{xx}^{\textrm{C}(\textrm{I})}$  [see Eq. (\ref{eq:g2-approx})].  
In addition, 
since the denominator of the leading terms of 
$g_{2;acdb}^{ss^{\prime\prime}s^{\prime\prime\prime}s^{\prime}}(k;0)$ 
includes $\Delta \xi^{\ast}_{\beta\alpha}(\boldk)^{2}(\geq 0)$ [see Eq. (\ref{eq:g2-approx})], 
the dominant multiband excitations become intraband (i.e., $\beta=\alpha$). 
Furthermore, 
due to that property, 
$\sigma_{xx}^{\textrm{C}(\textrm{I})}$ is always $O(\gamma^{-1})$ 
because  
we can approximate $\sigma_{xx}^{\textrm{C}(\textrm{I})}$ as 
\begin{align}
\sigma_{xx}^{\textrm{C}}&\approx 
\frac{(-e)^{2}}{N}
\sum\limits_{\boldk}
\sum\limits_{\alpha}
\frac{1}
{2\gamma_{\alpha}^{\ast}(\boldk)}\notag\\
&\times
\textrm{Re}[\tilde{\Lambda}_{x;2;\alpha\alpha}^{\textrm{C}(0)}(\boldk,\xi_{\alpha}^{\ast}(\boldk))
\tilde{\Lambda}_{x;2;\alpha\alpha}^{\textrm{C}}(\boldk,\xi_{\alpha}^{\ast}(\boldk))]\label{eq:sigCXX-limit}
\end{align}
with Eqs. (\ref{eq:LambC0-tild}) and (\ref{eq:LambC-tild}). 
Then, 
the dominance of the intraband excitations for $\sigma_{xx}^{\textrm{C}(\textrm{I})}$ 
indicates that 
\'{E}liashberg's approximation 
is always applicable 
in the left triangle region of Fig. \ref{fig:Fig2} 
because for the intraband excitations 
the overlap between the QP spectra is unimportant. 

Combining the above properties of $\sigma_{xx}^{\textrm{C}}$ 
with the corresponding properties of $\sigma_{xy}^{\textrm{C}}$, 
we can clarify the origin of the differences between 
$\sigma_{xx}^{\textrm{C}}$ and $\sigma_{xy}^{\textrm{C}}$. 
Namely, 
the origin 
is the difference in the dominant multiband excitations. 

In addition, 
we can deduce the general principles 
in formulating transport coefficients of an interacting multiorbital metal. 
If the dominant multiband excitations are intraband, 
we can sufficiently treat the electron-electron interaction 
in \'{E}liashberg's approximation. 
If the interband excitations are dominant, 
we need to use, instead of \'{E}liashberg's approximation, 
an approximation beyond it 
only in the low-$T$ case. 

Then, 
we argue 
the differences between 
the present formalism and Haldane's formalism~\cite{Haldane-AHE}. 
Assuming that 
$\sigma_{xy}^{\textrm{C}}$ is given only by the Berry-curvature term, 
Haldane proposed that 
the term of the Berry-curvature term 
after partial integral about $\epsilon$ 
could describe the excitations near the Fermi level~\cite{Haldane-AHE}. 
However, 
the exact $\sigma_{xy}^{\textrm{C}}$ 
includes the Fermi surface term which qualitatively differs from 
the Berry-curvature term [see Eq. (\ref{eq:sigXY-C}) with Eq. (\ref{eq:KC-real2})]; 
the difference arises from 
the effects of a retarded-advanced product of two single-particle Green's functions, 
which are important in the resistivity~\cite{Eliashberg,Yamada-Yosida} 
and the weak-field usual Hall conductivity~\cite{Fukuyama,Kohno-Yamada} 
in the Fermi-liquid. 
Thus, if the Fermi surface term is dominant, 
Haldane's formalism is inapplicable. 
Since 
we find the dominance of the Fermi surface term 
in the high-$T$ and the intermediate-$T$ region of Fig. \ref{fig:Fig2} 
even without impurities, 
the present formalism reveals 
the remarkable interaction effects arising from the non-Berry-curvature term 
outside the applicable region of Haldane's formalism~\cite{Haldane-AHE}. 

Finally, 
we discuss the correspondences between our results and experiments. 
First, 
we can check the interaction-driven mechanism of 
the damping dependence of $\sigma_{xy}^{\textrm{C}}$ or $\sigma_{xy}^{\textrm{S}}$ 
and crossover between damping-dependent to damping-independent 
$\sigma_{xy}^{\textrm{C}}$ or $\sigma_{xy}^{\textrm{S}}$ 
by measuring its temperature dependence 
in a clean system. 
This is because that temperature dependence is induced by 
the temperature dependence of the interaction-induced QP damping, 
as explained in Section III A 2. 
Also, 
we may observe 
the difference of the form of the red dotted line in Fig. \ref{fig:Fig2} 
between weakly-interacting and strongly-interacting metals 
because the Fermi liquid and the nearly-antiferromagnetic or nearly-ferromagnetic metal 
show the different temperature dependence of the QP damping~\cite{AGD,Hlubina-Rice}. 
Moreover, 
although it is difficult to detect 
the crossover between the damping-independent Fermi surface and Fermi sea terms 
only by experiments, 
we can check its existence 
by combination of experiments 
and first-principle calculations  
if we find the material in which 
the sign of $\sigma_{xy}^{\textrm{C}}$ or $\sigma_{xy}^{\textrm{S}}$ 
changes at the crossover line: 
to find such material, 
we need to systematically analyze 
the intrinsic AHE or SHE on the basis of a realistic band structure 
in the presence of the electron-electron interaction 
by using the first-principle calculation; 
after the finding,  
we need to experimentally analyze the sign of $\sigma_{xy}^{\textrm{C}}$ or $\sigma_{xy}^{\textrm{S}}$ 
as a function of temperature around the crossover temperature. 
Then, 
the results about the SCD indicate, first, 
that in a measurement of the SHE in the low-$T$ case, 
$\sigma_{xy}^{\textrm{S}}$ behaves as if 
the non-conservation of the spin current is not important; 
second, that 
we may observe the effects of the SCD on the intrinsic SHE 
at high or slightly-low temperatures 
where the Fermi-surface term is dominant. 
However, 
it remains a challenging issue to clarify 
how large its effects are among several transition metals 
and transition-metal oxides. 

\section{Summary}
In summary, 
we have constructed the general formalism 
for the intrinsic AHE and SHE of the interacting multiorbital metal 
by using the linear-response theory 
with the appropriate approximations, 
and have clarified the roles of the Fermi surface term and Fermi sea term 
of the dc conductivity 
and the effects of the SCD on these terms. 
In the high-$T$ and the intermediate-$T$ region of Fig. \ref{fig:Fig2}, 
we have used \'{E}liashberg's approximation, 
and in the low-$T$ region, 
we have constructed 
the approximation beyond \'{E}liashberg's approximation. 
Most importantly, 
we highlight the important roles of the Fermi surface term, 
a non-Berry-curvature term, 
even without impurities 
in the high-$T$ and the intermediate-$T$ region. 
Actually, 
this Fermi surface term leads to 
the interaction-driven temperature dependence of 
$\sigma_{xy}^{\textrm{C}}$ or $\sigma_{xy}^{\textrm{S}}$ in the high-$T$ region 
and the SCD-induced correction of $\sigma_{xy}^{\textrm{S}}$. 
These results considerably develop 
our understanding of the intrinsic AHE and SHE. 
In addition to those achievements, 
we have found that 
the differences between $\sigma_{xy}^{\textrm{C}}$ and $\sigma_{xx}^{\textrm{C}}$ 
arise from the difference in the dominant multiband excitations. 
Namely, 
due to the dominance of the interband excitations in $\sigma_{xy}^{\textrm{C}}$, 
the Fermi sea term such as the Berry-curvature term 
becomes dominant in clean and low-$T$ case, 
while due to the dominance of the intraband excitations in $\sigma_{xx}^{\textrm{C}}$, 
the Fermi surface term is always dominant. 
This answers how to construct the FL theory 
for the intrinsic AHE or SHE. 
Moreover, 
we have shown the principles 
to construct general formalism 
of transport coefficients including 
the interaction effects and the multiband effects. 
This may be useful for further research of charge, spin, and heat transports 
for an interacting multiorbital metal.

\appendix
%\begin{widetext}
\begin{figure*}[tb]
\includegraphics[width=160mm]{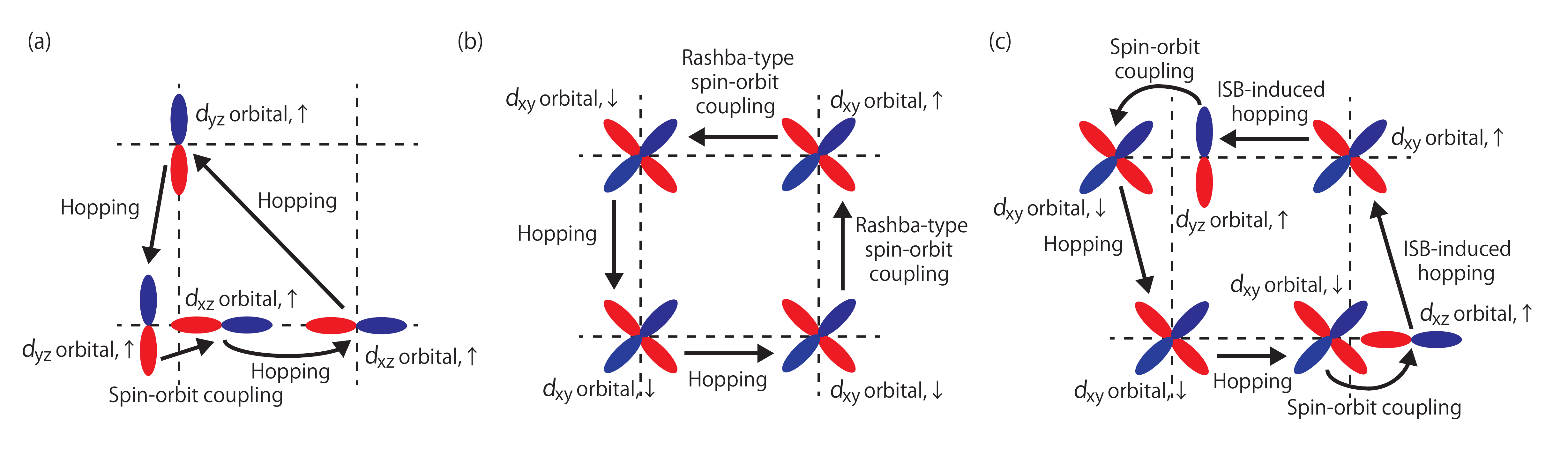}
\vspace{-5pt}
\caption{ 
Schematic pictures about the motion of an electron 
that gives a finite term of $\sigma_{xy}^{\textrm{C}}$ or $\sigma_{xy}^{\textrm{S}}$ 
for (a) SrRuO$_{3}$~\cite{Kontani-AHE} or Sr$_{2}$RuO$_{4}$~\cite{Kontani-OrbitalAB}, 
and (b) an effective single-orbital metal~\cite{Sinova-SHE} 
without the inversion symmetry at an $ab$-plane, 
and (c) the schematic picture for the $t_{2g}$-orbital metal~\cite{Mizoguchi-SHE} 
whose single-orbital limit corresponds to case of (b). 
In those panels, 
$\uparrow$ and $\downarrow$ denote spin-up and spin-down, respectively, 
each black arrow denotes each motions due to the single-body operator, 
and the color difference of an orbital denotes the sign difference 
of its wave function. 
ISB-induced hopping in panel (c) denotes the hopping integral 
induced by the inversion symmetry breaking 
at an $ab$ plane~\cite{Yanase-Rashba}. 
Several similarities in panels (c) and (a) should be noted: 
the two ISB-induced hoppings in panel (c) 
play a similar role for the next-nearest-neighbor hopping 
between the $d_{yz}$ and $d_{xz}$ orbitals in panel (a); 
a sequence of the SOCs between 
the spin-down $d_{xy}$ and the spin-up $d_{xz}$ orbital 
and between the spin-up $d_{yz}$ and the spin-down $d_{xy}$ orbital 
in panel (c)
play a similar role for the SOC 
between the spin-up $d_{xz}$ and the spin-up $d_{yz}$ orbital in panel (a). 
}
\label{fig:Supp-Fig1}
\end{figure*}
\section{Understanding of the intrinsic AHE or SHE 
  as orbital Aharanov-Bohm effect}

In this Appendix, 
we see that 
the origin of finite terms of $\sigma_{xy}^{\textrm{C}}$ or $\sigma_{xy}^{\textrm{S}}$ 
can be understood by analyzing the corresponding motion of 
an electron or a QP in real space, 
and that 
the origin of the intrinsic AHE or SHE 
in several metals is orbital Aharanov-Bohm (AB) effect~\cite{Kontani-OrbitalAB}. 

First, we can obtain 
the intuitive insight of the origin 
of finite $\sigma_{xy}^{\textrm{C}}$ or $\sigma_{xy}^{\textrm{S}}$ 
by expressing its finite term as the corresponding motion of 
an electron or a QP in real space~\cite{Kontani-OrbitalAB,Mizoguchi-SHE}. 
For simplicity of arguments, 
let us argue noninteracting case of $\sigma_{xy}^{\textrm{C}}$ 
because that argument for an electron is similarly applicable for $\sigma_{xy}^{\textrm{S}}$ 
and because the similar argument holds even for a QP in interacting case. 
In the linear response theory, 
$\sigma_{xy}^{\textrm{C}}$ has four matrix elements, 
the $x$ and the $y$ component of the charge current and 
two single-particle Green's functions~\cite{Streda,Kontani-AHE} 
[see Eqs. (\ref{eq:SigXYC0-I}) and (\ref{eq:SigXYC0-II})]; 
each term is the matrix element of the corresponding operator. 
Then, 
the charge current operator is single-body [see Eq. (\ref{eq:JC})], 
and the operator of the retarded or advanced 
noninteracting single-particle Green's function 
is given by the inverse matrix of 
$(\omega \hat{1}-\hat{H}_{0}-\hat{H}_{\textrm{LS}}+i\delta \hat{1})$ 
for $\delta = +0$ or $-0$, respectively. 
Since we can express it 
in terms of the series of $\frac{\hat{H}_{0}+\hat{H}_{\textrm{LS}}}{\omega}$, 
a single-body operator, 
we can decompose the terms of $\sigma_{xy}^{\textrm{C}}$ 
into the corresponding motion of an electron in real space~\cite{Kontani-OrbitalAB,Mizoguchi-SHE}. 
That motion helps understand 
which terms in the Hamiltonian are essential to 
obtain finite $\sigma_{xy}^{\textrm{C}}$. 
Thus, the analysis of that motion helps understand 
the origin of the finite terms of $\sigma_{xy}^{\textrm{C}}$~\cite{Kontani-OrbitalAB,Mizoguchi-SHE}. 

Then, 
we consider three examples, 
and see the finite terms of $\sigma_{xy}^{\textrm{C}}$ or $\sigma_{xy}^{\textrm{S}}$ 
arise from the acquisition of the AB-type phase factor of an electron 
because of the onsite SOC and several hopping integrals. 
The following arguments are applicable to other cases 
of the intrinsic AHE or SHE of a metal. 

The first example is case of a $t_{2g}$-orbital metal on a square lattice, 
corresponding to the AHE~\cite{Kontani-AHE,SrRuO3-AHE} in SrRuO$_{3}$ 
and the SHE~\cite{Kontani-OrbitalAB} in Sr$_{2}$RuO$_{4}$. 
By the analysis of the motions 
for the finite terms of $\sigma_{xy}^{\textrm{C}}$ or $\sigma_{xy}^{\textrm{S}}$, 
we find that 
one of the finite terms in this case 
arises from the motion shown in Fig. \ref{fig:Supp-Fig1}(a)~\cite{Kontani-OrbitalAB}. 
This figure shows that 
the SOC from the spin-up $d_{yz}$ orbital to the spin-up $d_{xz}$ orbital 
causes $-\frac{\pi}{2}$ rotation, 
resulting in a complex phase factor of the wave function of an electron, 
$\exp \frac{i\pi l_{z}}{2}=i$~\cite{Kontani-OrbitalAB}. 
This phase factor is similar to the AB phase factor~\cite{JJSakurai} 
in the presence of an external magnetic field. 
Thus, we can regard the acquisition of such phase factor 
using orbital degrees of freedom 
as the orbital AB effect~\cite{Kontani-OrbitalAB}. 
Namely, 
the orbital AB effect causes 
the intrinsic AHE or SHE in this case. 
In addition to the onsite SOC, 
the direct hopping integral between the $d_{yz}$ and $d_{xz}$ orbitals 
is important to obtain finite $\sigma_{xy}^{\textrm{C}}$.  

Second, 
we can apply the similar mechanism to case of 
the intrinsic SHE in Pt~\cite{Kontani-SHE,SHE-Pt-Nagaosa}. 
In this case, 
we can acquire the AB-type phase factor 
by using several hopping integrals and 
the onsite SOC; e.g., 
the onsite SOC 
from the spin-up $d_{xy}$ orbital to the spin-up $d_{x^{2}-y^{2}}$ orbital 
leads to $-\frac{\pi}{4}$ rotation, resulting in a complex phase factor, 
$\exp \frac{i\pi l_{z}}{4}=i$~\cite{Kontani-SHE}. 

Third, 
we can similarly understand the intrinsic AHE or SHE 
in an effective single-orbital metal~\cite{Sinova-SHE} 
without the inversion symmetry at an $ab$-plane. 
For the explicit argument, 
let us consider the situation of the $d_{xy}$-orbital system on a square lattice. 
(The following argument is applicable even for other single-orbital systems 
without the inversion symmetry.) 
Since the electronic structure in this situation may be described 
by the single-orbital Rashba model~\cite{Rashba}, 
we can determine the motion 
which gives the finite term of $\sigma_{xy}^{\textrm{C}}$ or $\sigma_{xy}^{\textrm{S}}$ 
in the Rashba model~\cite{Sinova-SHE} [see Fig. \ref{fig:Supp-Fig1}(b)]. 
Although that motion seems to be not categorized as 
the orbital AB effect, 
that motion can also be understood as 
the orbital AB effect~\cite{Mizoguchi-SHE}. 
This is 
because a $t_{2g}$-orbital model with the onsite SOC 
without the inversion symmetry at an $ab$-plane 
becomes an effective single-orbital Rashba model 
for a large difference of the single-body energy level 
between the $d_{xy}$ and $d_{xz/yz}$ orbitals~\cite{Yanase-Rashba}: 
the microscopic origin of the Rashba-type SOC 
is the combination of 
the transverse components of the onsite SOC 
and the hopping integral induced by the inversion symmetry breaking 
in the presence of the large single-body energy difference 
between the $d_{xy}$ and $d_{xz/yz}$ orbitals [see Fig. \ref{fig:Supp-Fig1}(c)].  
Note that 
except case for the large single-body energy difference 
the $t_{2g}$-orbital model qualitatively differs from 
the effective single-orbital Rashba model~\cite{Yanase-Rashba,Mizoguchi-SHE}, 
and that 
the differences play important roles in obtaining 
the intrinsic term, which defeats 
the extrinsic term in the Born approximation~\cite{Mizoguchi-CVC}.  

\section{Derivation of Eq. (\ref{eq:K-4VC})}
In this appendix, we derive Eq. (\ref{eq:K-4VC}) 
by carrying out the analytic continuations of 
the first and the second term of Eq. (\ref{eq:K}). 

First, we can carry out the analytic continuation of 
the first term of Eq. (\ref{eq:K}) as follows: 
\begin{align}
&-\delta_{\boldk,\boldk^{\prime}}
T\sum\limits_{m}
G_{ac}^{ss^{\prime\prime}}(\tilde{k_{+}})
G_{db}^{s^{\prime\prime}s}(\tilde{k})\notag\\
=&
-\delta_{\boldk,\boldk^{\prime}}
\int_{\textrm{C}}\frac{d\epsilon}{4\pi i}
\tanh \frac{\epsilon}{2T}
G_{ac}^{ss^{\prime\prime}}(\tilde{k_{+}})
G_{db}^{s^{\prime\prime}s}(\tilde{k})\notag\\
\rightarrow &
-\delta_{\boldk,\boldk^{\prime}}
\int_{-\infty}^{\infty}\frac{d\epsilon}{4\pi i}
\tanh \frac{\epsilon}{2T}
G_{ac}^{(\textrm{R})ss^{\prime\prime}}(\boldk,\epsilon+\omega)\notag\\
&\times 
[G_{db}^{(\textrm{R})s^{\prime\prime}s}(\boldk,\epsilon)
-G_{db}^{(\textrm{A})s^{\prime\prime}s}(\boldk,\epsilon)]\notag\\
&-\delta_{\boldk,\boldk^{\prime}}
\int_{-\infty}^{\infty}\frac{d\epsilon}{4\pi i}
\tanh \frac{\epsilon+\omega}{2T}\notag\\
&\times 
[G_{ac}^{(\textrm{R})ss^{\prime\prime}}(\boldk,\epsilon+\omega)
-G_{ac}^{(\textrm{A})ss^{\prime\prime}}(\boldk,\epsilon+\omega)]
G_{db}^{(\textrm{A})s^{\prime\prime}s}(\boldk,\epsilon)\notag\\  
&=
-\delta_{\boldk,\boldk^{\prime}}
\int_{-\infty}^{\infty}\frac{d\epsilon}{4\pi i}
\sum\limits_{l=1}^{3}
T_{l}(\epsilon,\omega)
g_{l;acdb}^{ss^{\prime\prime}s^{\prime\prime\prime}s^{\prime}}(k;\omega),\label{eq:1st-analytic}
\end{align}
where 
the contour $C$ consists of the three parts 
into which 
the circle is divided by inserting two horizontal lines 
$\textrm{Im}(\epsilon+i\Omega_{n})=0$ and $\textrm{Im}\epsilon=0$, 
$\rightarrow$ represents 
taking $i\Omega_{n}\rightarrow \omega+i0+$, 
$T_{l}(\epsilon,\omega)$ 
are given by Eqs. (\ref{eq:T1}){--}(\ref{eq:T3}), 
and 
$g_{l;acdb}^{ss^{\prime\prime}s^{\prime\prime\prime}s^{\prime}}(k;\omega)$ 
are given by Eqs. (\ref{eq:g1}){--}(\ref{eq:g3}). 

Second, we can similarly carry out 
the analytic continuation of 
the second term of Eq. (\ref{eq:K}): 
\begin{widetext}
\begin{align}
&-\frac{T^{2}}{N}
\sum\limits_{m,m^{\prime}}
\sum\limits_{\{A\}}
\sum\limits_{\{s_{1}\}}
G_{aA}^{ss_{1}}(\tilde{k_{+}})
G_{dD}^{s^{\prime\prime}s_{4}}(\tilde{k^{\prime}})
G_{Cc}^{s_{3}s^{\prime\prime}}(\tilde{k^{\prime}_{+}})
G_{Bb}^{s_{2}s}(\tilde{k})
\Gamma_{\{A\}}^{\{s_{1}\}}(\tilde{k},\tilde{k^{\prime}},i\epsilon_{m^{\prime}};\boldzero,i\Omega_{n})\notag\\
=&
-\frac{1}{N}
\int_{\textrm{C}}\frac{d\epsilon}{4\pi i}
\tanh \frac{\epsilon}{2T}
\sum\limits_{\{A\}}
\sum\limits_{\{s_{1}\}}
G_{aA}^{ss_{1}}(\boldk,\epsilon+i\Omega_{n})
G_{Bb}^{s_{2}s}(\boldk,\epsilon)
\notag\\
&\  
\times \Bigl[\int_{\textrm{C}^{\prime}}\frac{d\epsilon^{\prime}}{4\pi i}
\tanh \frac{\epsilon^{\prime}}{2T}
G_{dD}^{s^{\prime\prime}s_{4}}(\boldk^{\prime},\epsilon^{\prime})
G_{Cc}^{s_{3}s^{\prime\prime}}(\boldk^{\prime},\epsilon^{\prime}+i\Omega_{n})
\Gamma_{\{A\}}^{\{s_{1}\}}(\boldk,\epsilon,\boldk^{\prime},\epsilon^{\prime}; \boldzero,i\Omega_{n})\notag\\
&\ \ \ \ 
+TG_{dD}^{s^{\prime\prime}s_{4}}(\boldk^{\prime},\epsilon)
G_{Cc}^{s_{3}s^{\prime\prime}}(\boldk^{\prime},\epsilon+i\Omega_{n})
\Gamma_{\{A\}}^{\{s_{1}\}}(\boldk,\epsilon,\boldk^{\prime},\epsilon;\boldzero,i\Omega_{n})\notag\\
&\ \ \ \ 
+TG_{dD}^{s^{\prime\prime}s_{4}}(\boldk^{\prime},-\epsilon-i\Omega_{n})
G_{Cc}^{s_{3}s^{\prime\prime}}(\boldk^{\prime},-\epsilon)
\Gamma_{\{A\}}^{\{s_{1}\}}(\boldk,\epsilon,\boldk^{\prime},-\epsilon-i\Omega_{n};\boldzero,i\Omega_{n})
\Bigr]\notag\\
\rightarrow 
&-
\frac{1}{N}
\sum\limits_{\{A \}}
\sum\limits_{\{s_{1}\}}
\int^{\infty}_{-\infty}\frac{d\epsilon}{4\pi i}
P\int^{\infty}_{-\infty}\frac{d\epsilon^{\prime}}{4\pi i}
\coth \dfrac{\epsilon^{\prime}-\epsilon}{2T}\notag\\
&\times 
\Bigl\{\Bigl(\tanh \dfrac{\epsilon+\omega}{2T}
-\tanh \dfrac{\epsilon}{2T}\Bigr)
g_{2;aABb}^{ss_{1}s_{2}s}(k;\omega)
g_{2;CcdD}^{s_{3}s^{\prime\prime}s^{\prime\prime}s_{4}}(k^{\prime};\omega)
\Bigl[
\Gamma_{22\textrm{-II};\{A\}}^{\{s_{1}\}}(k,k^{\prime};\boldzero,\omega)
-\Gamma_{22\textrm{-III};\{A\}}^{\{s_{1}\}}(k,k^{\prime};\boldzero,\omega)
\Bigr]\notag\\
&-\tanh \dfrac{\epsilon+\omega}{2T}
g_{3;aABb}^{ss_{1}s_{2}s}(k;\omega)
g_{3;CcdD}^{s_{3}s^{\prime\prime}s^{\prime\prime}s_{4}}(k^{\prime};\omega)
\Bigl[
\Gamma_{33\textrm{-I};\{A\}}^{\{s_{1}\}}(k,k^{\prime};\boldzero,\omega)
-\Gamma_{33\textrm{-II};\{A\}}^{\{s_{1}\}}(k,k^{\prime};\boldzero,\omega)
\Bigr]\notag\\
&+\tanh \dfrac{\epsilon}{2T}
g_{1;aABb}^{ss_{1}s_{2}s}(k;\omega)
g_{1;CcdD}^{s_{3}s^{\prime\prime}s^{\prime\prime}s_{4}}(k^{\prime};\omega)
\Bigl[
\Gamma_{11\textrm{-II};\{A\}}^{\{s_{1}\}}(k,k^{\prime};\boldzero,\omega)
-\Gamma_{11\textrm{-I};\{A\}}^{\{s_{1}\}}(k,k^{\prime};\boldzero,\omega)
\Bigr]\Bigr\}\notag\\
&-
\frac{1}{N}
\sum\limits_{\{A \}}
\sum\limits_{\{s_{1}\}}
\int^{\infty}_{-\infty}\dfrac{d\epsilon}{4\pi i}
\int^{\infty}_{-\infty}\dfrac{d\epsilon^{\prime}}{4\pi i}
\tanh \dfrac{\epsilon^{\prime}}{2T}\notag\\
&\times 
\Bigl\{\Bigl(\tanh \dfrac{\epsilon+\omega}{2T}
-\tanh \dfrac{\epsilon}{2T}\Bigr)
g_{2;aABb}^{ss_{1}s_{2}s}(k;\omega)
\Bigl[
\Gamma_{21;\{A\}}^{\{s_{1}\}}(k,k^{\prime};\boldzero,\omega)
g_{1;CcdD}^{s_{3}s^{\prime\prime}s^{\prime\prime}s_{4}}(k^{\prime};\omega)
-\Gamma_{22\textrm{-II};\{A\}}^{\{s_{1}\}}(k,k^{\prime};\boldzero,\omega)
g_{2;CcdD}^{s_{3}s^{\prime\prime}s^{\prime\prime}s_{4}}(k^{\prime};\omega)
\Bigr]\notag\\
&-\tanh \dfrac{\epsilon+\omega}{2T}
g_{3;aABb}^{ss_{1}s_{2}s}(k;\omega)
\Bigl[
\Gamma_{31\textrm{-I};\{A\}}^{\{s_{1}\}}(k,k^{\prime};\boldzero,\omega)
g_{1;CcdD}^{s_{3}s^{\prime\prime}s^{\prime\prime}s_{4}}(k^{\prime};\omega)
-\Gamma_{32;\{A\}}^{\{s_{1}\}}(k,k^{\prime};\boldzero,\omega)
g_{2;CcdD}^{s_{3}s^{\prime\prime}s^{\prime\prime}s_{4}}(k^{\prime};\omega)
\Bigr]\notag\\
&+\tanh \frac{\epsilon}{2T}
g_{1;aABb}^{ss_{1}s_{2}s}(k;\omega)
\Bigl[
\Gamma_{11\textrm{-I};\{A\}}^{\{s_{1}\}}(k,k^{\prime}\boldzero,\omega)
g_{1;CcdD}^{s_{3}s^{\prime\prime}s^{\prime\prime}s_{4}}(k^{\prime};\omega)
-\Gamma_{12;\{A\}}^{\{s_{1}\}}(k,k^{\prime};\boldzero,\omega)
g_{2;CcdD}^{s_{3}s^{\prime\prime}s^{\prime\prime}s_{4}}(k^{\prime};\omega)
\Bigr] \Bigr\}\notag\\
&-
\frac{1}{N}
\sum\limits_{\{A \}}
\sum\limits_{\{s_{1}\}}
\int^{\infty}_{-\infty}\dfrac{d\epsilon}{4\pi i}
\int^{\infty}_{-\infty}\dfrac{d\epsilon^{\prime}}{4\pi i}
\tanh \dfrac{\epsilon^{\prime}+\omega}{2T}\notag\\
&\times 
\Bigl\{\Bigl(\tanh \dfrac{\epsilon+\omega}{2T}
-\tanh \dfrac{\epsilon}{2T}\Bigr)
g_{2;aABb}^{ss_{1}s_{2}s}(k;\omega)
\Bigl[
\Gamma_{22\textrm{-IV};\{A\}}^{\{s_{1}\}}(k,k^{\prime};\boldzero,\omega)
g_{2;CcdD}^{s_{3}s^{\prime\prime}s^{\prime\prime}s_{4}}(k^{\prime};\omega)
-\Gamma_{23;\{A\}}^{\{s_{1}\}}(k,k^{\prime};\boldzero,\omega)
g_{3;CcdD}^{s_{3}s^{\prime\prime}s^{\prime\prime}s_{4}}(k^{\prime};\omega)
\Bigr]\notag\\
&-\tanh \dfrac{\epsilon+\omega}{2T}
g_{3;aABb}^{ss_{1}s_{2}s}(k;\omega)
\Bigl[
\Gamma_{32;\{A\}}^{\{s_{1}\}}(k,k^{\prime};\boldzero,\omega)
g_{2;CcdD}^{s_{3}s^{\prime\prime}s^{\prime\prime}s_{4}}(k^{\prime};\omega)
-\Gamma_{33\textrm{-I};\{A\}}^{\{s_{1}\}}(k,k^{\prime};\boldzero,\omega)
g_{3;CcdD}^{s_{3}s^{\prime\prime}s^{\prime\prime}s_{4}}(k^{\prime};\omega)
\Bigr]\notag\\
&+\tanh \dfrac{\epsilon}{2T}
g_{1;aABb}^{ss_{1}s_{2}s}(k;\omega)
\Bigl[
\Gamma_{12;\{A\}}^{\{s_{1}\}}(k,k^{\prime};\boldzero,\omega)
g_{2;CcdD}^{s_{3}s^{\prime\prime}s^{\prime\prime}s_{4}}(k^{\prime};\omega)
-\Gamma_{13\textrm{-I};\{A\}}^{\{s_{1}\}}(k,k^{\prime};\boldzero,\omega)
g_{3;CcdD}^{s_{3}s^{\prime\prime}s^{\prime\prime}s_{4}}(k^{\prime};\omega)
\Bigr] \Bigr\}\notag\\
&-
\frac{1}{N}
\sum\limits_{\{A \}}
\sum\limits_{\{s_{1}\}}
\int^{\infty}_{-\infty}\dfrac{d\epsilon}{4\pi i}
P\int^{\infty}_{-\infty}\dfrac{d\epsilon^{\prime}}{4\pi i}
\coth \dfrac{\epsilon^{\prime}+\epsilon+\omega}{2T}\notag\\
&\times \Bigl\{\Bigl(\tanh \dfrac{\epsilon+\omega}{2T}
-\tanh \dfrac{\epsilon}{2T}\Bigr)
g_{2;aABb}^{ss_{1}s_{2}s}(k;\omega)
g_{2;CcdD}^{s_{3}s^{\prime\prime}s^{\prime\prime}s_{4}}(k^{\prime};\omega)
\Bigl[
\Gamma_{22\textrm{-III};\{A\}}^{\{s_{1}\}}(k,k^{\prime};\boldzero,\omega)
-\Gamma_{22\textrm{-IV};\{A\}}^{\{s_{1}\}}(k,k^{\prime};\boldzero,\omega)
\Bigr]\notag\\
&-\tanh \dfrac{\epsilon+\omega}{2T}
g_{3;aABb}^{ss_{1}s_{2}s}(k;\omega)
g_{1;CcdD}^{ss_{1}s_{2}s}(k^{\prime};\omega)
\Bigl[
\Gamma_{31\textrm{-II};\{A\}}^{\{s_{1}\}}(k,k^{\prime};\boldzero,\omega)
-\Gamma_{31\textrm{-I};\{A\}}^{\{s_{1}\}}(k,k^{\prime};\boldzero,\omega)
\Bigr]\notag\\
&+\tanh \dfrac{\epsilon}{2T}
g_{1;aABb}^{ss_{1}s_{2}s}(k;\omega)
g_{3;CcdD}^{s_{3}s^{\prime\prime}s^{\prime\prime}s_{4}}(k^{\prime};\omega)
\Bigl[
\Gamma_{13\textrm{-I};\{A\}}^{\{s_{1}\}}(k,k^{\prime};\boldzero,\omega)
-\Gamma_{13\textrm{-II};\{A\}}^{\{s_{1}\}}(k,k^{\prime};\boldzero,\omega)
\Bigr]\Bigr\}\notag\\
&=-
\frac{1}{N}
\int^{\infty}_{-\infty}\frac{d\epsilon}{4\pi i}
\int^{\infty}_{-\infty}\frac{d\epsilon^{\prime}}{4\pi i}
\sum\limits_{l=1}^{3}
T_{l}(\epsilon,\omega)
\sum\limits_{\{A\}}
\sum\limits_{\{s_{1}\}}
g_{l;aABb}^{ss_{1}s_{2}s}(k;\omega)
\sum\limits_{l^{\prime}=1}^{3}
\mathcal{J}_{ll^{\prime};\{A\}}^{\{s_{1}\}}(k,k^{\prime};\omega)
g_{l^{\prime};CcdD}^{s_{3}s^{\prime\prime}s^{\prime\prime}s_{4}}(k^{\prime};\omega).
\label{eq:2nd-analytic}
\end{align}
\end{widetext}
In Eq. (\ref{eq:2nd-analytic}), 
the contour $C$ is the same as that used for the first term of Eq. (\ref{eq:K}), 
the contour $C^{\prime}$ consists of 
the five parts into which 
the circle is divided by inserting four horizontal lines, 
$\textrm{Im}(\epsilon^{\prime}-i\epsilon_{m})=0$, 
$\textrm{Im}\epsilon^{\prime}=0$, $\textrm{Im}(\epsilon^{\prime}+i\Omega_{n})=0$, 
and $\textrm{Im}(\epsilon^{\prime}+i\epsilon_{m}+i\Omega_{n})=0$, 
without the terms corresponding to $\textrm{Im}\epsilon^{\prime}=\epsilon_{m}$ 
and $\textrm{Im}\epsilon^{\prime}=-\epsilon_{m}-\Omega_{n}$, 
$P$ represents taking the Cauchy principal value of the integral,  
and $\mathcal{J}_{ll^{\prime};\{A\}}^{\{s_{1}\}}(k,k^{\prime};\omega)$ 
is 
connected with the reducible four-point vertex functions 
in real-frequency representation as follows: 
\begin{widetext}
\begin{align}
\hspace{-5pt}
\mathcal{J}_{11;\{a\}}^{\{s_{1}\}}(k,k^{\prime};\omega)
&=\
\tanh \frac{\epsilon^{\prime}}{2T} 
\Gamma_{11\textrm{-I};\{a\}}^{\{s_{1}\}}(k,k^{\prime};\boldzero,\omega)
+\coth \frac{\epsilon^{\prime}-\epsilon}{2T} 
\Bigl[\Gamma_{11\textrm{-II};\{a\}}^{\{s_{1}\}}(k,k^{\prime};\boldzero,\omega)-
\Gamma_{11\textrm{-I};\{a\}}^{\{s_{1}\}}(k,k^{\prime};\boldzero,\omega)\Bigr],\label{eq:4VC-11}\\
\hspace{-5pt}
\mathcal{J}_{12;\{a\}}^{\{s_{1}\}}(k,k^{\prime};\omega)
&=\
\Bigl(\tanh \frac{\epsilon^{\prime}+\omega}{2T} 
-\tanh \frac{\epsilon^{\prime}}{2T}\Bigr)
\Gamma_{12;\{a\}}^{\{s_{1}\}}(k,k^{\prime};\boldzero,\omega),\label{eq:4VC-12}\\
\hspace{-5pt}
\mathcal{J}_{13;\{a\}}^{\{s_{1}\}}(k,k^{\prime};\omega)
&=
-\tanh \frac{\epsilon^{\prime}+\omega}{2T} 
\Gamma_{13\textrm{-I};\{a\}}^{\{s_{1}\}}(k,k^{\prime};\boldzero,\omega)
-\coth \frac{\epsilon+\epsilon^{\prime}+\omega}{2T} 
\Bigl[\Gamma_{13\textrm{-II};\{a\}}^{\{s_{1}\}}(k,k^{\prime};\boldzero,\omega)-
\Gamma_{13\textrm{-I};\{a\}}^{\{s_{1}\}}(k,k^{\prime};\boldzero,\omega)\Bigr],\label{eq:4VC-13}\\
\hspace{-5pt}
\mathcal{J}_{21;\{a\}}^{\{s_{1}\}}(k,k^{\prime};\omega)
&=\
\tanh \frac{\epsilon^{\prime}}{2T} 
\Gamma_{21;\{a\}}^{\{s_{1}\}}(k,k^{\prime};\boldzero,\omega),\label{eq:4VC-21}\\
\hspace{-5pt}
\mathcal{J}_{22;\{a\}}^{\{s_{1}\}}(k,k^{\prime};\omega)
&=\
\coth \frac{\epsilon^{\prime}-\epsilon}{2T}
\Bigl[
\Gamma_{22\textrm{-II};\{a\}}^{\{s_{1}\}}(k,k^{\prime};\boldzero,\omega)
-\Gamma_{22\textrm{-III};\{a\}}^{\{s_{1}\}}(k,k^{\prime};\boldzero,\omega)
\Bigr]
-\tanh \frac{\epsilon^{\prime}}{2T}
\Gamma_{22\textrm{-II};\{a\}}^{\{s_{1}\}}(k,k^{\prime};\boldzero,\omega)\notag\\
\hspace{-5pt}
+&
\coth \frac{\epsilon^{\prime}+\epsilon+\omega}{2T}
\Bigl[
\Gamma_{22\textrm{-III};\{a\}}^{\{s_{1}\}}(k,k^{\prime};\boldzero,\omega)
-
\Gamma_{22\textrm{-IV};\{a\}}^{\{s_{1}\}}(k,k^{\prime};\boldzero,\omega)
\Bigr]
+\tanh \frac{\epsilon^{\prime}+\omega}{2T}
\Gamma_{22\textrm{-IV};\{a\}}^{\{s_{1}\}}(k,k^{\prime};\boldzero,\omega),\label{eq:4VC-22}\\
\hspace{-5pt}
\mathcal{J}_{23;\{a\}}^{\{s_{1}\}}(k,k^{\prime};\omega)
&=
-\tanh \frac{\epsilon^{\prime}+\omega}{2T} 
\Gamma_{23;\{a\}}^{\{s_{1}\}}(k,k^{\prime};\boldzero,\omega),\label{eq:4VC-23}\\
\hspace{-5pt}
\mathcal{J}_{31;\{a\}}^{\{s_{1}\}}(k,k^{\prime};\omega)
&=\
\tanh \frac{\epsilon^{\prime}}{2T} 
\Gamma_{31\textrm{-I};\{a\}}^{\{s_{1}\}}(k,k^{\prime};\boldzero,\omega)
+\coth \frac{\epsilon+\epsilon^{\prime}+\omega}{2T} 
\Bigl[\Gamma_{31\textrm{-II};\{a\}}^{\{s_{1}\}}(k,k^{\prime};\boldzero,\omega)-
\Gamma_{31\textrm{-I};\{a\}}^{\{s_{1}\}}(k,k^{\prime};\boldzero,\omega)\Bigr],\label{eq:4VC-31}\\
\hspace{-5pt}
\mathcal{J}_{32;\{a\}}^{\{s_{1}\}}(k,k^{\prime};\omega)
&=\
\Bigl(\tanh \frac{\epsilon^{\prime}+\omega}{2T} 
-\tanh \frac{\epsilon^{\prime}}{2T}\Bigr)
\Gamma_{32;\{a\}}^{\{s_{1}\}}(k,k^{\prime};\boldzero,\omega),\label{eq:4VC-32}
\end{align}
and 
\begin{align}
\hspace{-18pt}
\mathcal{J}_{33;\{a\}}^{\{s_{1}\}}(k,k^{\prime};\omega)
=&
-\tanh \frac{\epsilon^{\prime}+\omega}{2T} 
\Gamma_{33\textrm{-I};\{a\}}^{\{s_{1}\}}(k,k^{\prime};\omega)
-\coth \frac{\epsilon^{\prime}-\epsilon}{2T} 
\Bigl[\Gamma_{33\textrm{-II};\{a\}}^{\{s_{1}\}}(k,k^{\prime};\omega)-
\Gamma_{33\textrm{-I};\{a\}}^{\{s_{1}\}}(k,k^{\prime};\omega)\Bigr],\label{eq:4VC-33}
\end{align}
\end{widetext}
where the subscript $X$ in $\Gamma_{X;\{a\}}^{\{s_{1}\}}(k,k^{\prime};\boldzero,\omega)$ 
represents the inequalities about $\epsilon$, $\epsilon^{\prime}$, 
and $\omega$ of the reducible four-point vertex functions 
in real-frequency representation: 
$\Gamma_{X;\{a\}}^{\{s_{1}\}}(k,k^{\prime};\boldzero,\omega)$ 
for $X=$ 11-I, 11-II, 21, 31-II, 31-I, 32, 33-I, 
33-II, 23, 13-II, 13-I, 12, 22-III, 22-II, 22-I, and 22-IV 
satisfy, respectively, 
\begin{widetext}
\begin{align}
&\textrm{Im}\epsilon>0,\ 
\textrm{Im}\epsilon+\textrm{Im}\omega>0,\
\textrm{Im}\epsilon^{\prime}>0,\
\textrm{Im}\epsilon^{\prime}+\textrm{Im}\omega>0,\
\textrm{Im}\epsilon+\textrm{Im}\epsilon^{\prime}+\textrm{Im}\omega>0,\
\textrm{Im}\epsilon-\textrm{Im}\epsilon^{\prime}>0,\\
&\textrm{Im}\epsilon>0,\ 
\textrm{Im}\epsilon+\textrm{Im}\omega>0,\
\textrm{Im}\epsilon^{\prime}>0,\
\textrm{Im}\epsilon^{\prime}+\textrm{Im}\omega>0,\
\textrm{Im}\epsilon+\textrm{Im}\epsilon^{\prime}+\textrm{Im}\omega>0,\
\textrm{Im}\epsilon-\textrm{Im}\epsilon^{\prime}<0,\\
&\textrm{Im}\epsilon<0,\
\textrm{Im}\epsilon+\textrm{Im}\omega>0,\
\textrm{Im}\epsilon^{\prime}>0,\
\textrm{Im}\epsilon^{\prime}+\textrm{Im}\omega>0,\
\textrm{Im}\epsilon+\textrm{Im}\epsilon^{\prime}+\textrm{Im}\omega>0,\
\textrm{Im}\epsilon-\textrm{Im}\epsilon^{\prime}<0,\\
&\textrm{Im}\epsilon<0,\
\textrm{Im}\epsilon+\textrm{Im}\omega<0,\
\textrm{Im}\epsilon^{\prime}>0,\
\textrm{Im}\epsilon^{\prime}+\textrm{Im}\omega>0,\
\textrm{Im}\epsilon+\textrm{Im}\epsilon^{\prime}+\textrm{Im}\omega>0,\
\textrm{Im}\epsilon-\textrm{Im}\epsilon^{\prime}<0,\\
&\textrm{Im}\epsilon<0,\
\textrm{Im}\epsilon+\textrm{Im}\omega<0,\
\textrm{Im}\epsilon^{\prime}>0,\
\textrm{Im}\epsilon^{\prime}+\textrm{Im}\omega>0,\
\textrm{Im}\epsilon+\textrm{Im}\epsilon^{\prime}+\textrm{Im}\omega<0,\
\textrm{Im}\epsilon-\textrm{Im}\epsilon^{\prime}<0,\\
&\textrm{Im}\epsilon<0,\
\textrm{Im}\epsilon+\textrm{Im}\omega<0,\
\textrm{Im}\epsilon^{\prime}<0,\
\textrm{Im}\epsilon^{\prime}+\textrm{Im}\omega>0,\
\textrm{Im}\epsilon+\textrm{Im}\epsilon^{\prime}+\textrm{Im}\omega<0,\
\textrm{Im}\epsilon-\textrm{Im}\epsilon^{\prime}<0,\\
&\textrm{Im}\epsilon<0,\
\textrm{Im}\epsilon+\textrm{Im}\omega<0,\
\textrm{Im}\epsilon^{\prime}<0,\
\textrm{Im}\epsilon^{\prime}+\textrm{Im}\omega<0,\
\textrm{Im}\epsilon+\textrm{Im}\epsilon^{\prime}+\textrm{Im}\omega<0,\
\textrm{Im}\epsilon-\textrm{Im}\epsilon^{\prime}<0,\\
&\textrm{Im}\epsilon<0,\
\textrm{Im}\epsilon+\textrm{Im}\omega<0,\
\textrm{Im}\epsilon^{\prime}<0,\
\textrm{Im}\epsilon^{\prime}+\textrm{Im}\omega<0,\
\textrm{Im}\epsilon+\textrm{Im}\epsilon^{\prime}+\textrm{Im}\omega<0,\
\textrm{Im}\epsilon-\textrm{Im}\epsilon^{\prime}>0,\\
&\textrm{Im}\epsilon<0,\
\textrm{Im}\epsilon+\textrm{Im}\omega>0,\
\textrm{Im}\epsilon^{\prime}<0,\
\textrm{Im}\epsilon^{\prime}+\textrm{Im}\omega<0,\
\textrm{Im}\epsilon+\textrm{Im}\epsilon^{\prime}+\textrm{Im}\omega<0,\
\textrm{Im}\epsilon-\textrm{Im}\epsilon^{\prime}>0,\\
&\textrm{Im}\epsilon>0,\
\textrm{Im}\epsilon+\textrm{Im}\omega>0,\
\textrm{Im}\epsilon^{\prime}<0,\
\textrm{Im}\epsilon^{\prime}+\textrm{Im}\omega<0,\
\textrm{Im}\epsilon+\textrm{Im}\epsilon^{\prime}+\textrm{Im}\omega<0,\
\textrm{Im}\epsilon-\textrm{Im}\epsilon^{\prime}>0,\\
&\textrm{Im}\epsilon>0,\
\textrm{Im}\epsilon+\textrm{Im}\omega>0,\
\textrm{Im}\epsilon^{\prime}<0,\
\textrm{Im}\epsilon^{\prime}+\textrm{Im}\omega<0,\
\textrm{Im}\epsilon+\textrm{Im}\epsilon^{\prime}+\textrm{Im}\omega>0,\
\textrm{Im}\epsilon-\textrm{Im}\epsilon^{\prime}>0,\\
&\textrm{Im}\epsilon>0,\
\textrm{Im}\epsilon+\textrm{Im}\omega>0,\
\textrm{Im}\epsilon^{\prime}<0,\
\textrm{Im}\epsilon^{\prime}+\textrm{Im}\omega>0,\
\textrm{Im}\epsilon+\textrm{Im}\epsilon^{\prime}+\textrm{Im}\omega>0,\
\textrm{Im}\epsilon-\textrm{Im}\epsilon^{\prime}>0,\\
&\textrm{Im}\epsilon<0,\
\textrm{Im}\epsilon+\textrm{Im}\omega>0,\
\textrm{Im}\epsilon^{\prime}<0,\
\textrm{Im}\epsilon^{\prime}+\textrm{Im}\omega>0,\
\textrm{Im}\epsilon+\textrm{Im}\epsilon^{\prime}+\textrm{Im}\omega>0,\
\textrm{Im}\epsilon-\textrm{Im}\epsilon^{\prime}>0,\\
&\textrm{Im}\epsilon<0,\
\textrm{Im}\epsilon+\textrm{Im}\omega>0,\
\textrm{Im}\epsilon^{\prime}<0,\
\textrm{Im}\epsilon^{\prime}+\textrm{Im}\omega>0,\
\textrm{Im}\epsilon+\textrm{Im}\epsilon^{\prime}+\textrm{Im}\omega>0,\
\textrm{Im}\epsilon-\textrm{Im}\epsilon^{\prime}<0,\\
&\textrm{Im}\epsilon<0,\
\textrm{Im}\epsilon+\textrm{Im}\omega>0,\
\textrm{Im}\epsilon^{\prime}<0,\
\textrm{Im}\epsilon^{\prime}+\textrm{Im}\omega>0,\
\textrm{Im}\epsilon+\textrm{Im}\epsilon^{\prime}+\textrm{Im}\omega<0,\
\textrm{Im}\epsilon-\textrm{Im}\epsilon^{\prime}<0,
\end{align}
and
\begin{align}
&\textrm{Im}\epsilon<0,\
\textrm{Im}\epsilon+\textrm{Im}\omega>0,\
\textrm{Im}\epsilon^{\prime}<0,\
\textrm{Im}\epsilon^{\prime}+\textrm{Im}\omega>0,\
\textrm{Im}\epsilon+\textrm{Im}\epsilon^{\prime}+\textrm{Im}\omega<0,\
\textrm{Im}\epsilon-\textrm{Im}\epsilon^{\prime}>0.
\end{align} 
\end{widetext}
We also obtain 
the connections between 
$\mathcal{J}_{ll^{\prime};\{a\}}^{(1)\{s_{1}\}}(k,k^{\prime};\omega)$ 
and $\Gamma_{X;\{a\}}^{(1)\{s_{1}\}}(k,k^{\prime};\boldzero,\omega)$, 
the irreducible four-point vertex functions 
in real-frequency representation, 
by adding the superscript $(1)$ in both 
$\mathcal{J}_{ll^{\prime};\{a\}}^{\{s_{1}\}}(k,k^{\prime};\omega)$ 
and $\Gamma_{X;\{a\}}^{\{s_{1}\}}(k,k^{\prime};\boldzero,\omega)$ 
in Eqs. (\ref{eq:4VC-11}){--}(\ref{eq:4VC-33}). 

Combining Eqs. (\ref{eq:1st-analytic}) and (\ref{eq:2nd-analytic}) 
with Eqs. (\ref{eq:K-analytic}) and (\ref{eq:K}), 
we obtain Eq. (\ref{eq:K-4VC}). 

\section{Derivation of Eq. (\ref{eq:KC-real2})}
In this appendix, 
we derive Eq. (\ref{eq:KC-real2}). 
The derivation is in the following way: 
First, by using Eq. (\ref{eq:relation_L0-L}), 
we can rewrite the terms for $l=1$ and $3$ in Eq. (\ref{eq:KC-real}) as 
\begin{widetext}
\begin{align}
&-
\frac{(-e)^{2}}{2i}
\sum\limits_{k}
\sum\limits_{\{a\}}
\sum\limits_{\{s\}}
\delta_{s^{\prime},s}
(v_{\boldk x})_{ba}^{ss}
\sum\limits_{l=1,3}
T_{l}(\epsilon,\omega)
g_{l;acdb}^{ss^{\prime\prime}s^{\prime\prime\prime}s^{\prime}}(k;\omega)
\Lambda_{y;l;cd}^{\textrm{C};s^{\prime\prime}s^{\prime\prime\prime}}(k;\omega)\notag\\
=&-\frac{(-e)^{2}}{2i}
\sum\limits_{k}
\sum\limits_{\{a\}}
\sum\limits_{\{s\}}
\delta_{s^{\prime},s}
(v_{\boldk x})_{ba}^{ss}
\tanh \frac{\epsilon}{2T}
g_{1;acdb}^{ss^{\prime\prime}s^{\prime\prime\prime}s^{\prime}}(k;\omega)
\Lambda_{y;1;cd}^{\textrm{C}(0)s^{\prime\prime}s^{\prime\prime\prime}}(k;\omega)\notag\\
&-\frac{(-e)^{2}}{(2i)^{2}}
\sum\limits_{k}
\sum\limits_{\{a\}}
\sum\limits_{\{s\}}
\delta_{s^{\prime},s}
(v_{\boldk x})_{ba}^{ss}
\tanh \frac{\epsilon}{2T}
g_{1;acdb}^{ss^{\prime\prime}s^{\prime\prime\prime}s^{\prime}}(k;\omega)
\sum\limits_{k^{\prime}}
\sum\limits_{\{A\}}
\sum\limits_{\{s_{1}\}}
\mathcal{J}_{12;cdCD}^{(0)s^{\prime\prime}s^{\prime\prime\prime}s_{3}s_{4}}(k,k^{\prime};\omega)
g_{2;CABD}^{s_{3}s_{1}s_{2}s_{4}}(k^{\prime};\omega)
\Lambda_{y;2;AB}^{\textrm{C}; s_{1}s_{2}}(k^{\prime};\omega)\notag\\
&+\frac{(-e)^{2}}{2i}
\sum\limits_{k}
\sum\limits_{\{a\}}
\sum\limits_{\{s\}}
\delta_{s^{\prime},s}
(v_{\boldk x})_{ba}^{ss}
\tanh \frac{\epsilon+\omega}{2T}
g_{3;acdb}^{ss^{\prime\prime}s^{\prime\prime\prime}s^{\prime}}(k;\omega)
\Lambda_{y;3;cd}^{\textrm{C}(0)s^{\prime\prime}s^{\prime\prime\prime}}(k;\omega)\notag\\
&+\frac{(-e)^{2}}{(2i)^{2}}
\sum\limits_{k}
\sum\limits_{\{a\}}
\sum\limits_{\{s\}}
\delta_{s^{\prime},s}
(v_{\boldk x})_{ba}^{ss}
\tanh \frac{\epsilon+\omega}{2T}
g_{3;acdb}^{ss^{\prime\prime}s^{\prime\prime\prime}s^{\prime}}(k;\omega)
\sum\limits_{k^{\prime}}
\sum\limits_{\{A\}}
\sum\limits_{\{s_{1}\}}
\mathcal{J}_{32;cdCD}^{(0)s^{\prime\prime}s^{\prime\prime\prime}s_{3}s_{4}}(k,k^{\prime};\omega)
g_{2;CABD}^{s_{3}s_{1}s_{2}s_{4}}(k^{\prime};\omega)\notag\\
&\times 
\Lambda_{\nu;2;AB}^{\textrm{C}; s_{1}s_{2}}(k^{\prime};\omega).\label{eq:rewrite-g1g3terms-pre}
\end{align}
\end{widetext}
Furthermore, 
the second and the fourth term in Eq. (\ref{eq:rewrite-g1g3terms-pre}) 
can be rewritten as, respectively, 
\begin{widetext}
\begin{align}
&\frac{(-e)^{2}}{4}
\sum\limits_{k}
\sum\limits_{k^{\prime}}
\sum\limits_{\{a\}}
\sum\limits_{\{A\}}
\sum\limits_{\{s\}}
\sum\limits_{\{s_{1}\}}
\delta_{s^{\prime},s}
(v_{\boldk x})_{ba}^{ss}
\tanh \frac{\epsilon}{2T}
g_{1;acdb}^{ss^{\prime\prime}s^{\prime\prime\prime}s^{\prime}}(k;\omega)
\mathcal{J}_{12;cdCD}^{(0)s^{\prime\prime}s^{\prime\prime\prime}s_{3}s_{4}}(k,k^{\prime};\omega)
g_{2;CABD}^{s_{3}s_{1}s_{2}s_{4}}(k^{\prime};\omega)
\Lambda_{y;2;AB}^{\textrm{C}; s_{1}s_{2}}(k^{\prime};\omega)\notag\\
=&
\frac{(-e)^{2}}{4}
\sum\limits_{k^{\prime}}
\sum\limits_{k}
\sum\limits_{\{a\}}
\sum\limits_{\{A\}}
\sum\limits_{\{s\}}
\sum\limits_{\{s_{1}\}}
\delta_{s^{\prime},s}
(v_{\boldk^{\prime} x})_{ba}^{ss}
\tanh \frac{\epsilon^{\prime}}{2T}
g_{1;acdb}^{ss^{\prime\prime}s^{\prime\prime\prime}s^{\prime}}(k^{\prime};\omega)
\Bigl(\tanh \frac{\epsilon+\omega}{2T} 
-\tanh \frac{\epsilon}{2T}\Bigr)
\Gamma_{12;cdCD}^{(0)s^{\prime\prime}s^{\prime\prime\prime}s_{3}s_{4}}
(k^{\prime},k;\boldzero,\omega)\notag\\
&\times 
g_{2;CABD}^{s_{3}s_{1}s_{2}s_{4}}(k;\omega)
\Lambda_{y;2;AB}^{\textrm{C}; s_{1}s_{2}}(k;\omega)\notag\\
=&
\frac{(-e)^{2}}{4}
\sum\limits_{k}
\sum\limits_{k^{\prime}}
\sum\limits_{\{a\}}
\sum\limits_{\{A\}}
\sum\limits_{\{s\}}
\sum\limits_{\{s_{1}\}}
\mathcal{J}_{21;baCD}^{(0)s^{\prime}ss_{3}s_{4}}
(\boldk,\epsilon,\boldk,\epsilon+\omega,
\boldk^{\prime},\epsilon^{\prime},\boldk^{\prime},\epsilon^{\prime}+\omega)
g_{1;CABD}^{s_{3}s_{1}s_{2}s_{4}}
(\boldk^{\prime},\epsilon^{\prime},\boldk^{\prime},\epsilon^{\prime}+\omega)
\delta_{s_{1},s_{2}}
(v_{\boldk^{\prime} x})_{AB}^{s_{1}s_{1}}\notag\\
&\times 
\Bigl(\tanh \frac{\epsilon+\omega}{2T} 
-\tanh \frac{\epsilon}{2T}\Bigr)
g_{2;acdb}^{ss^{\prime\prime}s^{\prime\prime\prime}s^{\prime}}(k;\omega)
\Lambda_{y;2;cd}^{\textrm{C}; s^{\prime\prime}s^{\prime\prime\prime}}(k;\omega),
\label{eq:rewrite-g1term-2nd}
\end{align}
and 
\begin{align}
&-\frac{(-e)^{2}}{4}
\sum\limits_{k}
\sum\limits_{k^{\prime}}
\sum\limits_{\{a\}}
\sum\limits_{\{A\}}
\sum\limits_{\{s\}}
\sum\limits_{\{s_{1}\}}
\delta_{s^{\prime},s}
(v_{\boldk x})_{ba}^{ss}
\tanh \frac{\epsilon+\omega}{2T}
g_{3;acdb}^{ss^{\prime\prime}s^{\prime\prime\prime}s^{\prime}}(k;\omega)
\mathcal{J}_{32;cdCD}^{(0)s^{\prime\prime}s^{\prime\prime\prime}s_{3}s_{4}}(k,k^{\prime};\omega)
g_{2;CABD}^{s_{3}s_{1}s_{2}s_{4}}(k^{\prime};\omega)
\Lambda_{y;2;AB}^{\textrm{C}; s_{1}s_{2}}(k^{\prime};\omega)\notag\\
=&
-\frac{(-e)^{2}}{4}
\sum\limits_{k^{\prime}}
\sum\limits_{k}
\sum\limits_{\{a\}}
\sum\limits_{\{A\}}
\sum\limits_{\{s\}}
\sum\limits_{\{s_{1}\}}
\delta_{s^{\prime},s}
(v_{\boldk^{\prime} x})_{ba}^{ss}
\tanh \frac{\epsilon^{\prime}+\omega}{2T}
g_{3;acdb}^{ss^{\prime\prime}s^{\prime\prime\prime}s^{\prime}}(k^{\prime};\omega)
\Bigl(\tanh \frac{\epsilon+\omega}{2T} 
-\tanh \frac{\epsilon}{2T}\Bigr)
\Gamma_{32;cdCD}^{(0)s^{\prime\prime}s^{\prime\prime\prime}s_{3}s_{4}}
(k^{\prime},k;\boldzero,\omega)\notag\\
&\times 
g_{2;CABD}^{s_{3}s_{1}s_{2}s_{4}}(k;\omega)
\Lambda_{y;2;AB}^{\textrm{C}; s_{1}s_{2}}(k;\omega)\notag\\
=&
\frac{(-e)^{2}}{4}
\sum\limits_{k}
\sum\limits_{k^{\prime}}
\sum\limits_{\{a\}}
\sum\limits_{\{A\}}
\sum\limits_{\{s\}}
\sum\limits_{\{s_{1}\}}
\mathcal{J}_{23;baCD}^{(0)s^{\prime}ss_{3}s_{4}}
(\boldk,\epsilon,\boldk,\epsilon+\omega,
\boldk^{\prime},\epsilon^{\prime},\boldk^{\prime},\epsilon^{\prime}+\omega)
g_{3;CABD}^{s_{3}s_{1}s_{2}s_{4}}
(\boldk^{\prime},\epsilon^{\prime},\boldk^{\prime},\epsilon^{\prime}+\omega)
\delta_{s_{1},s_{2}}
(v_{\boldk^{\prime} x})_{AB}^{s_{1}s_{1}}\notag\\
&\times 
\Bigl(\tanh \frac{\epsilon+\omega}{2T} 
-\tanh \frac{\epsilon}{2T}\Bigr)
g_{2;acdb}^{ss^{\prime\prime}s^{\prime\prime\prime}s^{\prime}}(k;\omega)
\Lambda_{y;2;cd}^{\textrm{C}; s^{\prime\prime}s^{\prime\prime\prime}}(k;\omega).
\label{eq:rewrite-g3term-2nd}
\end{align}
\end{widetext}
In deriving Eq. (\ref{eq:rewrite-g1term-2nd}), 
we have used Eqs. (\ref{eq:4VC-12}) and (\ref{eq:4VC-21}) 
and an identity~\cite{Eliashberg,NA-full},  
\begin{align}
&\Gamma_{12;abcd}^{(0)ss^{\prime}s^{\prime\prime}s^{\prime\prime\prime}}
(\boldk^{\prime},\epsilon^{\prime}+\omega,\boldk^{\prime},\epsilon^{\prime},
\boldk,\epsilon+\omega,\boldk,\epsilon)\notag\\
=\
&
\Gamma_{21;dcba}^{(0)s^{\prime\prime\prime}s^{\prime\prime}s^{\prime}s}
(\boldk,\epsilon,\boldk,\epsilon+\omega,
\boldk^{\prime},\epsilon^{\prime},\boldk^{\prime},\epsilon^{\prime}+\omega).
\end{align} 
In addition, 
to derive Eq. (\ref{eq:rewrite-g3term-2nd}), 
we have used Eqs. (\ref{eq:4VC-23}) and (\ref{eq:4VC-32}) 
and another identity~\cite{Eliashberg,NA-full},
\begin{align}
&\Gamma_{32;abcd}^{(0)ss^{\prime}s^{\prime\prime}s^{\prime\prime\prime}}
(\boldk^{\prime},\epsilon^{\prime}+\omega,\boldk^{\prime},\epsilon^{\prime},
\boldk,\epsilon+\omega,\boldk,\epsilon)\notag\\
=\ 
&
\Gamma_{23;dcba}^{(0)s^{\prime\prime\prime}s^{\prime\prime}s^{\prime}s}
(\boldk,\epsilon,\boldk,\epsilon+\omega,
\boldk^{\prime},\epsilon^{\prime},\boldk^{\prime},\epsilon^{\prime}+\omega).
\end{align}
Returning Eqs. (\ref{eq:rewrite-g1term-2nd}) and (\ref{eq:rewrite-g3term-2nd}) 
to Eq. (\ref{eq:rewrite-g1g3terms-pre}), 
we obtain
\begin{widetext}
\begin{align}
&-\frac{(-e)^{2}}{2i}
\sum\limits_{k}
\sum\limits_{\{a\}}
\sum\limits_{\{s\}}
\delta_{s^{\prime},s}
(v_{\boldk x})_{ba}^{ss}
\sum\limits_{l=1,3}
T_{l}(\epsilon,\omega)
g_{l;acdb}^{ss^{\prime\prime}s^{\prime\prime\prime}s^{\prime}}(k;\omega)
\Lambda_{y;l;cd}^{\textrm{C};s^{\prime\prime}s^{\prime\prime\prime}}(k^{\prime};\omega)\notag\\
=&-\frac{(-e)^{2}}{2i}
\sum\limits_{k}
\sum\limits_{\{a\}}
\sum\limits_{\{s\}}
\delta_{s^{\prime},s}
(v_{\boldk x})_{ba}^{ss}
\sum\limits_{l=1,3}
T_{l}(\epsilon,\omega)
g_{l;acdb}^{ss^{\prime\prime}s^{\prime\prime\prime}s^{\prime}}(k;\omega)
\Lambda_{y;l;cd}^{\textrm{C}(0)s^{\prime\prime}s^{\prime\prime\prime}}(k^{\prime};\omega)\notag\\
&-
\frac{(-e)^{2}}{2i}
\sum\limits_{k}
\sum\limits_{\{a\}}
\sum\limits_{\{s\}}
\Bigl[\frac{1}{2i}
\sum\limits_{k^{\prime}}
\sum\limits_{\{A\}}
\sum\limits_{\{s_{1}\}}
\sum\limits_{l^{\prime}=1,3}
\mathcal{J}_{2l^{\prime};baCD}^{(0)s^{\prime}ss_{3}s_{4}}
(\boldk,\epsilon,\boldk,\epsilon+\omega,
\boldk^{\prime},\epsilon^{\prime},\boldk^{\prime},\epsilon^{\prime}+\omega)
g_{l^{\prime};CABD}^{s_{3}s_{1}s_{2}s_{4}}
(\boldk^{\prime},\epsilon^{\prime},\boldk^{\prime},\epsilon^{\prime}+\omega)
\notag\\
&\ \ \ \ \ \ \ \ \ \ \ \ \ \ \ \ \ \ \ \ \ \ \ \ \ \ \ \times 
\delta_{s_{1},s_{2}}
(v_{\boldk^{\prime} x})_{AB}^{s_{1}s_{1}}\Bigr]
T_{2}(\epsilon,\omega)
g_{2;acdb}^{ss^{\prime\prime}s^{\prime\prime\prime}s^{\prime}}(k;\omega)
\Lambda_{y;2;cd}^{\textrm{C}; s^{\prime\prime}s^{\prime\prime\prime}}(k;\omega)\notag\\
=&-\frac{(-e)^{2}}{2i}
\sum\limits_{k}
\sum\limits_{\{a\}}
\sum\limits_{\{s\}}
\delta_{s^{\prime},s}
(v_{\boldk x})_{ba}^{ss}
\sum\limits_{l=1,3}
T_{l}(\epsilon,\omega)
g_{l;acdb}^{ss^{\prime\prime}s^{\prime\prime\prime}s^{\prime}}(k;\omega)
\Lambda_{y;l;cd}^{\textrm{C}(0)s^{\prime\prime}s^{\prime\prime\prime}}(k^{\prime};\omega)\notag\\
&-
\frac{(-e)^{2}}{2i}
\sum\limits_{k}
\sum\limits_{\{a\}}
\sum\limits_{\{s\}}
\Bigl[
\Lambda_{x;2;ba}^{\textrm{C}(0)s^{\prime}s}(\boldk,\epsilon,\boldk,\epsilon+\omega)
-\delta_{s^{\prime},s}(v_{\boldk x})_{ba}^{ss}
\Bigr]
T_{2}(\epsilon,\omega)
g_{2;acdb}^{ss^{\prime\prime}s^{\prime\prime\prime}s^{\prime}}(k;\omega)
\Lambda_{y;2;cd}^{\textrm{C}; s^{\prime\prime}s^{\prime\prime\prime}}(k;\omega).\label{eq:rewrite-g1g3term}
\end{align}
\end{widetext}
Then, 
combining Eq. (\ref{eq:rewrite-g1g3term}) with Eq. (\ref{eq:KC-real}), 
we obtain Eq. (\ref{eq:KC-real2}). 
This is another exact expression of $\tilde{K}_{xy}^{\textrm{C}(\textrm{R})}(\omega)$. 

\begin{acknowledgments}
I thank E. Saitoh 
for a useful comment on a draft, 
Y. Yanase, H. Kontani, H. Kohno, and H. Fukuyama 
for valuable discussions and several useful comments, 
and H. Kurebayashi for a critical comment on the title. 
\end{acknowledgments}


\begin{thebibliography}{99}
\bibitem{AHE-exp-review}
E. M. Pugh and N. Rostoker, 
Rev. Mod. Phys. \textbf{25}, 151 (1953).

\bibitem{Karplus-Luttinger}
R. Karplus and J. M. Luttinger, 
Phys. Rev. \textbf{95}, 1154 (1954).

\bibitem{Smit}
J. Smit, 
Physica \textbf{21}, 877 (1955).

\bibitem{AnomalousNernst}
T. Miyasato \textit{et al.},  
Phys. Rev. Lett. \textbf{99}, 086602 (2007).

\bibitem{AHE-review-Nagaosa}
N. Nagaosa, J. Sinova, S. Onoda, 
A. H. MacDonald and N. P. Ong, 
Rev. Mod. Phys. \textbf{82}, 1539 (2010).

\bibitem{Dyakonov-Perel}
M. I. Dyakonov and V. I. Perel, 
Phys. Lett. A \textbf{35}, 459 (1971).  

\bibitem{Hirsch-SHE}
J. E. Hirsch,  
Phys. Rev. Lett. \textbf{83}, 1834 (1999).

\bibitem{Murakami-SHE}
S. Murakami, N. Nagaosa and S.-C. Zhang, 
Science \textbf{301}, 1348 (2003).

\bibitem{Sinova-SHE}
J. Sinova \textit{et al.}, 
Phys. Rev. Lett. \textbf{92}, 126603 (2004).

\bibitem{Kato-SHE}
Y. K. Kato, R. C. Myers, A. C. Gossard and D. D. Awschalom, 
Science \textbf{306} 1910 (2004).

\bibitem{Saitoh-SHE}
E. Saitoh, M. Ueda, H. Miyajima and G. Tatara, 
Appl. Phys. Lett. \textbf{88}, 182509 (2006).

\bibitem{Tinkham-SHE}
S. O. Valenzuela and M. Tinkham, 
Nature \textbf{442}, 176 (2006).

\bibitem{SHE-review-Sinova}
J. Sinova, S. O. Valenzuela, J. Wunderlich, C. H. Back and T. Jungwirth, 
arXiv:1411.3249.

\bibitem{Ashcroft-Mermin}
N. W. Ashcroft and N. D. Mermin,  
\textit{Solid State Physics} 
(Thomson Learning, Inc., New York, 1976).

\bibitem{SHE-review-Science}
S. A. Wolf \textit{et al.},  
Science \textbf{294}, 1488 (2001). 

\bibitem{SHE-review-Nature}
T. Jungwirth, J. Wunderlich and K. Olenjn\'{i}k, 
Nature Mater. \textbf{11}, 382 (2012).

\bibitem{Kontani-AHE}
H. Kontani, T. Tanaka and K. Yamada, 
Phys. Rev. B \textbf{75}, 184416 (2007).

\bibitem{SrRuO3-AHE}
Z. Fang \textit{et al.}, 
Science \textbf{302}, 92 (2003).

\bibitem{Kontani-SHE}
T. Tanaka \textit{et al.}, 
Phys. Rev. B \textbf{77}, 165117 (2008). 

\bibitem{SHE-Pt-Nagaosa}
G. Y. Guo, S. Murakami, T.-W. Chen and N. Nagaosa, 
Phys. Rev. Lett. \textbf{100}, 096401 (2008).

\bibitem{Mizoguchi-SHE}
T. Mizoguchi and N. Arakawa, 
arXiv:1411.5432. 

\bibitem{Berger}
L. Berger, 
Phys. Rev. B \textbf{2}, 4559 (1970).

\bibitem{Extrinsic}
A. Cr\'{e}pieux and P. Bruno, 
Phys. Rev. B \textbf{64}, 014416 (2001).

\bibitem{JJSakurai}
J. J. Sakurai, 
\textit{Modern Quantum Mechanics} 
(Benjamin/Cummings, Menlo park, 1985). 

\bibitem{Kontani-OrbitalAB}
H. Kontani, T. Tanaka, D. S. Hirashima, 
K. Yamada and J. Inoue, 
Phys. Rev. Lett. \textbf{100}, 096601 (2008).

\bibitem{Mizoguchi-CVC}
T. Mizoguchi, 
Master thesis, The University of Tokyo (2013); T. Mizoguchi and N. Arakawa, 
in preparation. 

\bibitem{imp-1stprinciple}
K. Nakamura, R. Arita and H. Ikeda, 
Phys. Rev. B \textbf{83}, 144512 (2011).

\bibitem{SHE-sytematic-exp}
M. Morota \textit{et al.}, 
Phys. Rev. B \textbf{83}, 174405 (2011).

\bibitem{Kimura-SHE}
T. Kimura, Y. Otani, T. Sato, S. Takahashi, and S. Maekawa, 
Phys. Rev. Lett \textbf{98}, 156601 (2007).

\bibitem{Streda}
P. St\v{r}eda, 
J.Phys.C: Solid State Phys. \textbf{15}, L717 (1982).

\bibitem{Onoda-Nagaosa}
M. Onoda and N. Nagaosa, 
J. Phys. Soc. Jpn. \textbf{71}, 19 (2002).

\bibitem{Haldane-AHE}
F. D. M. Haldane, 
Phys. Rev. Lett. \textbf{93}, 206602 (2004).

\bibitem{SCD-first}
I. D'Amico and G. Vignale, 
Phys. Rev. B \textbf{62}, 4853 (2000).

\bibitem{SCD-review}
E. M. Hankiewicz and G. Vignale, 
J.Phys.: Condens. Matter \textbf{21}, 253202 (2009).

\bibitem{Ziman}
J. M. Ziman, 
\textit{Principles of the Theory of Solids} 
(Cambridge University Press, Cambridge, 1979). 

\bibitem{Ziman2}
J. M. Ziman, 
\textit{Electrons and Phonons} 
(Oxford University Press, New York, 1960). 

\bibitem{Yamada-Yosida}
K. Yamada and K. Yosida, 
Prog. Theor. Phys. \textbf{76}, 621 (1986).

\bibitem{SCD-exp}
C. P. Weber \textit{et al.}, 
Nature \textbf{437}, 1330 (2005).

\bibitem{Kubo-formula}
R. Kubo, 
J. Phys. Soc. Jpn. \textbf{12}, 570 (1957).

\bibitem{MultiHubbard-Yanase}
Y. Yanase and M. Ogata, 
J. Phys. Soc. Jpn. \textbf{72}, 673 (2003).

\bibitem{Yanase-Rashba} 
Y. Yanase, 
J. Phys. Soc. Jpn. \textbf{82}, 044711 (2013).

\bibitem{AGD}
A. A. Abrikosov, L. P. Gor'kov and I. E. Dyaloshinski, 
\textit{Methods of Quantum Field Theory in Statistical Physics}
(Dover Publications, New York, 1963).

\bibitem{Kontani-review}
H. Kontani, 
Rep. Prog. Phys. \textbf{71}, 026501 (2008).

\bibitem{Takada-text-old}
Y. Takada, Tataimondai [in Japanese] 
(Asakura shoten, Tokyo, 1999).

\bibitem{JS-UsusalDef-Niu}
G. Y. Guo, Y. Yao and Q. Niu, 
Phys. Rev. Lett. \textbf{94}, 226601 (2005).

\bibitem{NA-review}
N. Arakawa, 
Mod. Phys. Lett. B \textbf{29}, 1530005 (2015).

\bibitem{Baym-Kadanoff}
G. Baym and L. P. Kadanoff, 
Phys. Rev. \textbf{124}, 287 (1961).


\bibitem{Fukuyama}
H. Fukuyama, H. Ebisawa and Y. Wada, 
Prog. Theor. Phys. \textbf{42}, 494 (1969).

\bibitem{Eliashberg} 
G. M. $\acute{\textrm{E}}$liashberg, 
Zh. Eksp. Teor. Fiz. \textbf{41}, 1241 (1962)
[Sov. Phys. JETP \textbf{14}, 886 (1962)]. 

\bibitem{Pines-Nozieres}
D. Pines and P. Nozi$\grave{\textrm{e}}$res, 
 \textit{Theory of Quantum Liquids} 
(Westview Press, Boulder, 1999).

\bibitem{Nozieres}
P. Nozi$\grave{\textrm{e}}$res,  
\textit{Theory of Interacting Fermi Systems} 
(Westview Press, Boulder, 1997). 

\bibitem{NA-full}
N. Arakawa, arXiv:1505.05274.

\bibitem{Takada-new}
Y. Takada, Tataimondai tokuron [in Japanese] 
(Asakura shoten, Tokyo, 2009).

\bibitem{Kohno-Yamada} 
H. Kohno and K. Yamada, 
Prog. Theor. Phys. \textbf{80}, 623 (1988).

\bibitem{Landau}
L. D. Landau, 
Zh. Eksp. Teor. Fiz. \textbf{30}, 1058 (1956)
[Sov. Phys. JETP \textbf{3}, 920 (1957)].

\bibitem{Sr2RuO4-APRES}
A. Damascelli \textit{et al.}, 
\textit{Phys. Rev. Lett.} \textbf{85}, 5194 (2000).

\bibitem{FMQCP-exp-QPspectrum}
A. Shimoyamada \textit{et al.}, 
\textit{Phys. Rev. Lett.} \textbf{102}, 086401 (2009).

\bibitem{Mott-exp-QPspectrum}
A. de la Torre \textit{et al.}, 
Phys. Rev. Lett. \textbf{113}, 256402 (2014).

\bibitem{Kuroda-Nagi}
Y. Kuroda and A. D. S. Nagi, 
Phys. Rev. \textbf{15}, 4460 (1977).

\bibitem{Hlubina-Rice}
R. Hlubina and T. M. Rice, 
Phys. Rev. B {\bf 51}, 9253 (1995).











\bibitem{Rashba} 
E. I. Rashba, 
Fiz. Tverd. Tela. \textbf{2}, 1224 (1960)
[Sov. Phys. Solid State \textbf{2}, 1109 (1960)].

\end{thebibliography}
\end{document}